\documentclass[floatfix,aps,prb,twocolumn,superscriptaddress,showpacs]{revtex4-1}
\usepackage{graphicx}
\usepackage{psfrag}
\usepackage{amsmath}
\usepackage{amssymb}
\usepackage{mathtools}
\usepackage{color}

\newcommand{\txtb}{\textcolor{black}}

\begin{document}


\title{Magnetic form factor, field map and field distribution for a BCS type-II superconductor near its $B_{\rm c2}(T)$ phase boundary}


\author{A. Maisuradze}
\affiliation{Physik-Institut der Universit\"at Z\"urich, Winterthuerstrasse 190, CH-8057 Z\"urich, Switzerland}
\affiliation{Laboratory for Muon-Spin Spectroscopy, Paul Scherrer Institute, 5232 Villigen-PSI, Switzerland}

\author {A.~Yaouanc}
\affiliation{Laboratory for Muon-Spin Spectroscopy, Paul Scherrer Institute, 5232 Villigen-PSI, Switzerland}
\affiliation{Institut Nanosciences et Cryog\'enie, SPSMS, CEA and University Joseph Fourier, F-38054 Grenoble, France}



\date{\today}

\begin{abstract}

We review the magnetic form factor deduced by Delrieu from the Gorkov's equation for a Bardeen-Cooper-Schrieffer (BCS) type-II
superconductor near its $B_{\rm c2}$ phase boundary, i.e. when its magnetization is small. A numerical study of the  form factor,
field map, and field distribution follows. The characteristics of the transition  from the low-temperature BCS to the high-temperature Ginzburg-Landau
vortex lattices is studied. The exotic shape of the component field distribution and the form factor at low
temperature and as a function of the external field intensity
are discussed. Our numerical work should be helpful for the analysing of small angle neutron scattering and muon spin rotation
vortex-lattice data  recorded for BCS superconductors and maybe other superconductors in the clean limit.

\end{abstract}

\pacs{74.25.Uv, 61.05.fg, 76.75.+i}

\maketitle

\section{Introduction}
\label{Introduction}

The bulk properties of a vortex lattice (VL) of a type II superconductor are studied experimentally, among other techniques,
by magnetization, small angle neutron scattering (SANS), muon spin rotation ($\mu$SR) and nuclear
magnetic resonance (NMR) measurements. To extract physical
information on the investigated compound modeling
of the VL magnetic properties is required. This is usually done using either the London or the
Ginzburg-Landau (GL) models.\cite{Tinkham96}
While the London model neglects the vortex cores altogether --- which is acceptable
for low fields only --- the GL model accounts for them.
Although the GL theory is strictly valid only near the superconducting critical
temperature at low field it turned out to be a good approximation for a number of
classical BCS superconductors.\cite{Landau07}
The GL model is usually found to provide a proper description of the VL thoughout the mixed phase for
unconventional superconductors also, assuming the London penetration depth and the Ginzburg-Landau
coherence length to be effective parameters.\cite{Sonier07}
This is further discussed by Landau and Keller.\cite{Landau07}

One of the interesting phenomena predicted yet in 1972 by Delrieu is the diffraction of the Cooper's
pairs on the periodic potential induced by the VL.\cite{Delrieu72}
He showed that for clean superconductors at low temperatures and fields close to the
upper critical field $B_{\rm c2}$ exotic behaviours of the VL may be observed due to
the Cooper's pair diffraction. The spatial field distribution around vortex cores obtains a conical shapes and
the positions of the minimal and saddle point fields interexchange. As a consequence the probability field
distribution shows a linear tail around the vortex core field. Nearly at the same time E. H. Brandt
came to the same conclusion based on a nonlocal theory of superconductivity.\cite{Brandt74p467} In the following
publications he presented analytical and numerical results for the nonlocal VL behaviour at fields close
to $B_{\rm c2}$ with an arbitrary impurity scattering,\cite{Brandt74}
and later for a broad range of fields.\cite{Brandt76p105}
Later on, an exact numerical solution of the Eilenberger's quasiclassical
equations by U. Klein allowed one to determine the microscopic structure of
the order parameter and magnetic field in the whole range of the applied fields.\cite{Klein87}

Writing a simplified Gorkov's\cite{Gorkov59} integral equation for the Green's function in terms of
a set of linear algebraic equations for the time and space Fourier components,
U.~Brandt {\it et al.} have been able to compute analytically the density of states
under high fields.\cite{Brandt67} Later on the magnetization\cite{Brandt69} as well as the field
distribution\cite{Delrieu72,Delrieu74} were also obtained by U.~Brandt {\it et al.} and Delrieu,
respectively, using the previously derived results for the
Green's function. The clean and dirty limits were considered by Delrieu in
his  PhD thesis.\cite{Delrieu74}

It is worth to note that the Cooper's pair diffraction is a property of clean superconductors.
Therefore, in most of high temperature superconductors these diffraction effects may be observable.

A VL field distribution which may exhibit a high-field linear tail  was reported from NMR measurement on
vanadium.\cite{Kung70} Later on the existence of the tail was confirmed by Herlach {\it et al.} for
niobium using $\mu$SR measurements.\cite{Herlach90}
They were performed at low temperature and for an external field $B_{\rm ext}$ relatively
close to $B_{\rm c2}$. The linear tail is qualitatively different from GL model expectation for which
a field cutoff should be present.\cite{Yaouanc97a}

The NMR and the $\mu$SR results seem to support Delrieu's predictions, in particular the linear tail.
SANS data may also be consistent with them.\cite{Delrieu72} However, after more than 40 years there
is still no definitive experimental observation of the predicted exotic VL at low temperature.
Probably the difficulty of reading the Delrieu's works has prevented experimentalists to perform the
required combined SANS and $\mu$SR measurements. Here we review this work and present a numerical analysis
of the form factor, field map, and field distribution.

The organization of this paper is as follows.  Section~\ref{geometry} recalls the geometry of a VL
and introduces useful reduced quantities. In Sec.~\ref{Correlation} the physical principles behind the
computations of the magnetization and form factor are given. The following sections, i.e. Sec.~\ref{Delrieu_m}
and Sec.~\ref{Delrieu_ff}, deal with the magnetization and form factor, respectively. In Sec.~\ref{Distribution}
the component field distribution is discussed. The numerical analysis of the form factor, field map and
field distribution is the subject of Sec.~\ref{Numerics_th_clean}. We end up with a discussion and the
conclusions in Sec.~\ref{Discussion_conclusions}. Analytical and numerical details can be found in five appendices.
In particular, the last appendix shows that the form factor can be expressed in terms of
a reduced number of parameters. This  result allows in Sec.~\ref{Numerics_th_clean}
to easily study the crossover from BCS to GL in the VL structures.

\section{Geometry}
\label{geometry}

We assume ${\bf B}_{\rm ext}$ to be applied along the Z axis of an orthogonal reference frame,
with the vortex tubes of a type II superconductor in its mixed phase running along that axis. The VL is taken to be composed of equilateral
triangles. Therefore in the direct space the unit cell is defined by the three vectors ${\bf v}_1 = X_1 {\hat {\bf x}}$,
${\bf v}_2= X_2 {\hat {\bf x}} + Y_2 {\hat {\bf y}}$ and ${\bf v}_3 = {\hat {\bf z}}$, where ${\hat {\bf x}}$
and ${\hat {\bf y}}$ are two mutually orthogonal unit vectors perpendicular to the Z axis,  ${\hat {\bf z}}$ is the unit vector of the Z axis,
and $\{X_1, X_2, Y_2 \}$ are coordinates with the relations
$X_2=X_1 /2$ and $Y_2 =\sqrt{3}X_1 /2$. Since the vortex tubes are taken as straight, a VL point is labeled by the
two dimensional  vector
\begin{equation}
{\bf R}_{p,q} = p {\bf v}_1 + q {\bf v}_2
= (p X_1 +  q X_2){\hat {\bf x}} + q Y_2{\hat {\bf y}}.
\label{eq:geo:1}
\end{equation}
Following Delrieu's convention a lattice point in the reciprocal lattice is
specified by the reciprocal vector
\begin{equation}
{\bf K}_{m,h} = - h {\bf v}^*_1 + m {\bf v}^*_2 =
\frac{2 \pi }{ s_{\rm c}}
[-hY_2{\hat {\bf x}} + (m X_1 + hX_2){\hat {\bf y}}],
\label{eq:geo:2}
\end{equation}
with the VL unit cell area
\begin{equation}
s_{\rm c} = X_1 Y_2 = {\Phi_0 \over \overline {B^Z}}.
\label{eq:geo:3}
\end{equation}
The mean value of the induction is denoted  $\overline {B^Z({\bf r})}$ or $\overline {B^Z}$ in short,
where ${\bf r}$ refers to a position in the direct space. We have introduced the magnetic flux quantum
$\Phi_0$ ($\Phi_0 = 2.06783 \times 10^{-15} \ {\rm T} {\rm m}^2$),
and the two vectors ${\bf v}^*_1$ and ${\bf v}^*_2$ which define the unit cell in the reciprocal lattice.

As examples, with an obvious notation, we have
${\bf R}_{1,0} = X_1 (1,0)$, ${\bf R}_{0,1} = X_1 (1/2, \sqrt{3}/2)$, and  ${\bf R}_{-1,1} = X_1 (-1/2, \sqrt{3}/2)$.
For the reciprocal lattice points, we compute
${\bf K}_{1,0} = [ 4 \pi/ (\sqrt{3}X_1)] (0,1)$, ${\bf K}_{0,1} = [ 4 \pi/ (\sqrt{3}X_1)](-\sqrt{3}/2, 1/2)$, and
${\bf K}_{-1,1} = [ 4 \pi/ (\sqrt{3}X_1)] (-\sqrt{3}/2, -1/2)$. The complementary three points in the direct and reciprocal lattices respectively
can be obtained by symmetry.

We introduce for convenience two reduced quantities. We define the magnetic length
\begin{equation}
\Lambda = \sqrt{ s_{\rm c}\over 2 \pi} = \sqrt{X_1 Y_2\over 2 \pi} = \sqrt{\Phi_0 \over 2 \pi \overline {B^Z} }.
\label{eq:geo:4}
\end{equation}
It is related to the lattice parameter of the VL:
\begin{equation}
X_1 = \left ({4 \over 3 } \right)^{1/4} \sqrt{\Phi_0 \over \overline {B^Z} } = 2.693 \times \Lambda.
\label{eq:geo:5}
\end{equation}
Unless $B_{\rm ext}$ is close to the lower critical field $B_{\rm c1}$, the magnetization of a superconductor is negligible,
and therefore $ \overline {B^Z} \simeq B_{\rm ext}$. We shall express the form factor with the unitless parameter $n_{{\bf K}_{m,h}}$
defined through the wave-vector scalar product
\begin{equation}
{\bf K}_{m,h} \cdot {\bf K}_{m,h} = K^2_{m,h} = {4 \over \Lambda^2} n^2_{{\bf K}_{m,h}}.
\label{eq:geo:6}
\end{equation}
From this definition, we derive the important formula:
\begin{equation}
n^2_{{\bf K}_{m,h}} = { \pi X_1 \over 2 Y_2} \left (m^2 + mh + h^2  \right) = { \pi  \over \sqrt{3}} \left (m^2 + mh + h^2  \right).
\label{eq:geo:7}
\end{equation}

\section{Physical principles for the description of the magnetization and form factor}
\label{Correlation}

The computations of the magnetization and form factor are based on an approximate form of the integral Gorkov's equation
for the temperature Green's function $ G_{\omega_\ell}({\bf r}, {\bf r}^\prime)$:\cite{Werthamer69}
\begin{align}
\label{eq:Correlation:1}
&G_{\omega_\ell}({\bf r}, {\bf r}^\prime)  =  G^{\rm n}_{\omega_\ell}({\bf r} - {\bf r}^\prime) -\\
&\int  G^{\rm n}_{\omega_\ell}({\bf r} - {\bf r}_1) V({\bf r}_1 , {\bf r}_2) G^{\rm n}_{-{\omega_\ell}}({\bf r}_2 - {\bf r}_1)
G_{\omega_\ell}({\bf r}_2, {\bf r}^\prime) {\rm d}^3 {\bf r}_1 {\rm d}^3 {\bf r}_2,\nonumber
\end{align}
where $V({\bf r}_1 , {\bf r}_2)$ is the correlation function of the order parameter $\Delta({\bf r})$:
\begin{eqnarray}
V({\bf r}_1 , {\bf r}_2) = \Delta({\bf r}_1) \Delta^*({\bf r}_2)  \exp \left (- {2i e  \over h} \int_{{\bf r}_1}^{{\bf r}_2} {\bf A} ({\bf l}) \cdot  {\rm d} {\bf l} \right ).
\label{eq:Correlation:2}
\end{eqnarray}
Note that  $V({\bf r}_1 , {\bf r}_2)$ describes the correlation function of the Cooper's pairs rather
than the correlation of the electrons. A justification of the correlation function nature of
$V({\bf r}_1,{\bf r}_2)$ is given after Eq.~\ref{eq:Correlation:5}.
We have introduced the Matsubara angular precession frequency $\omega_\ell$:
\begin{eqnarray}
\omega_\ell = (2 \ell +1 ) \pi k_{\rm B} T/\hbar.
\label{eq:Correlation:2:1}
\end{eqnarray}
The path of integration over the potential vector ${\bf A}$ is a straight line between ${{\bf r}_1}$ and ${{\bf r}_2}$. Within the  semiclassical approximation
the effect of the field on the material is entirely described by the phase integral in the correlation function and
$G^{\rm n}_{\omega_\ell}({\bf r} - {\bf r}^\prime)$ refers to the temperature Green's function of
the normal metal in the absence of a magnetic field.
The semiclassical approximation is
valid if the spacing between the Landau levels is small compared to the sum of their thermal
and collision broadenings,
i.e.\cite{Werthamer69}
\begin{eqnarray}
{\mathcal R}_\phi = {\mu_{\rm B} B_{\rm ext} \over 2 \pi k_{\rm B} T  + \hbar /{\tau_{\rm life}}} \ll 1.
\label{eq:Correlation:3}
\end{eqnarray}
Here $\tau_{\rm life}$ is the level lifetime which accounts for the finite electron mean-free path.
It is futher discussed in Sec.~\ref{Discussion_conclusions}.  Neglecting the
$\hbar /\tau_{\rm life}$ term in the ratio expression, we compute ${\mathcal R}_\phi = 1$ when $T=0.04$~K for  $B_{\rm ext} = 0.4$~T.
This $B_{\rm ext}$ value corresponds approximately to $B_{\rm c2}$ for a high quality niobium sample.\cite{Herlach90}
The theory for the form factor and resulting field map and field distribution discussed here is therefore
expected to be valid down to  0.04~K for simple superconductors
such as niobium when the VL is composed of equilateral triangles.

The correlation function has the periodicity of the VL with respect to the center of mass of a Cooper's pair, i.e.  $({\bf r}_1 + {\bf r}_2)/2$.\cite{Brandt69}
Therefore it can be Fourier expanded. Neglecting the spatial variation of the induction, a valid approximation in our case since we are
interested in the high-field VL for which the magnetization is negligible,\cite{Delrieu72,Delrieu74}
\begin{align}
V({\bf r}_1 , {\bf r}_2) &= \\
&\sum_{m,h} V_{{\bf K}_{m,h}}({\bf r}_1 - {\bf r}_2) \exp \left [i {\bf K}_{m,h} \cdot ({\bf r}_1 + {\bf r}_2)/2  \right ],\nonumber
\label{eq:Correlation:4}
\end{align}
with 
\begin{eqnarray}
V_{{\bf K}_{m,h}}({\bf r}) = (-1 )^{mh} V_0({\bf r} - {\bf R}_{m,h}).
\label{eq:Correlation:5}
\end{eqnarray}
Since $V({\bf r}_1,{\bf r}_2)$ depends on the difference
$({\bf r}_1 -{\bf r}_2)$ and the center of mass of the Cooper's pairs, and not on
${\bf r}_1$ and ${\bf r}_2$ individually, it is really a correlation function.
The Fourier components have a remarkable property: $V_{{\bf K}_{m,h}}({\bf R}_{m,h}) = (-1 )^{mh} V_0(0)$.
Therefore at the VL nodes, i.e. at position
${\bf R}_{m,h}$, within a phase the correlation function of the order parameter has a common single value.
In fact the phase changes from one position to the next
nearest neighbors in a coherent fashion. Because of this long-range coherence, we expect
Cooper's pair diffraction on the vortex cores.
However, the diffraction can only
be partial since the coherent nature of the correlation is only active for a finite
set of Cooper's pairs, i.e. the balistic ones with trajectory
through the vortex cores. It is the extended correlation which is
the origin of the slow power decay of the form factor at low temperature discussed latter on.
This property does not hold near $B_{\rm c1}$ since the spatial variation of the
induction has been neglected in deriving Eq.~\ref{eq:Correlation:5}.

As the correlation function, the Green's function is periodic with respect to
$({\bf r}_1 + {\bf r}_2)/2$. In addition to the half sum of the coordinates,
the two functions depend on the difference of the coordinates. To the difference corresponds the
continuous conjugate Fourier vector ${\bf p}$.  Performing the ${\bf K}_{m,h}$ and  ${\bf p}$ Fourier transforms on the
approximate Gorkov's equation given at Eq.~\ref{eq:Correlation:1}, an algebraic set of
equations is derived.
An approximate simple
solution has been proposed in Ref.~[\onlinecite{Brandt67}] based on the fact that the terms
of the set with ${\bf K}_{m,h} \neq {\bf 0}$ are negligible relative
to the ones with ${\bf K}_{m,h} ={\bf 0}$ . Latter on, it has been shown to be satisfactory only for the computation of
thermodynamic quantities. It cannot be used to describe the dynamics. It is valid even at low temperature. \cite{Delrieu74}

The free energy of a superconductor can be written in terms of $G_{\omega_\ell}({\bf r}, {\bf r}^\prime)$.\cite{Eilenberger65}
Focusing again on the case for which $B_{\rm ext}$ is near $B_{\rm c2}$, and therefore the spatial variation of the
order parameter can be
neglected, and assuming the Abrikosov's vortex solution,\cite{Brandt69,Delrieu72} 
\begin{equation}
\begin{aligned}
& F  = { \Delta^2_0 N_0 } \ln \left ( { T \over T_{\rm c0}} \right )  -  \label{eq:Dlr:M:Expr1:1}  \\  
&  8 \pi  k_{\rm B} T N_0 \sum_{\ell = 0}^\infty
\left[ \int_0^{\pi/2} \sin \theta \left ( { u_\ell \over 2 a} - \hbar \omega_\ell \right ){\rm d} \theta  - {\Delta_0^2 \over 4 \hbar |\omega_\ell |}\right]. 
\end{aligned}
\end{equation}
We have defined the function 
\begin{eqnarray}
a = a(\theta) = {\Lambda \over \hbar v_{\rm F} \sin \theta},
\label{eq:Dlr:M:Expr2}
\end{eqnarray}
and the variable $u_\ell$ which is the root of the equation
\begin{eqnarray}
u_\ell = 2 \hbar \omega_\ell a + \Delta^2_0 a^2 i v(i u_\ell).
\label{Delrieu_m_expression_3}
\end{eqnarray}
The Fermi surface has been assumed to be spherical. As we are using the BCS theory,
we are working in the weak-coupling limit.
The function $v(z)$ is defined in Appendix~\ref{Math}.

In addition to the temperature, the free energy depends on four parameters: the critical temperature
at low field $T_{\rm c0}$, the mean order parameter squared
$\Delta^2_0= \overline{|\Delta({\bf r})|^2}$ where the bar is for the spatial averaging, the density of states at the Fermi energy per spin,
volume and energy in the normal state $N_0$, and the Fermi velocity $v_{\rm F}$. The order parameter can be related to basic parameters of the
superconductor. Assuming that the average value $\Delta_0$ is equal to the value of the order parameter in zero field, according to BCS
\begin{eqnarray}
{ \Delta_0(0) \over k_{\rm B} T_{\rm c0}} = {\pi \over \exp(\gamma)} = 1.7639,
\label{Delrieu_m_expression_4}
\end{eqnarray}
where $\Delta_0(0)$ is the value of $\Delta_0$ at $T=0$ and $\gamma$ the Euler-Mascheroni constant, i.e. $\gamma = 0.57722$.
So at low temperature
only three material parameters are left if the spherical Fermi surface and weak-coupling aproximations are valid.

In addition to $\Delta_0(0)$, other parameters
characterize a superconductor. The Pippard-BCS coherence length $\xi_0(0)$ is related to
$\Delta_0(0)$,
\begin{eqnarray}
\xi_0(0) = {\hbar v_{\rm F} \over \pi \Delta_0(0)},
\label{Delrieu_m_expression_5}
\end{eqnarray}
and in the clean limit -- see for example
Ref.~[\onlinecite{Tinkham96}] at page 120,
\begin{eqnarray}
\xi_{\rm GL} = \xi_0/0.96.
\label{Delrieu_m_expression_5_1}
\end{eqnarray}
Here $\xi_{\rm GL}$ is the GL coherence length. We stress that Eq.~\ref{Delrieu_m_expression_5_1}
is derived in the $T=0$ limit.
We note the GL relation
\begin{eqnarray}
B_{\rm c2}(T) = {\Phi_0 \over 2 \pi \xi^2_{\rm GL}(T)}.
\label{Delrieu_m_expression_6}
\end{eqnarray}

It is also possible to derive information on the mean order parameter
from the minimisation of the free energy written above. This leads to a
formula of the Helfand-Werthamer type\cite{Helfand66} which
can be used to model
$B_{\rm c2} (T)$ for an isotropic Fermi surface in the weak coupling approximation. However,
that type of formula does not describe
$B_{\rm c2}(T)$ for niobium,\cite{Williamson70a,Williamson70b} one of the metal for which the
theory discussed here  may apply.
This is attributed to the strong anisotropy of the Fermi surface.\cite{Hohenberg67,Takanaka71}
Hence, we shall not pursue any longer our discussion of $\Delta_0$ in terms of the free energy
discussed in this work.

\section{Magnetization}
\label{Delrieu_m}

While our main purpose in this report is the study of the form factor, here
we discuss the magnetization $M$. This is justified since, thanks to Abrikosov, the relation between
$M$ and the induction is known for a temperature in the vicinity of $T_{\rm c0}$. Therefore the study of $M$ gives us the possibility to check the validity of the
formula for the induction, and therefore the form factor.

We recall that $M = - \partial F/ \partial {\overline {B^Z}}$.  In terms of the free energy expression given
at Eq.~\ref{eq:Dlr:M:Expr1:1},
$\partial F/ \partial {\overline {B^Z}} = - [a/(2 {\overline {B^Z}})](\partial F/ \partial a)$ and
therefore we need to evaluate $a (\partial/\partial a)(u_\ell/a)$.
This is done in Appendix~\ref{Cal}. We finally obtain
\begin{equation}
M = { \pi k_{\rm B} T N_0 \Delta^2_0 \over 2 {\overline {B^Z}}}\sum_{\ell = 0}^\infty
\int_0^{\pi/2} \sin (\theta)  i v^{\prime \prime}(iu_\ell) {\partial u_\ell \over \partial (\hbar \omega_\ell) } {\rm d}\theta .
\label{Delrieu_m_4}
\end{equation}
$M$ can be evaluated numerically and compared to experimental data. Ourselves we shall used it in our discussion of the form factor.
Interestingly, it can be drastically simplified in the $T=0$ limit. From Appendix~\ref{Delrieu_m_asym}
\begin{eqnarray}
M = -{ N_0 \Delta^2_0 \over 2 {\overline {B^Z}}} .
\label{Delrieu_m_6}
\end{eqnarray}
This result will be used in Sec.~\ref{Delrieu_ff}. As expected, $M$ is negative.

\section{Form factor}
\label{Delrieu_ff}

Here we consider the field Fourier component $B^Z_{{\bf K}_{m,h}}$ which is usually called the form factor in the SANS literature. It is related to the
space-dependent induction component $B^Z ({\bf r})$ by the Fourier relation:
\begin{eqnarray}
B^Z ({\bf r}) = \sum_{{\bf K}_{m,h}}  B^Z_{{\bf K}_{m,h}} \exp(i {\bf K}_{m,h} \cdot {\bf r}),
\label{Delrieu_ff_expression_1N1}
\end{eqnarray}
with
\begin{eqnarray}
B^Z_{{\bf K}_{m,h}} = {1 \over s_{\rm c}} \int_{s_{\rm c}} B^Z ({\bf r}) \exp(- i {\bf K}_{m,h} \cdot {\bf r}) {\rm d}^2 {\bf r}.
\label{Delrieu_ff_expression_1_2}
\end{eqnarray}
According to Refs.~[\onlinecite{Delrieu72,Delrieu74}] and correcting for misprints, 
%
%
\txtb{
\begin{align}
B^Z_{{\bf K}_{m,h}} & =  -{ \mu_0  N_0  \Delta_0^2\over 4 \overline {B^Z}} { (-1)^{mh} \exp \left ( - n^2_{{\bf K}_{m,h}}\right )
\over n^2_{{\bf K}_{m,h}}} 2 \pi k_{\rm B} T \times \nonumber \\
&\sum_{\ell =0}^\infty \int_0^{\pi/2}\sin (\theta) g_\ell(\theta) {\rm d} \theta, 
\label{eq:Dlr:FF:Expr1}
\end{align}
}
where we have introduced the auxiliary function  
\begin{multline}
g_\ell(\theta) =\\
{iv\left (iu_\ell + i n_{{\bf K}_{m,h}} \right ) + iv\left (iu_\ell - i n_{{\bf K}_{m,h}} \right ) -2 iv\left (iu_\ell \right )
\over  \left [1 + { \Delta^2_0 \Lambda^2\over \hbar^2 v^2_{\rm F}\sin^2 \theta} v^\prime\left (iu_\ell \right )\right ]^2}
{\partial u_\ell \over \partial (\hbar \omega_\ell)}.
\label{eq:Dlr:FF:Expr2}
\end{multline}
It is convenient to rewrite the variable $u_\ell$ without the intermediate function $a(\theta)$.
Refering to Eqs.~\ref{eq:Dlr:M:Expr2}
and \ref{Delrieu_m_expression_3},
\begin{eqnarray}
u_\ell = 2 \omega_\ell {\Lambda \over v_{\rm F} \sin \theta} +
{\Delta^2_0 \Lambda^2 \over \hbar^2 v^2_{\rm F} \sin^2 \theta} i v(i u_\ell).
\label{eq:Dlr:FF:Expr3}
\end{eqnarray}
Therefore $g_\ell(\theta)$ is indeed a function of the angle $\theta$ and the angular frequency $\omega_\ell$.

This formula for $B^Z_{{\bf K}_{m,h}}$ is extremely complicated.
Remarkably, as $M$ does, $B^Z_{{\bf K}_{m,h}}$ depends on three material
parameters, i.e. $\Delta^2_0$, $N_0$, and $v_{\rm F}$,  and $T$ and ${\overline B^Z}$.
This suggests that simple relations may exist betweem them. They do exist in
two limiting cases. Interestingly, $N_0$ only appears as a proportionality parameter.

In the proximity of $T_{\rm c0}$ the asymptotic limit of the form factor is (see Sec.~\ref{Delrieu_limits_Tc}):
\begin{eqnarray}
B^Z_{{\bf K}_{m,h}} & = & \mu_0 M (-1)^{mh} \exp \left ( - n^2_{{\bf K}_{m,h}}\right ) .
\label{eq:Dlr:FF:Expr4}
\end{eqnarray}
Recalling the definition of the Fourier transform of the induction, see Eq.~\ref{Delrieu_ff_expression_1N1},
\begin{eqnarray}
B^Z({\bf r}) & = & {\overline {B^Z}} + \sum_{(m,h) \neq (0,0)} B^Z_{{\bf K}_{m,h}} \exp(i {\bf K}_{m,h} \cdot {\bf r}).
\end{eqnarray}
This means that
\begin{align}
& B^Z({\bf r})  =  {\overline {B^Z}} - \label{eq:Dlr:FF:Expr4:1}\\
& \mu_0 |M|\sum_{(m,h) \neq (0,0)} (-1)^{mh} \exp \left ( - n^2_{{\bf K}_{m,h}}\right )\exp(i {\bf K}_{m,h} \cdot {\bf r}) ,\nonumber
\end{align}%
since $M < 0$, see Sec.~\ref{Delrieu_m}. This is the Abrikosov result.\cite{Abrikosov57}

An exotic behaviour is found in the $T \rightarrow 0$ limit. From Sec.~\ref{Delrieu_limits_zero} for $n_{{\bf K}_{m,h}}\gg 1$,
%
%
\txtb{
\begin{eqnarray}
B^Z_{{\bf K}_{m,h}} & = & -{\mu_0  \sqrt {\pi} N_0  \Delta_0^2\over 4 \overline {B^Z}} { (-1)^{mh} \over n^3_{{\bf K}_{m,h}}}. 
\label{eq:Dlr:FF:Expr5}
\end{eqnarray}
}
In terms of the magnetization, see Eq.~\ref{Delrieu_m_6}, it gives  
%
%
\txtb{
\begin{eqnarray}
B^Z_{{\bf K}_{m,h}} & = & - { \mu_0 |M| } {\sqrt {\pi} \over 2}
{ (-1)^{mh} \over n^3_{{\bf K}_{m,h}}}. 
\label{eq:Dlr:FF:Expr5:1}
\end{eqnarray}
}

Whereas $B^Z_{{\bf K}_{m,h}}$ has a Gaussian  wave-vector dependence near $T_{\rm c0}$, it decays much slowly with a wave-vector power law
at low temperature. This is quantified in Sec.~\ref{Numerics_th_clean}.
Hence, whereas only the low indices Bragg spots might
be observed by small angle neutron scattering (SANS) near $T_{\rm c0}$, more spots should be detected at low temperature.
In fact, in addition to the usual six $ {\bf K}_{1,0}$ Bragg reflections,  $ {\bf K}_{1,1}$  and $ {\bf K}_{2,0}$ spots
have already been observed for niobium at 4.2~K for ${\bf B}_{\rm ext}$ parallel to the $[111]$ crystal direction.\cite{Schelten71}
Only for that field direction is the VL triangular, and hence is of interest here. Because of possible double Bragg scattering effects
and large distortions in the VL, it is not easy to derive information from the published form-factor data on niobium to compare with the
power law prediction.

\section{Field distribution}
\label{Distribution}

It is difficult to analyse analytically the spatial dependence of the induction. Here we shall focus
on the field distribution as measured  by $\mu$SR
and NMR techniques. We shall assume a disorder-free VL and the field width of the distribution to be small relative
to ${\overline {B^Z}}$, i.e. only the distribution of the field component along ${\bf B}_{\rm ext}$ is
measured.\cite{Yaouanc11} This latter condition can be checked to be fulfilled by looking at the computed distribution.
For simplicity we shall write the distribution $D_{\rm c}(B^Z)$ without specifying that
it does depend on $B_{\rm ext}$.

Mathematically, the distribution can be expressed in terms of a two-dimensional Dirac function:
\begin{eqnarray}
D_{\rm c}(B^Z) = \int_{s_{\rm c}} \delta \left [B^Z({\bf r}) - B^Z  \right] {{\rm d}^2 {\bf r} \over s_{\rm c}},
\label{eq:Distribution:2}
\end{eqnarray}
where the integral only extends over the unit cell.
An important characterisation of a distribution is its variance:\cite{Yaouanc11}
\begin{eqnarray}
\Delta^2_{Z,{\rm v}} = \sum _{\bf K \neq 0} |B^Z_{\bf K}|^2.
\label{eq:Distribution:3}
\end{eqnarray}
For the simple asymptotic form factor given at Eq.~\ref{eq:Dlr:FF:Expr5}, we derive for the standard deviation
%
%
\txtb{
\begin{align}
\Delta_{Z,{\rm v}} = &{ 3^{3/4} \mu_0 N_0 \Delta_0^2 \over 4\pi {\overline {B^Z}} }
\sqrt{\sum_{(m,h) \neq (0,0)}{1 \over \left (m^2 + mh + h^2  \right )^3} }
\nonumber \\ \label{eq:Distribution:4}
&=0.4581 {\mu_0 N_0 \Delta_0^2 \over {\overline {B^Z}}}.
\end{align}
}

\section{Numerical analysis of the  form factor, field map and field distribution}
\label{Numerics_th_clean}

As shown  in  Appendix~\ref{Reduced_clean}, the form factor $B^Z_{{\bf K}_{m,h}}$ depends in
fact only on three parameters noted $\tilde{a}$, $\tilde{b}$ and $\tilde{c}$.
The parameter \txtb{$\tilde{a}=-\mu_0  N_0  \Delta_0^2 \tilde{c}/2 \overline{B^Z} $}
stands as a proportionality coefficient
while $\tilde{b}$ and $\tilde{c}$ are dimensionless [$\tilde{b}=( \Lambda/\pi\xi^B)^2$ and
$\tilde{c}=\Lambda/\xi^T$ with the temperature and field dependent length scale parameters
$\xi^T =\hbar v_{\rm F}/ (2 \pi k_{\rm B}T)$ and
$\xi^B={\hbar v_{\rm F} / \pi \Delta_0}$, see Appendix \ref{Reduced_clean}]. Interestingly,
for $B_{\rm ext}\simeq B_{\rm c2}$ $\tilde{b} \ll 1$ and $\tilde{c}$ changes notably as $T$ varies.
As shown below, this enables us to easily discuss the physics of the form factor,
field map and field distribution.

\begin{figure}
\includegraphics[scale=0.40]{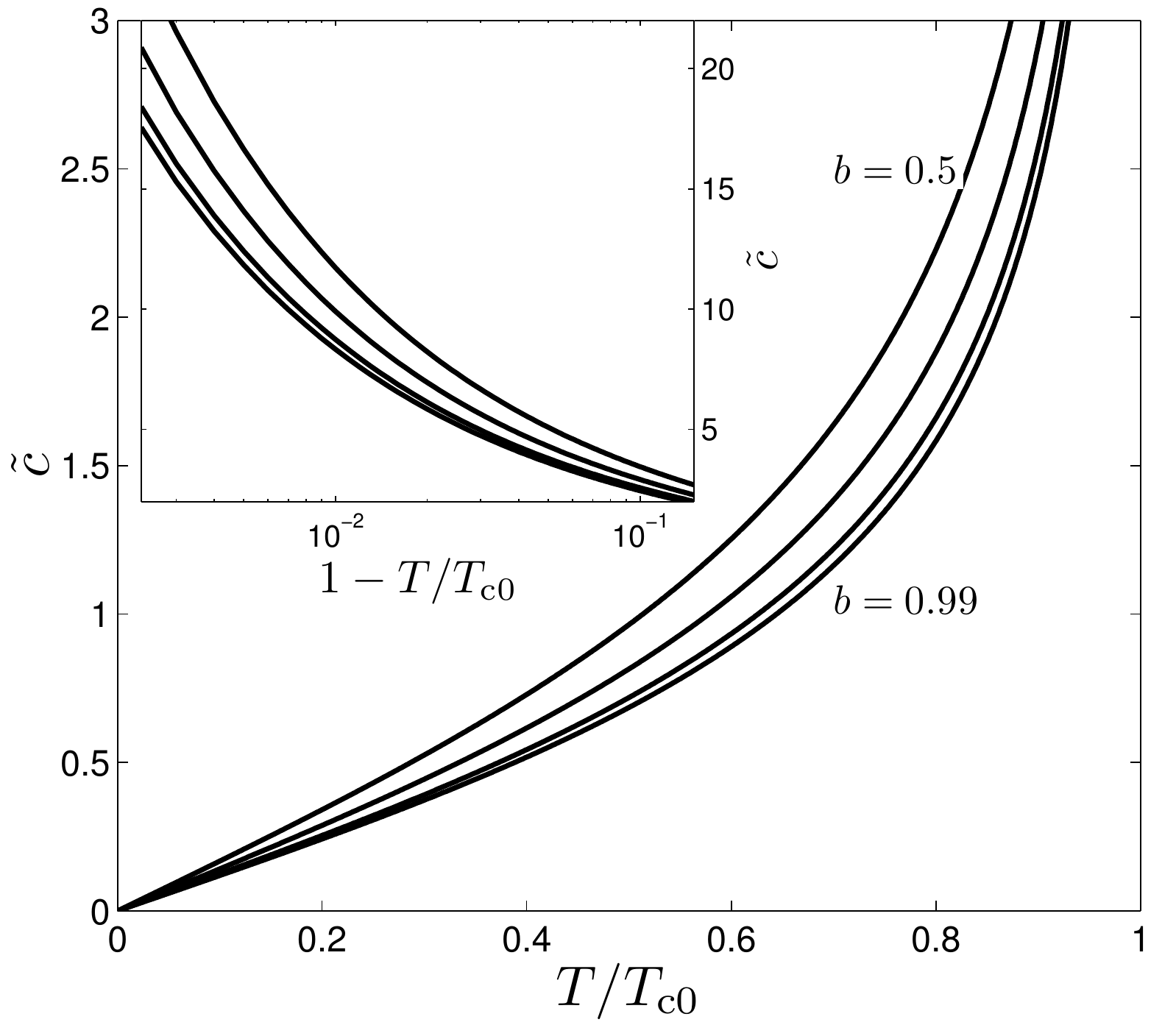}
\caption{Temperature dependence of the parameter $\tilde{c}$  in reduced temperature scale
for four different values of the reduced field
$b=B_{\rm ext} /B_{\rm c2}$, i.e. $b = 0.5$, 0.7, 0.9, and 0.99. The insert shows
the same but on semilogarithmic scale in the vicinity of $T_{\rm c0}$. }
\label{fig:TildeC}
\end{figure}

Let us consider a classical BCS superconductor with $T_{\rm c0}=10$~K. From
Eq.~\ref{Delrieu_m_expression_4} we compute $\Delta_0(0) \simeq 1.52$~meV. Assuming
$B_{\rm c2}(0) =0.4$~T, from the GL relation (Eq.~\ref{Delrieu_m_expression_6})
we find $\xi_{\rm GL} (0)=28.7$~nm. Using the BCS relation $v_{\rm F} = \xi_0 \pi \Delta_0(0)/\hbar$,
see Eq.~\ref{Delrieu_m_expression_5}, and the approximation of $\xi_0$ in terms of $\xi_{\rm GL}$
given at Eq.~\ref{Delrieu_m_expression_5_1},  we estimate $v_{\rm F} \simeq 2.00\times 10^5$~m/s.
Let us assume for the second critical field the simple temperature dependence
$B_{\rm c2}(T)=B_{\rm c2}(0)(1-t^{2})$, with  the reduced temperature $t=T/T_{\rm c0}$.
The field and temperature dependencies of the gap are taken to be described by the compact
formula
$\Delta_0 =\Delta_0(T)\sqrt{1-b} = \Delta_0(0)\sqrt{1-b}\sqrt{1-t^{2}}$,
with the reduced field $b=B_{\rm ext}/B_{\rm c2}(T)$.
Here we note that $\xi_0 \neq \xi^B$ since $\xi_0$ only depends on $\Delta_0(0)$  while $\xi^B$
is a function of $\Delta_0$.
Under these assumptions:
\begin{align}
{\tilde b} =& \left({\Lambda \Delta_0 \over \hbar v_{\rm F} }\right)^2 =
{\Delta^2_0(T)\times(1-b)\over \hbar^2 v^2_{\rm F}}{\Phi_0 \over 2\pi B_{\rm c2}(T) } {B_{\rm c2}(T)\over \overline{B^Z}} \nonumber \\
=&{1 \over \pi^2}  {\xi^2_{\rm GL}(T)\over \xi^2_0(T)}\cdot{1-b \over b}
           \simeq 0.110 {1 - b \over b}.
\label{eq:Numerics:th:clean:1}
\end{align}
Above we used Eqs.~\ref{Delrieu_m_expression_5}-\ref{Delrieu_m_expression_6}.
The parameter $\tilde{c}$ is field and temperature dependent:
\begin{equation}
\tilde{c} = 1.1811\frac{t}{\sqrt{b(1-t^2)}}.
 \label{eq:Numerics:th:clean:tildCvsT}
\end{equation}
Interestingly, $\tilde c$ is only written in terms of two reduced variables.
Numerically, $\tilde{b} = 0.110,$ 0.073, 0.047, 0.027, 0.012, and 0.0011 for $b=0.5$, 0.6, 0.7,
0.8, 0.9, and 0.99, respectively.
$\tilde{c}$ is linear in temperature in the low temperature limit and diverges as $t\rightarrow 1$.
The temperature dependence of $\tilde{c}$ for
$b=0.5$, 0.7, 0.9, and 0.99 is given in Fig.~\ref{fig:TildeC}.
In agreement with our
discussion in  Appendix~\ref{Reduced_clean}, while the  $\tilde{c}$ thermal dependence is quite
strong and it gets large as the critical temperature is approached,
$\tilde{b}$ is weakly temperature dependent and has a negligibly small value
in the  $B_{\rm ext} \rightarrow B_{\rm c2}$ limit.

\begin{figure*}
\includegraphics[width=0.3\linewidth]{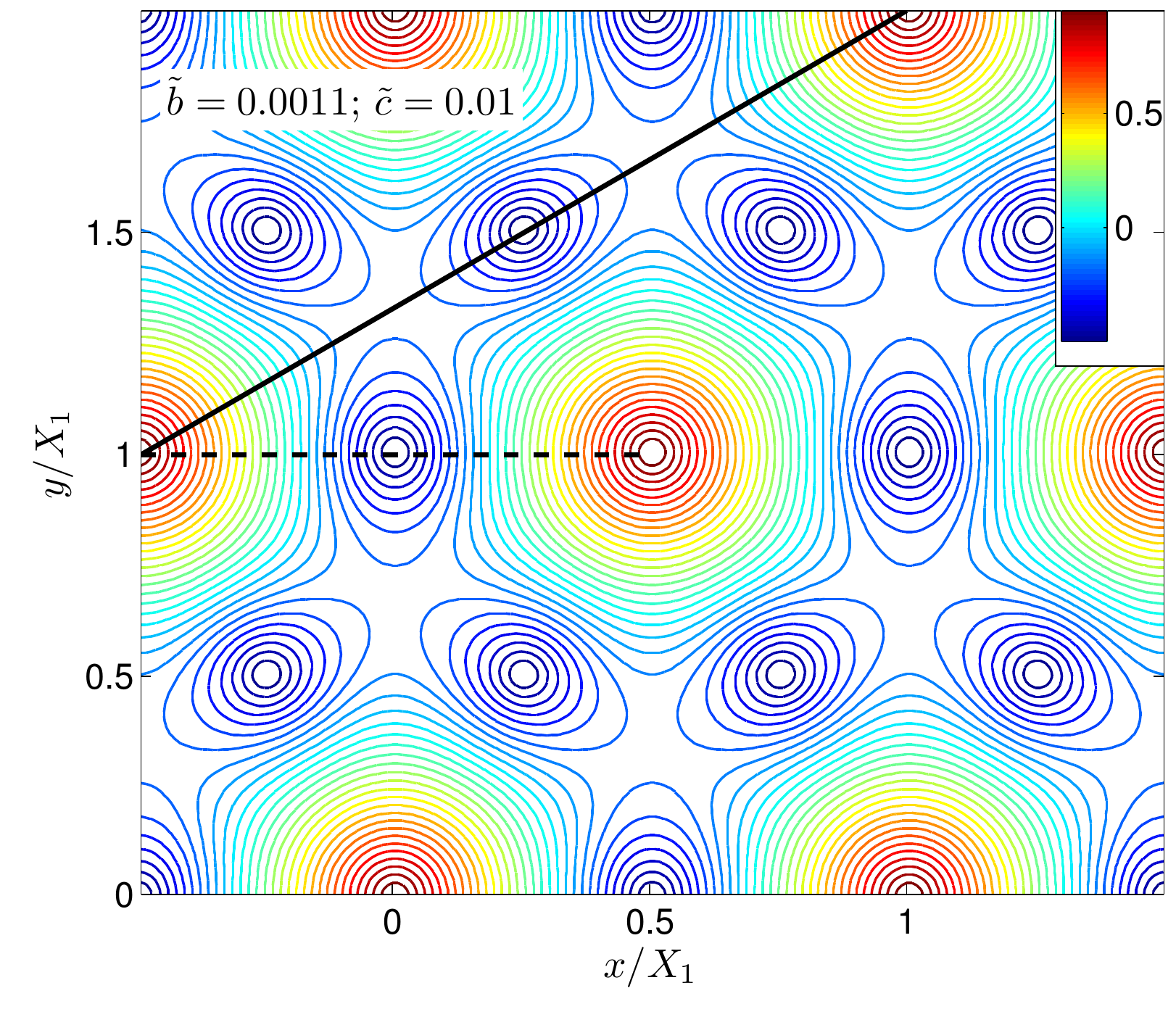}
\includegraphics[width=0.3\linewidth]{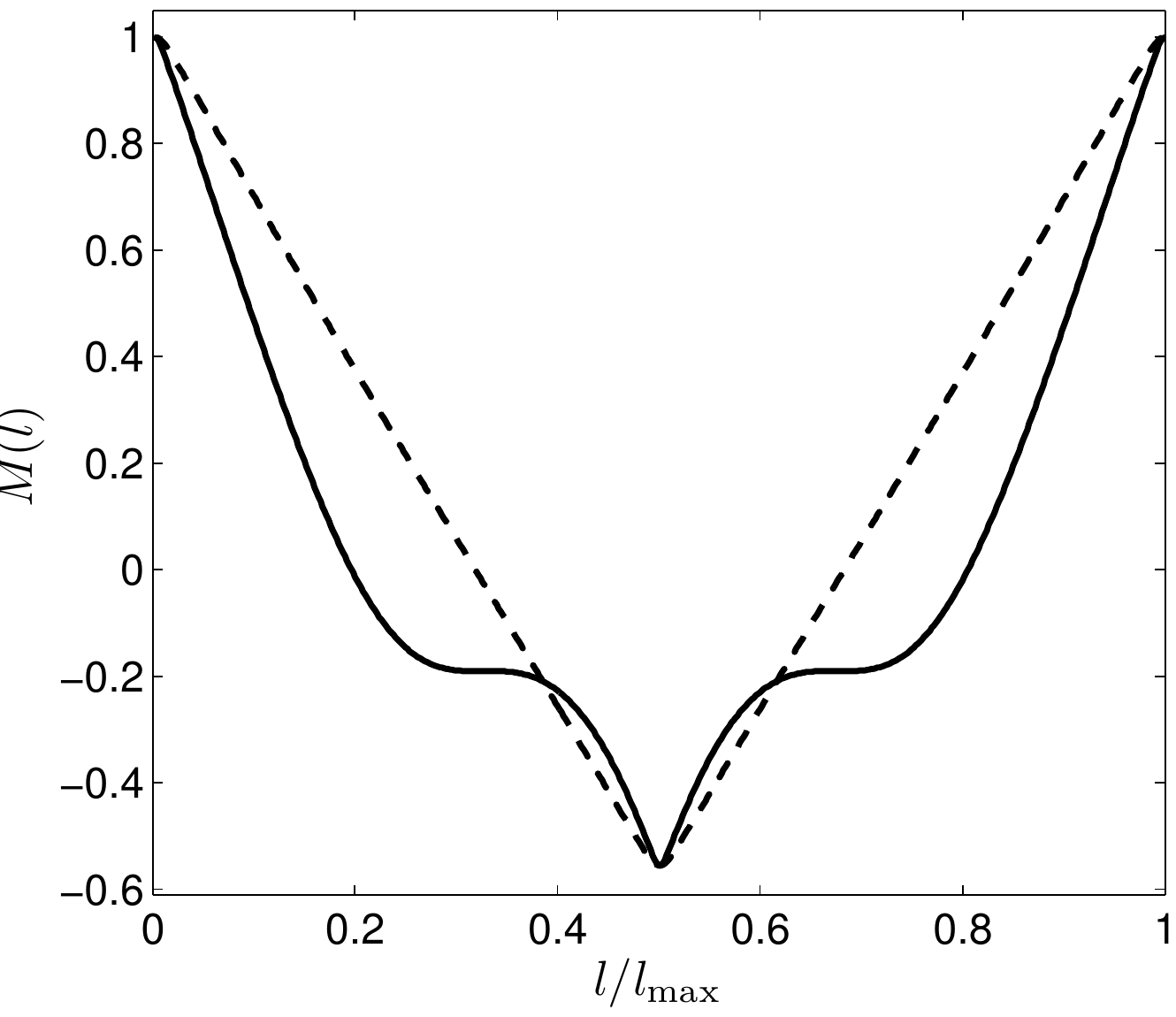}
\includegraphics[width=0.3\linewidth]{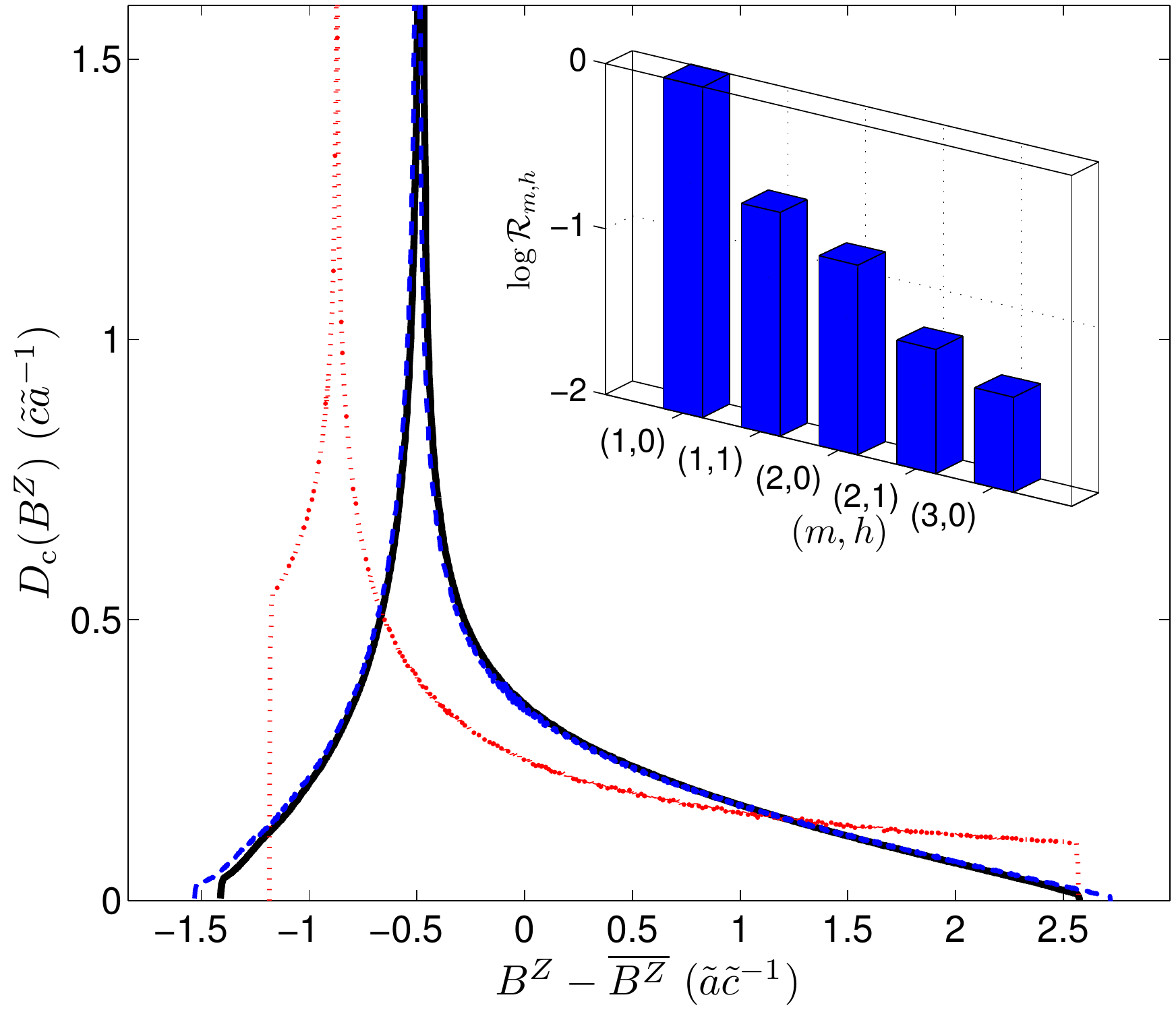}\\
\includegraphics[width=0.3\linewidth]{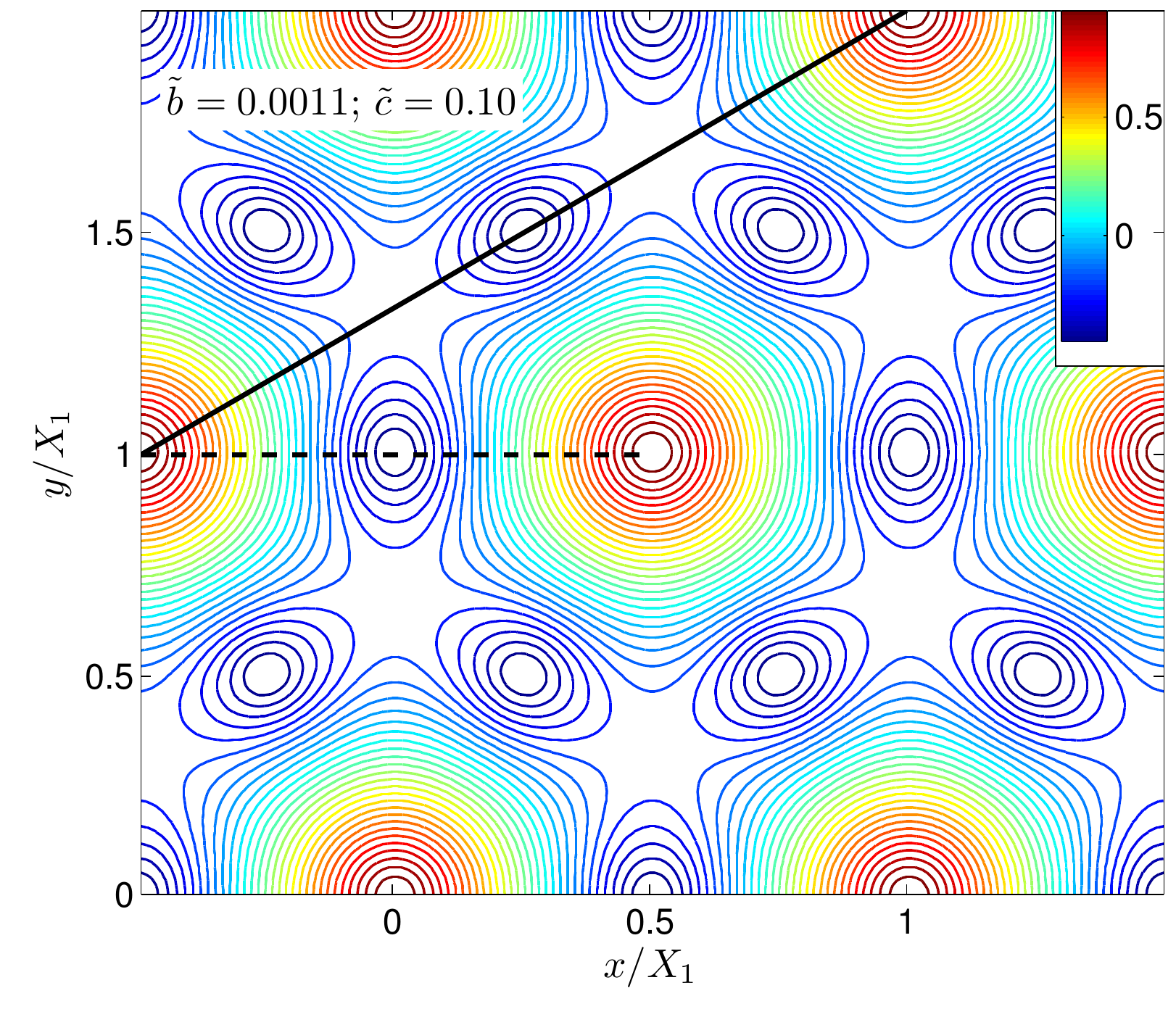}
\includegraphics[width=0.3\linewidth]{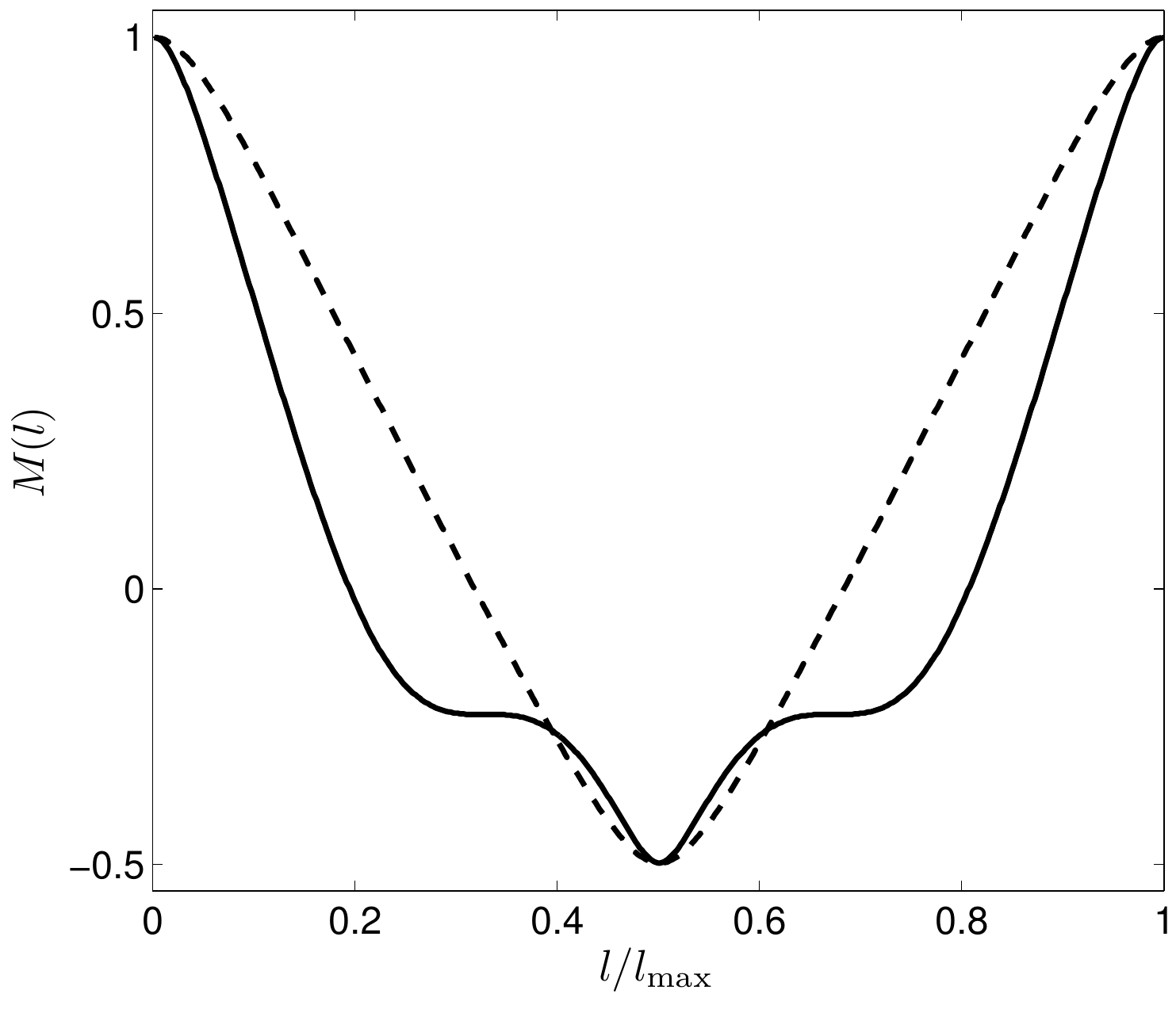}
\includegraphics[width=0.3\linewidth]{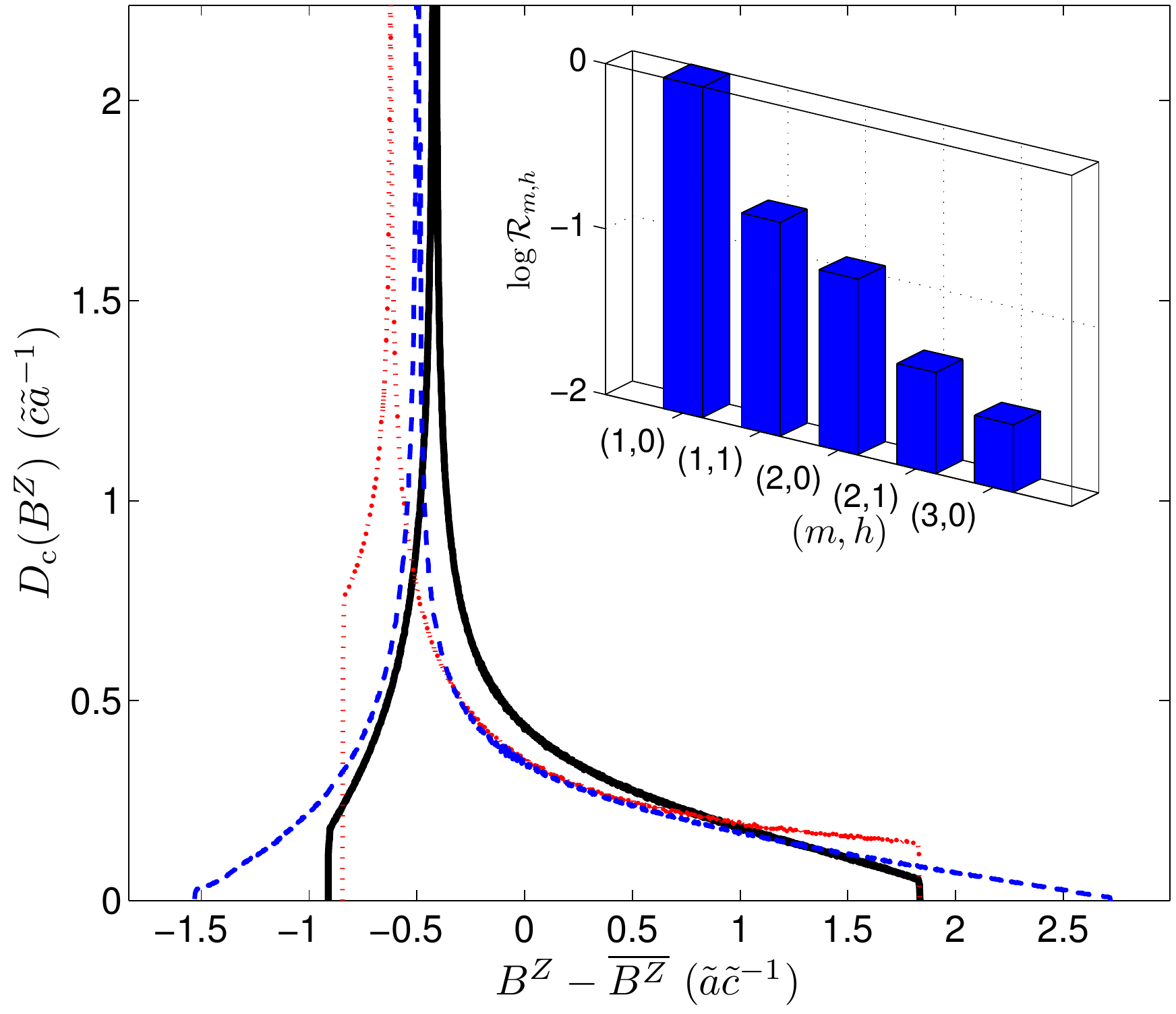}\\
\includegraphics[width=0.3\linewidth]{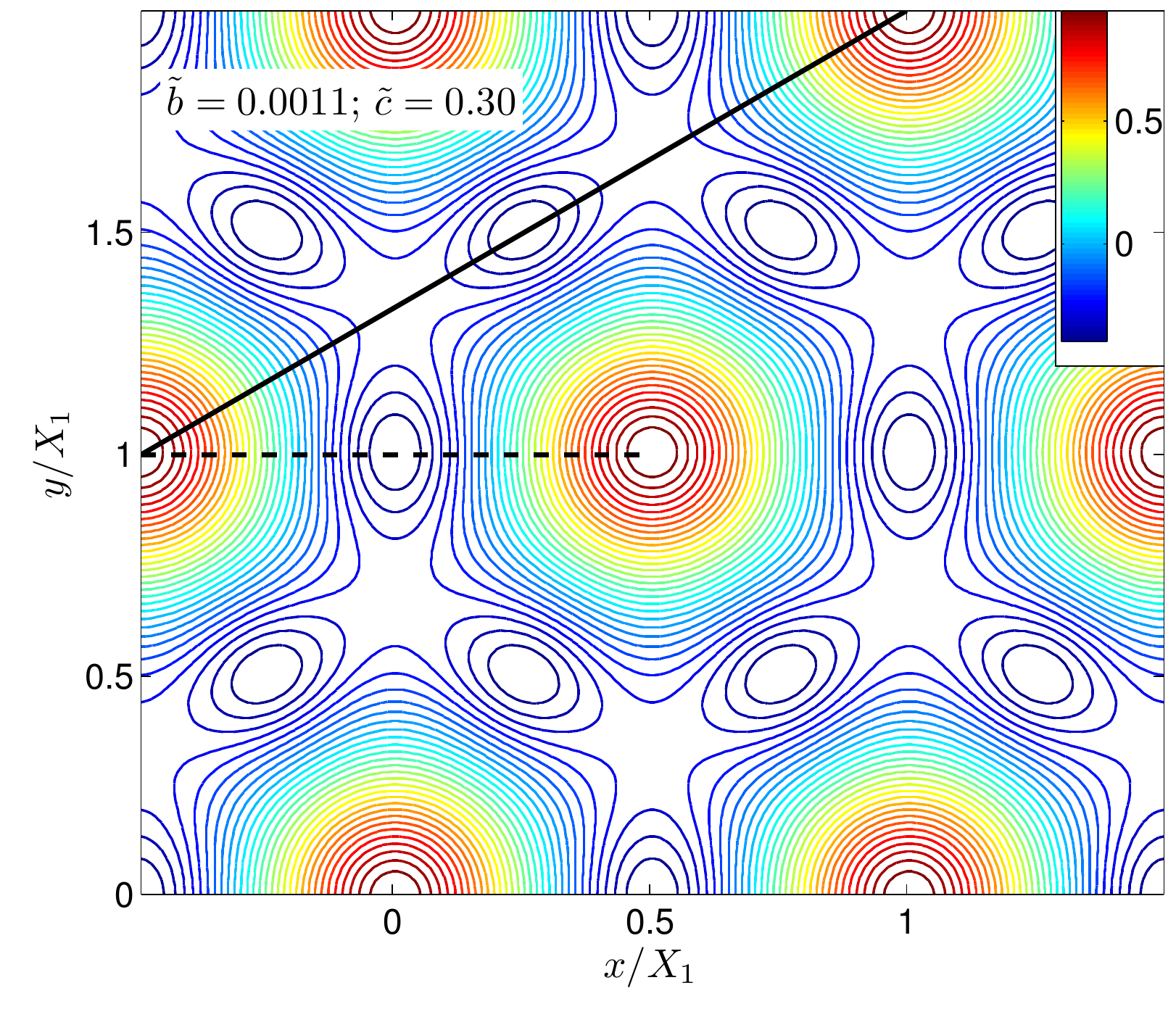}
\includegraphics[width=0.3\linewidth]{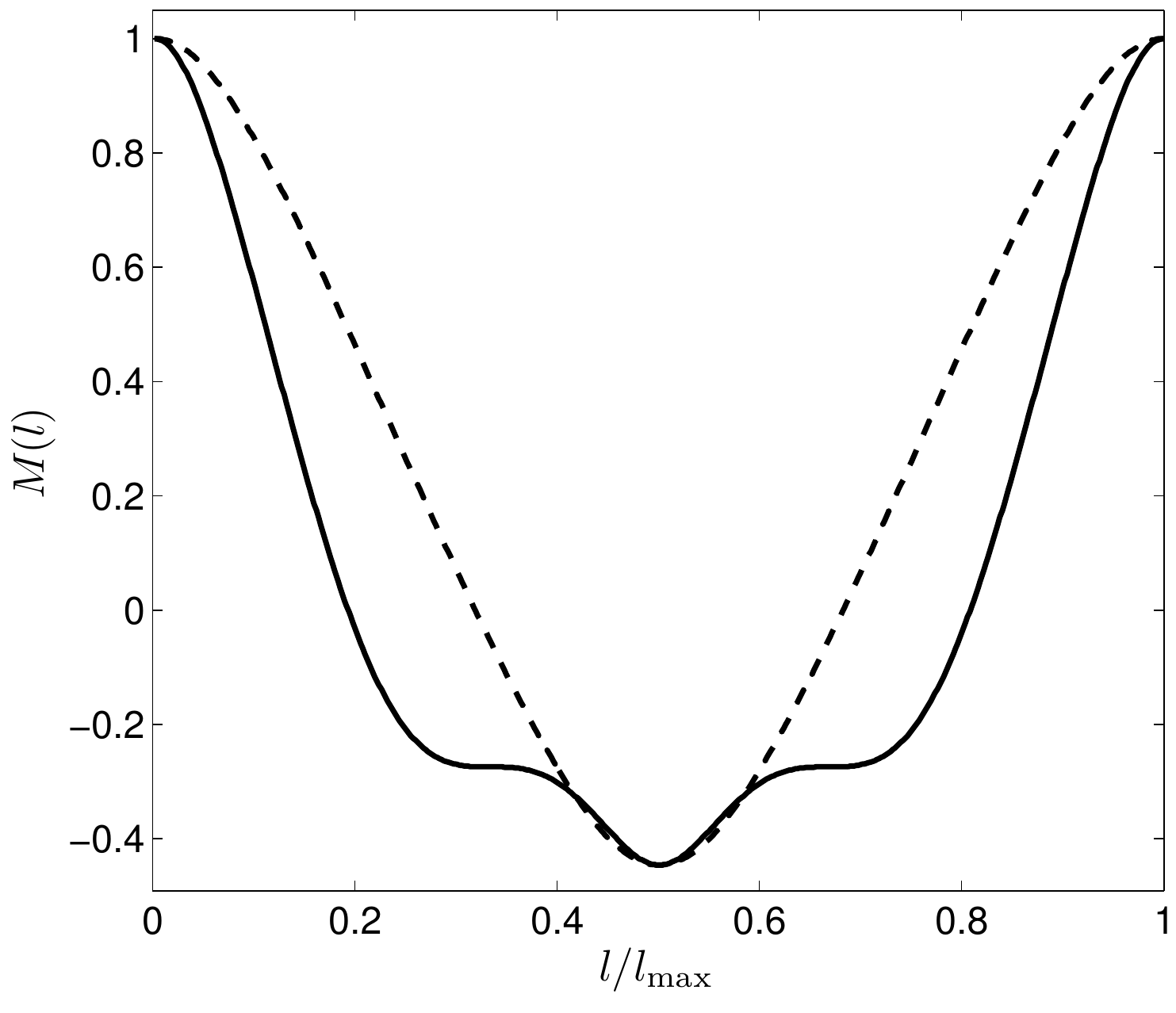}
\includegraphics[width=0.3\linewidth]{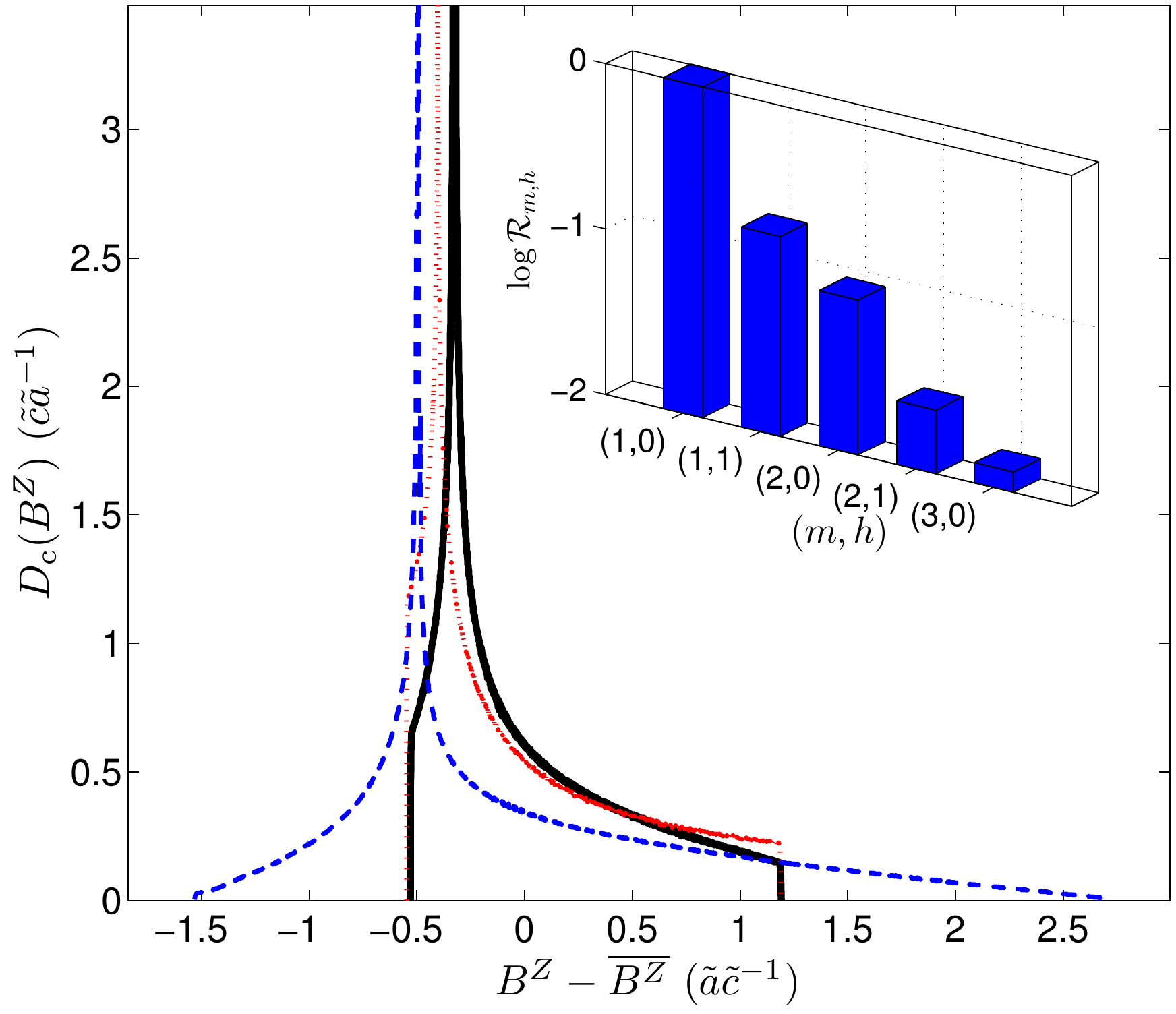}\\
\includegraphics[width=0.3\linewidth]{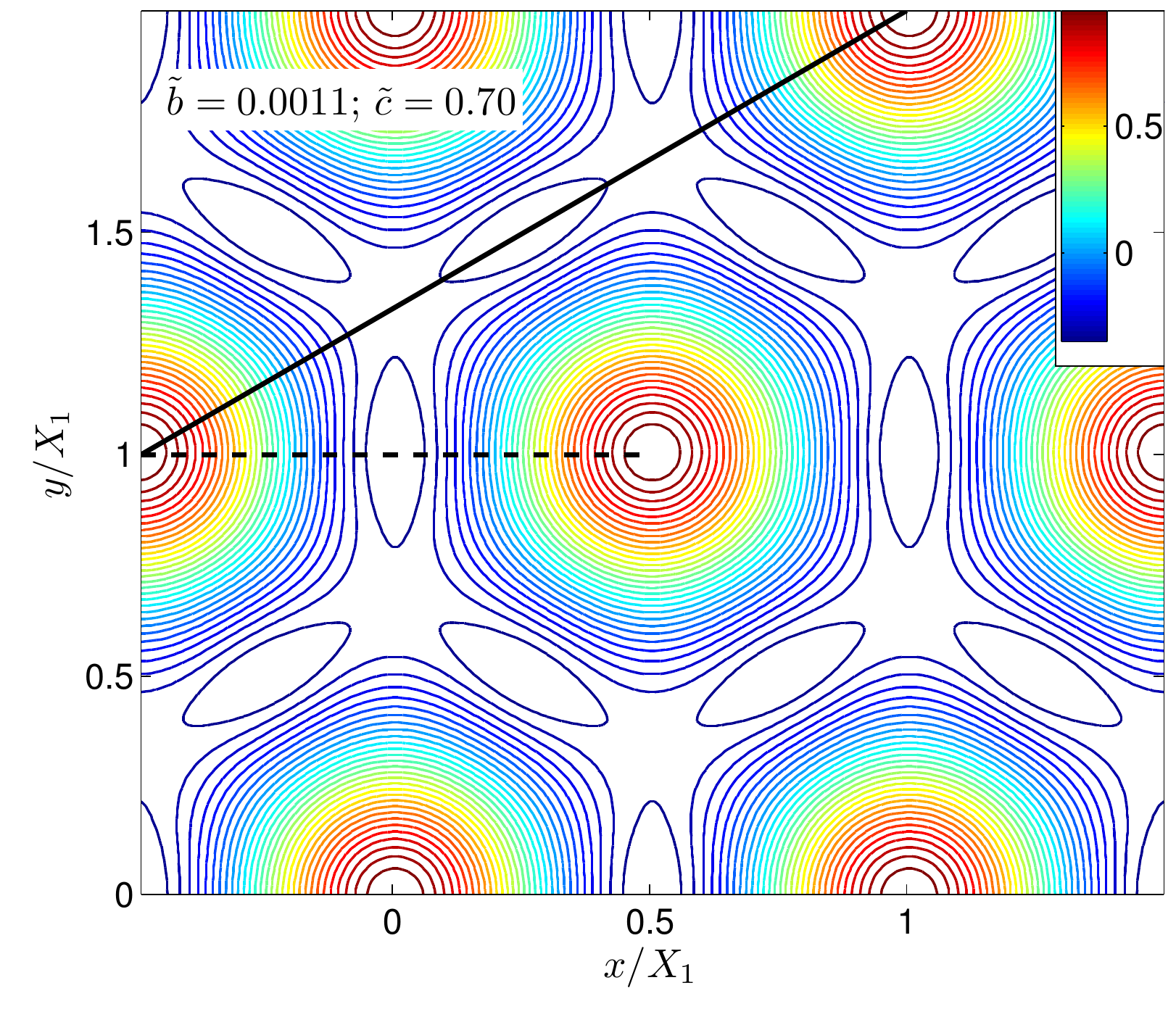}
\includegraphics[width=0.3\linewidth]{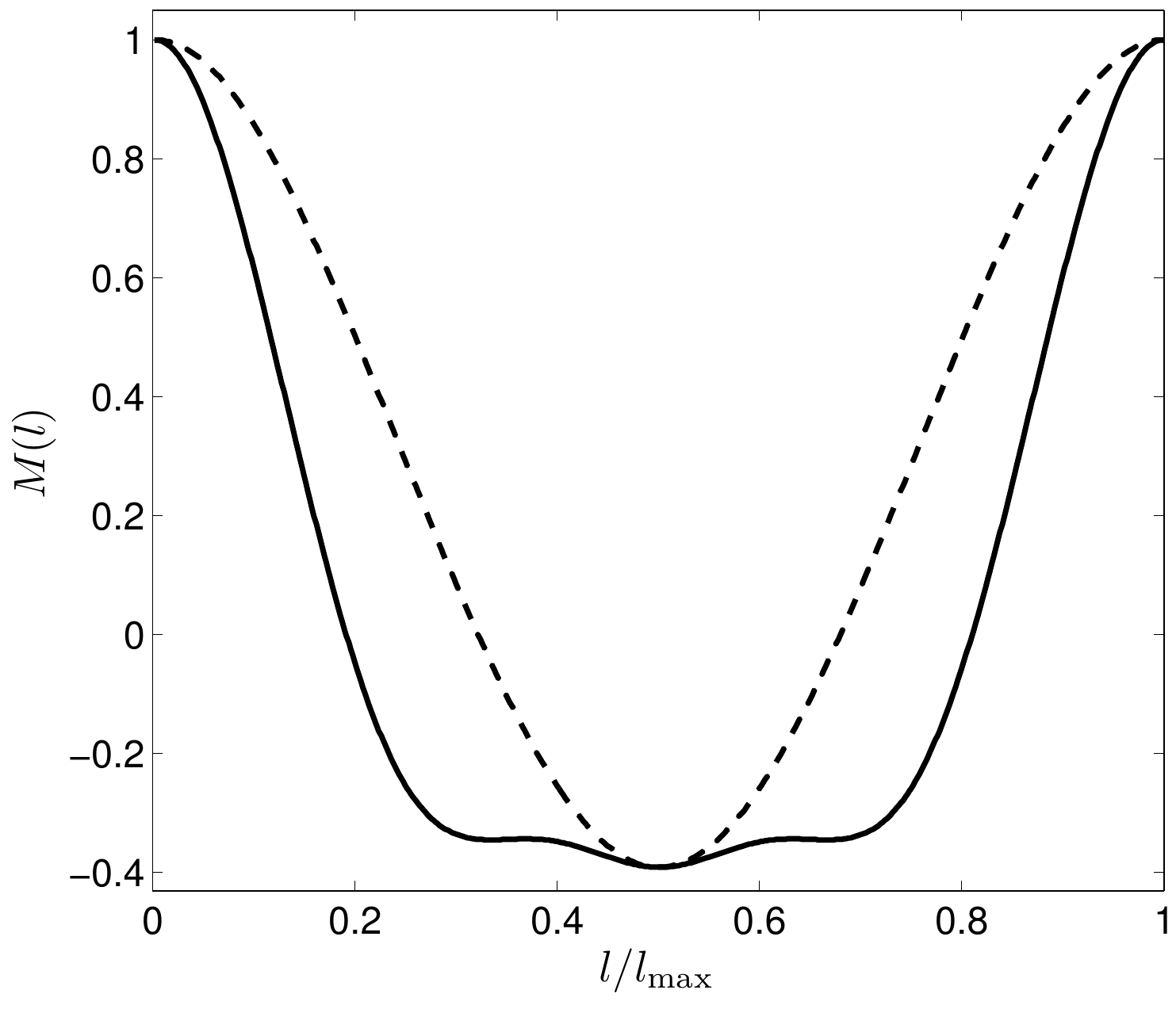}
\includegraphics[width=0.3\linewidth]{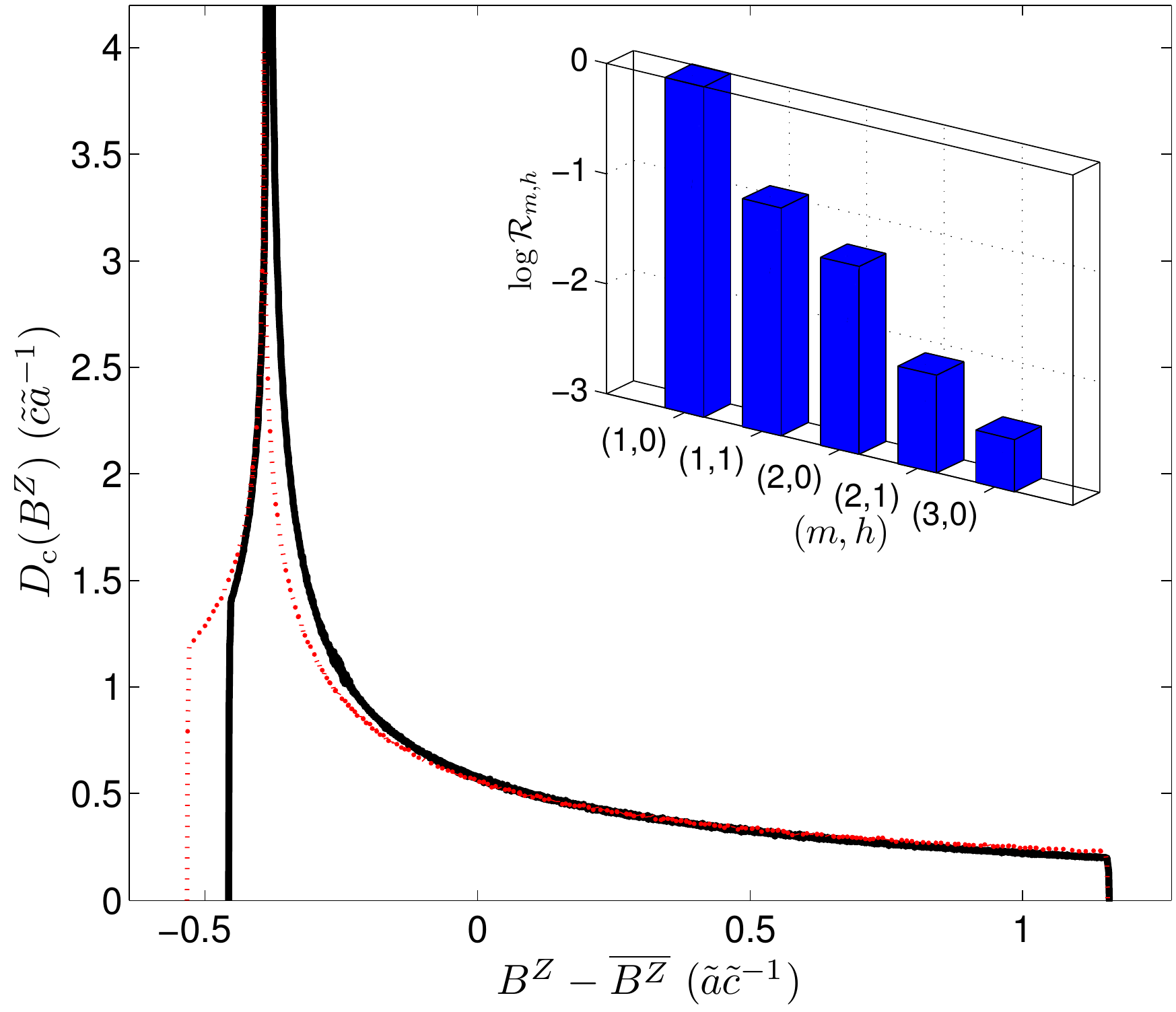}\\
\caption{(Color online) (left panel) Contour plot
$M(x,y)=[B^Z(x,y)- {\overline {B^Z}}]/[B^Z_{\rm vc}-{\overline {B^Z} }]$
for $\tilde{b}=0.0011$ and four different $\tilde{c}$ values, i.e.
$\tilde{c}=0.01$, 0.10, 0.30, and 0.70,  from top to bottom.
(middle panel) Field profile along the solid and the dashed lines shown in the left panel.
(right panel) The corresponding component field distribution $D_{\rm c}(B^Z)$ is shown as a solid black
line. For comparison, we present in dashed blue and dotted red lines $D_{\rm c}(B^Z)$ for
$\tilde{c}\rightarrow 0$ and $\tilde{c}\rightarrow \infty$,
respectively. In order to match the vortex core field $B^Z_{\rm vc}$, the
horizontal axis for $D_{\rm c}(B^Z)$ shown with the dotted red line
($\tilde{c}\rightarrow \infty$) has been  scaled (${\overline {B^Z} }$ is identical for all the curves).
The insert shows the values of
${\mathcal R}_{m,h} = \left| B^Z_{{\bf K}_{m,h}}/B^Z_{{\bf K}_{1,0}}\right|$ for the first five form factors
in logarithm scale. To simplify the drawings, and without any lost of information,
the dashed blue line for the $D_{\rm c}(B^Z)$ plots (corresponding to $\tilde{c}\rightarrow 0$) is presented
only when ${\tilde c} \leq 0.30$.}
\label{fig1}
\end{figure*}

\begin{figure*}
\includegraphics[width=0.3\linewidth]{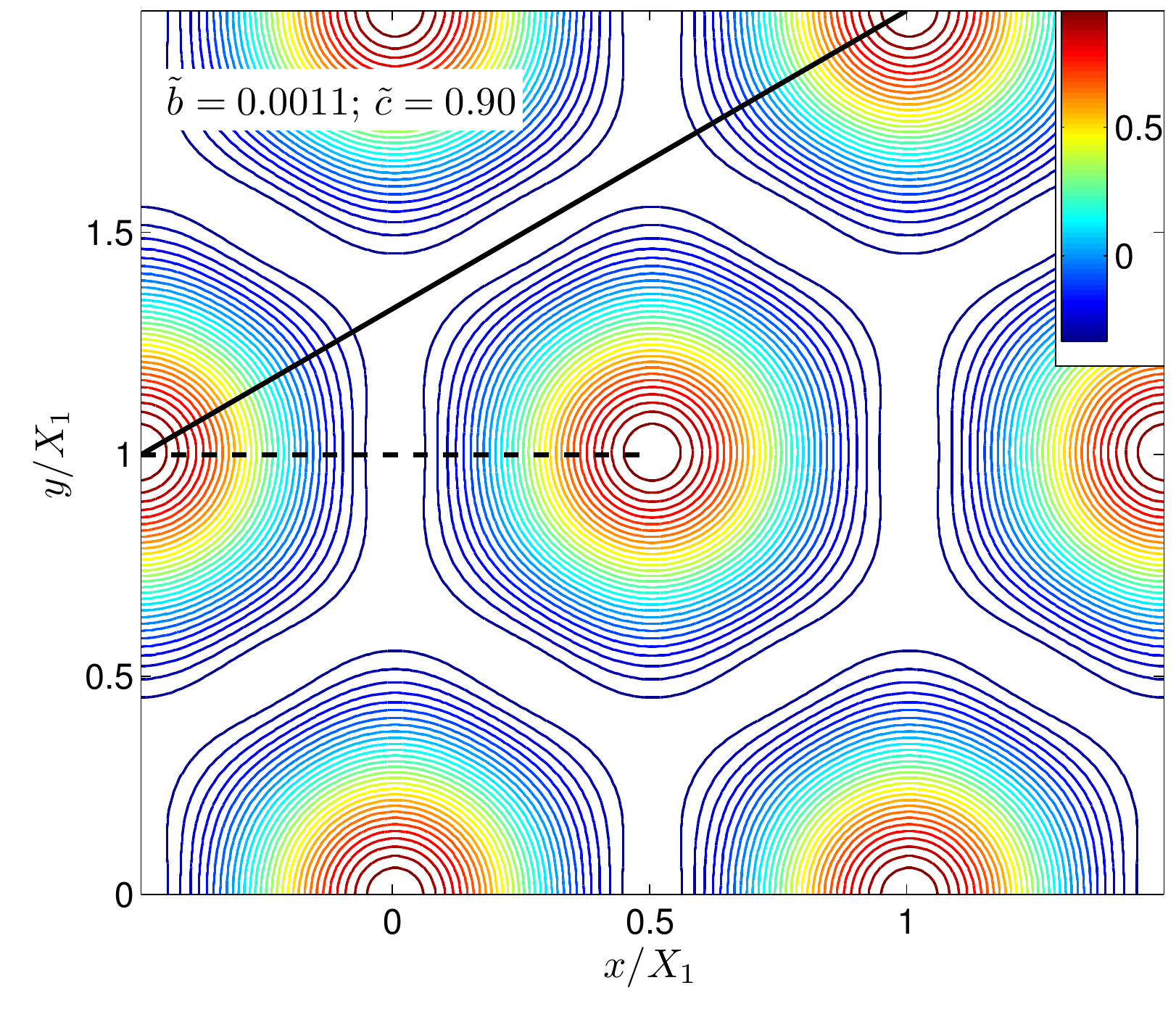}
\includegraphics[width=0.3\linewidth]{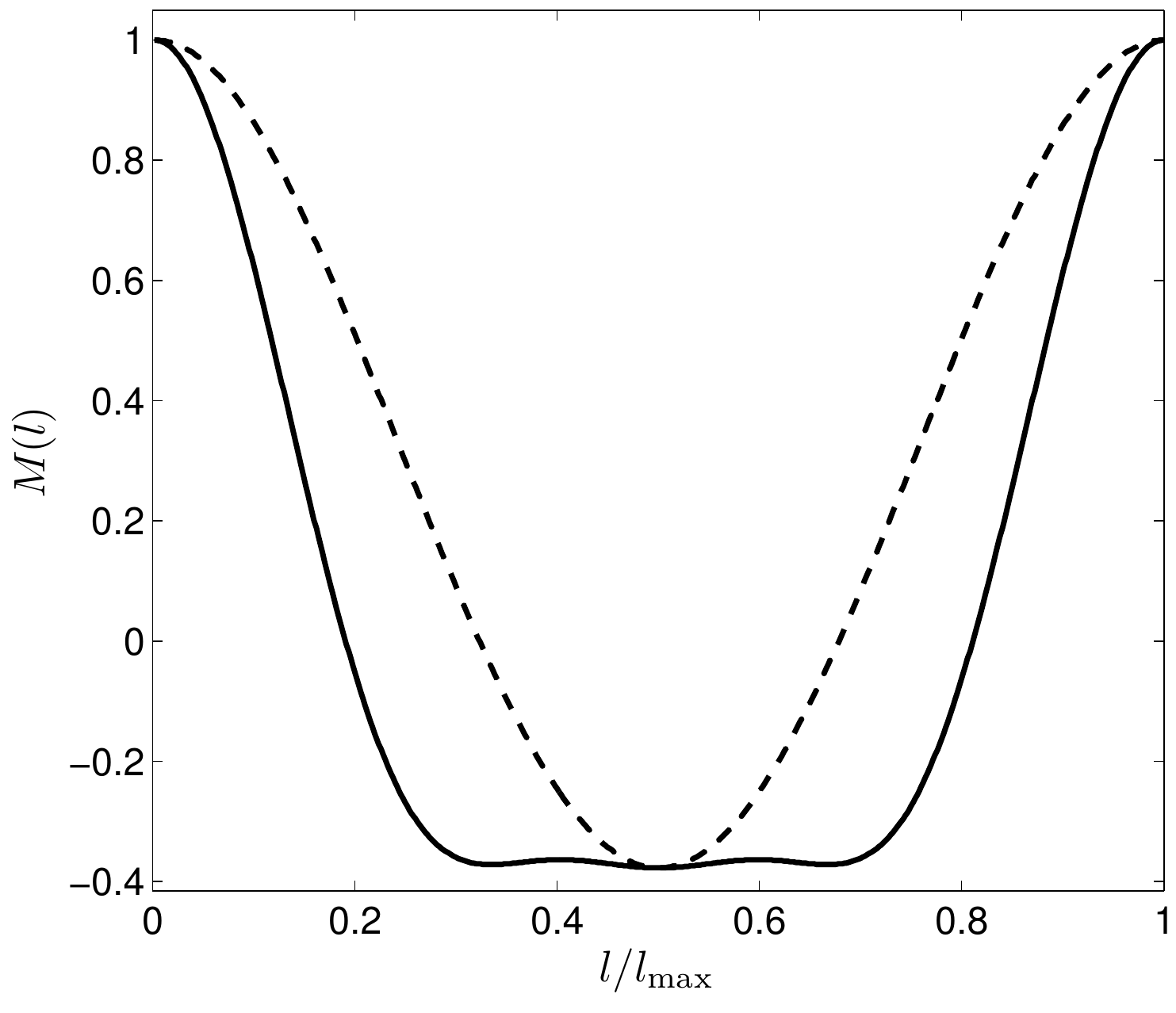}
\includegraphics[width=0.3\linewidth]{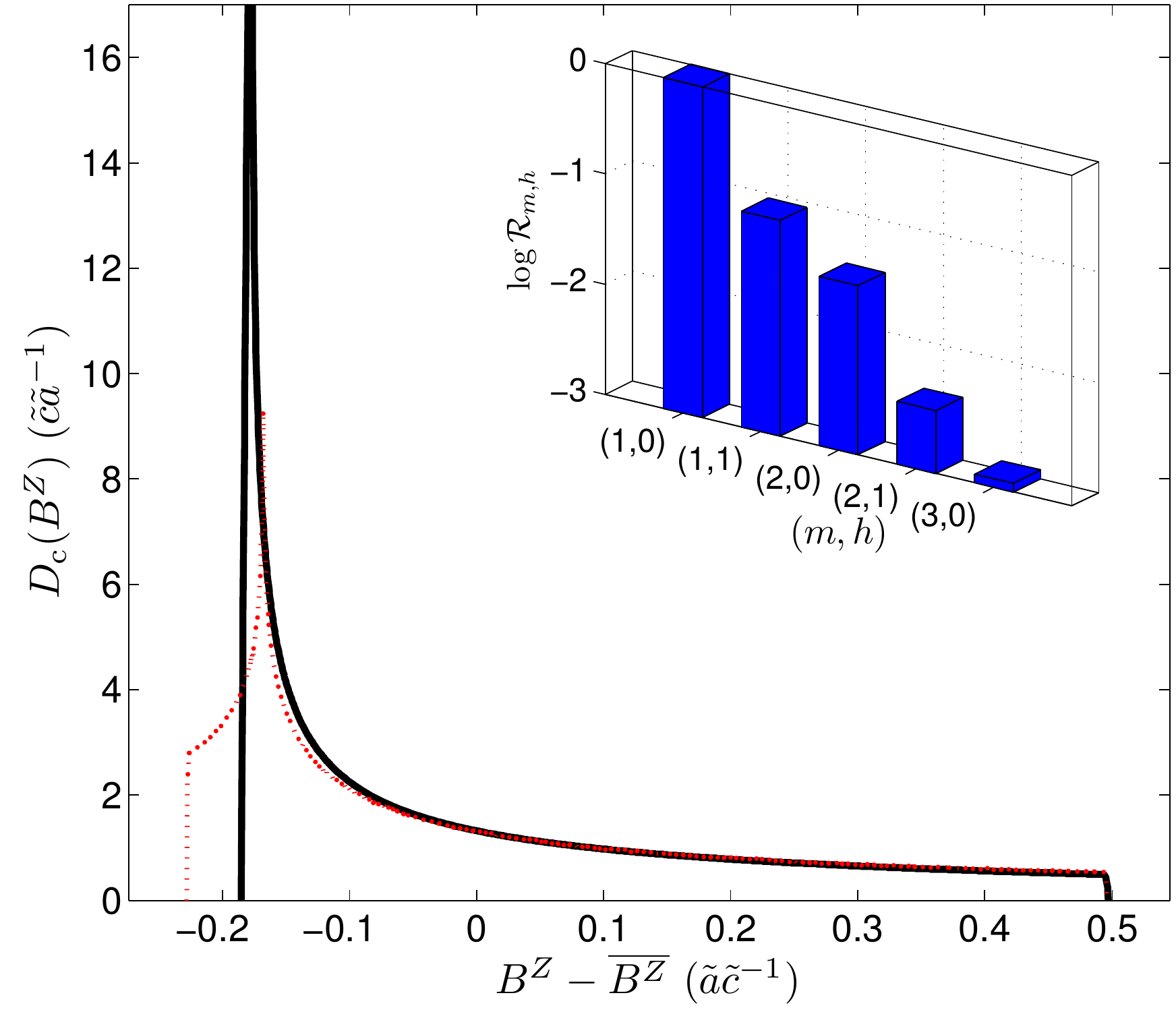}\\
\includegraphics[width=0.3\linewidth]{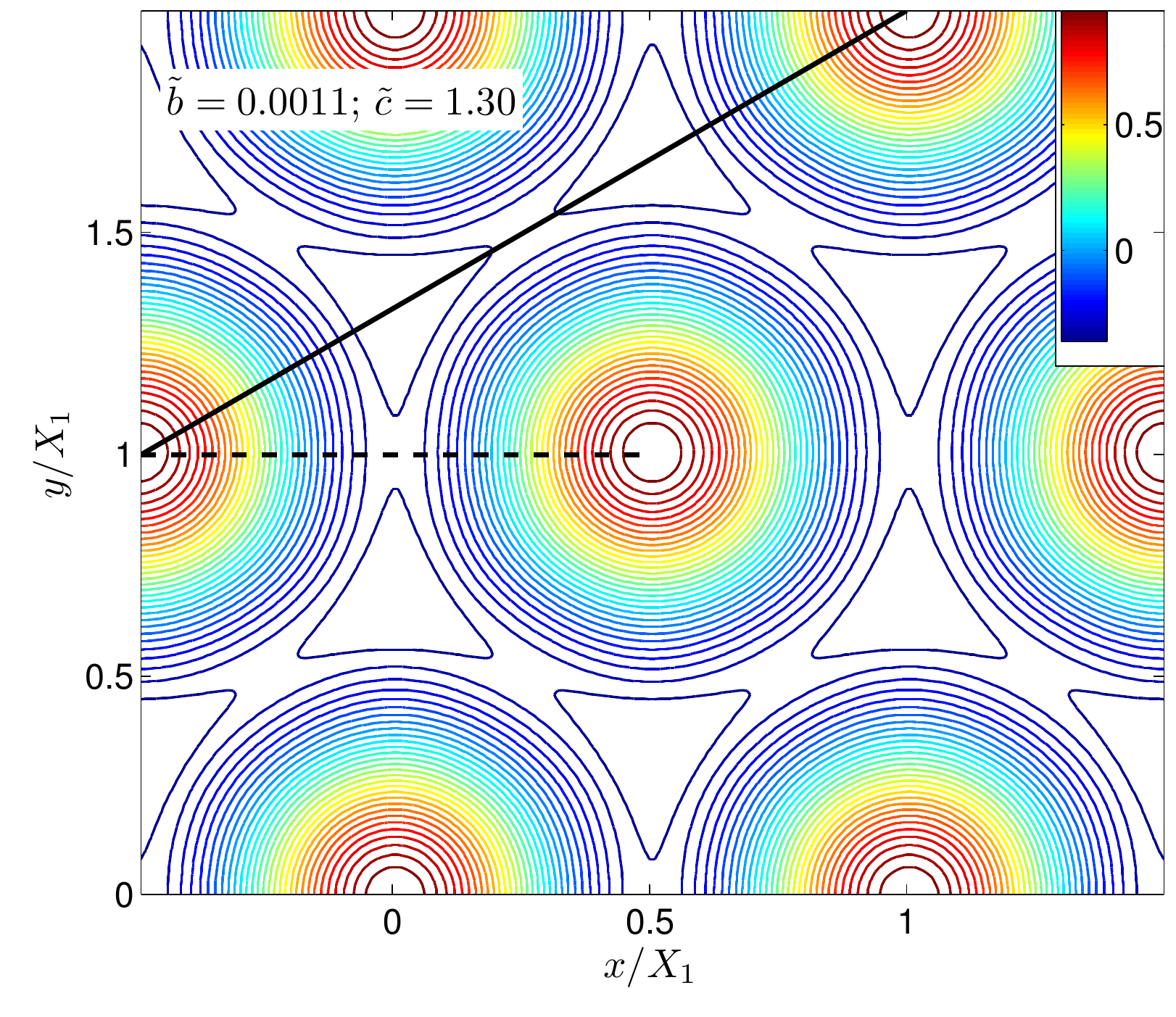}
\includegraphics[width=0.3\linewidth]{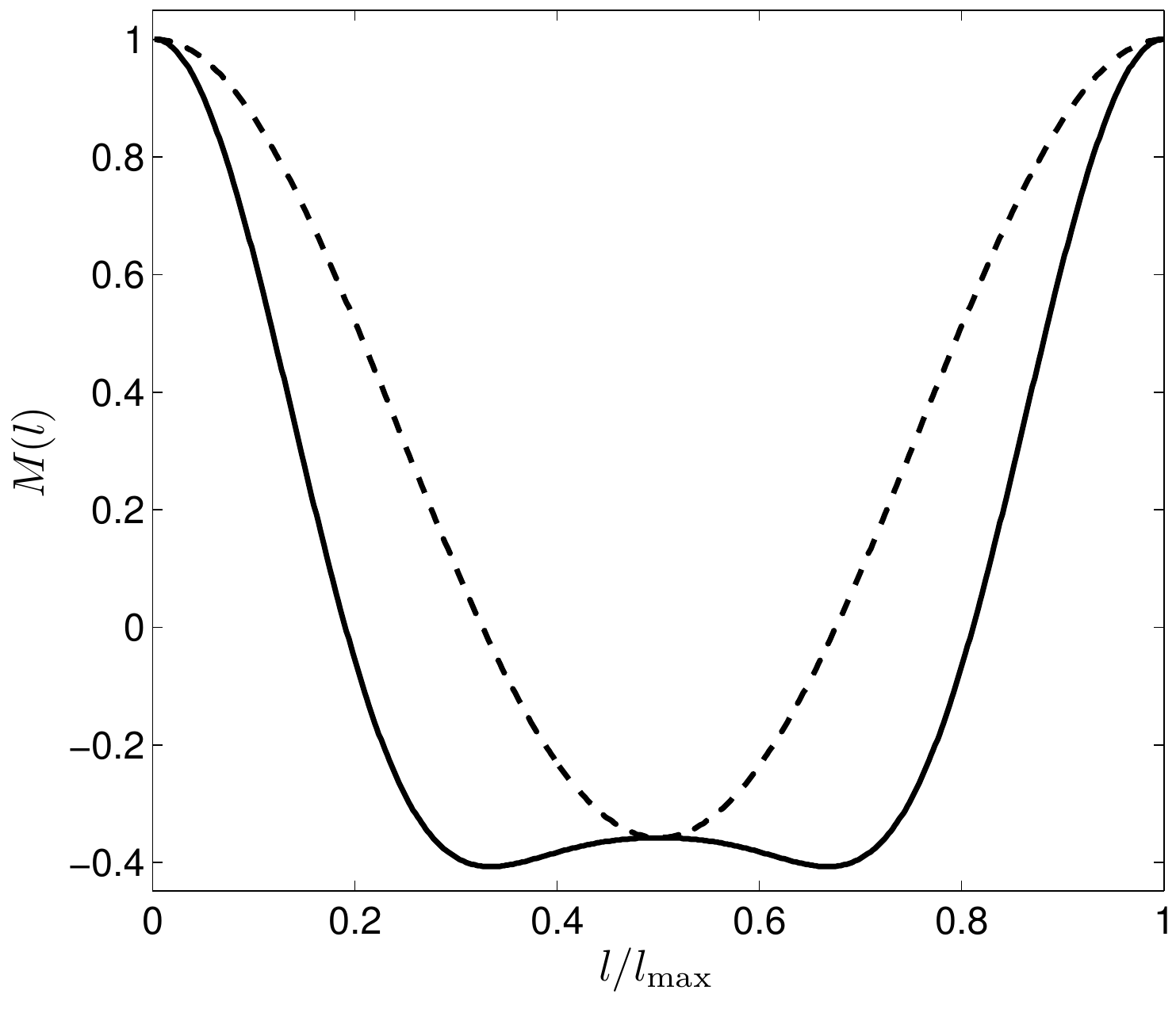}
\includegraphics[width=0.3\linewidth]{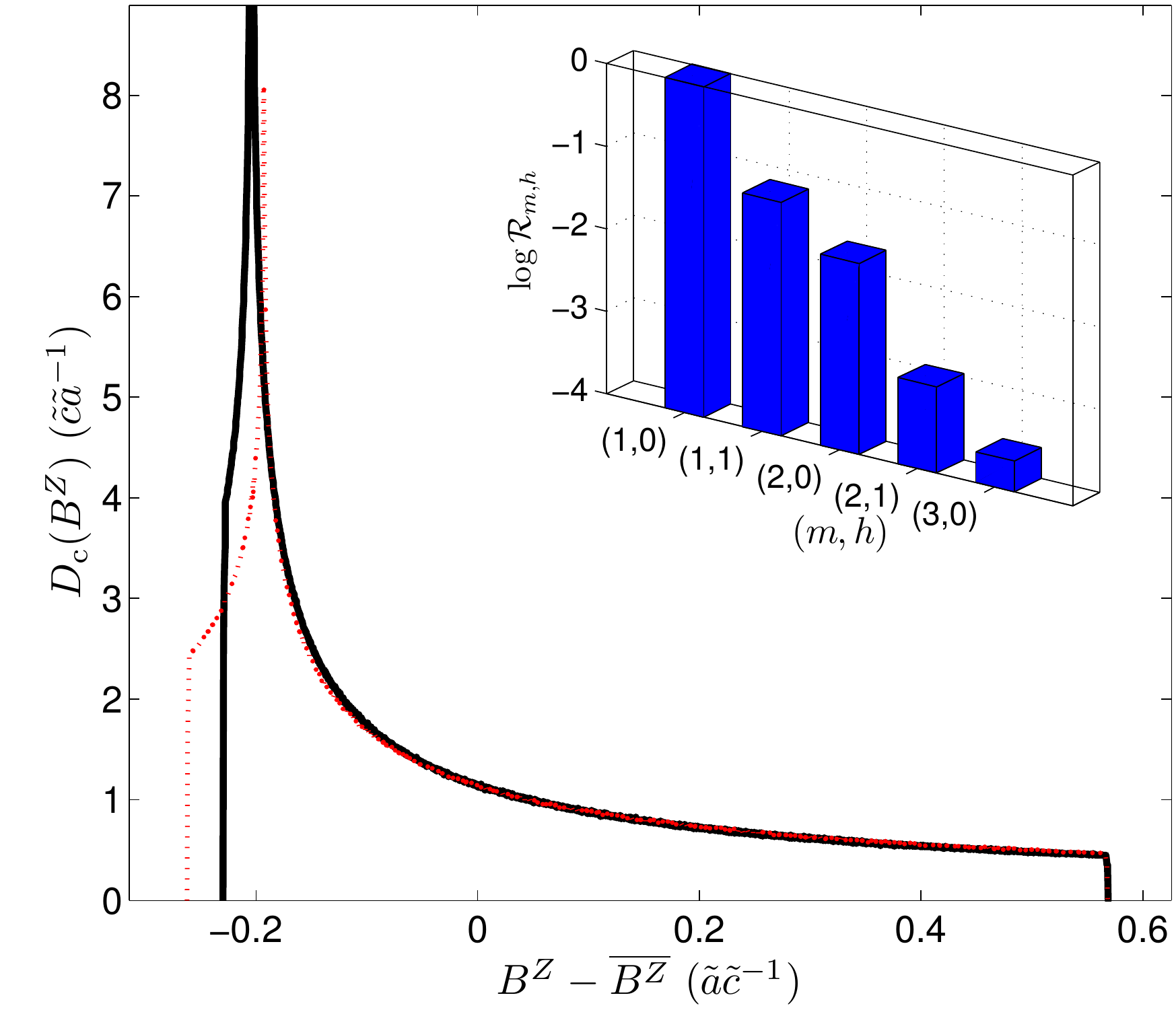}\\
\includegraphics[width=0.3\linewidth]{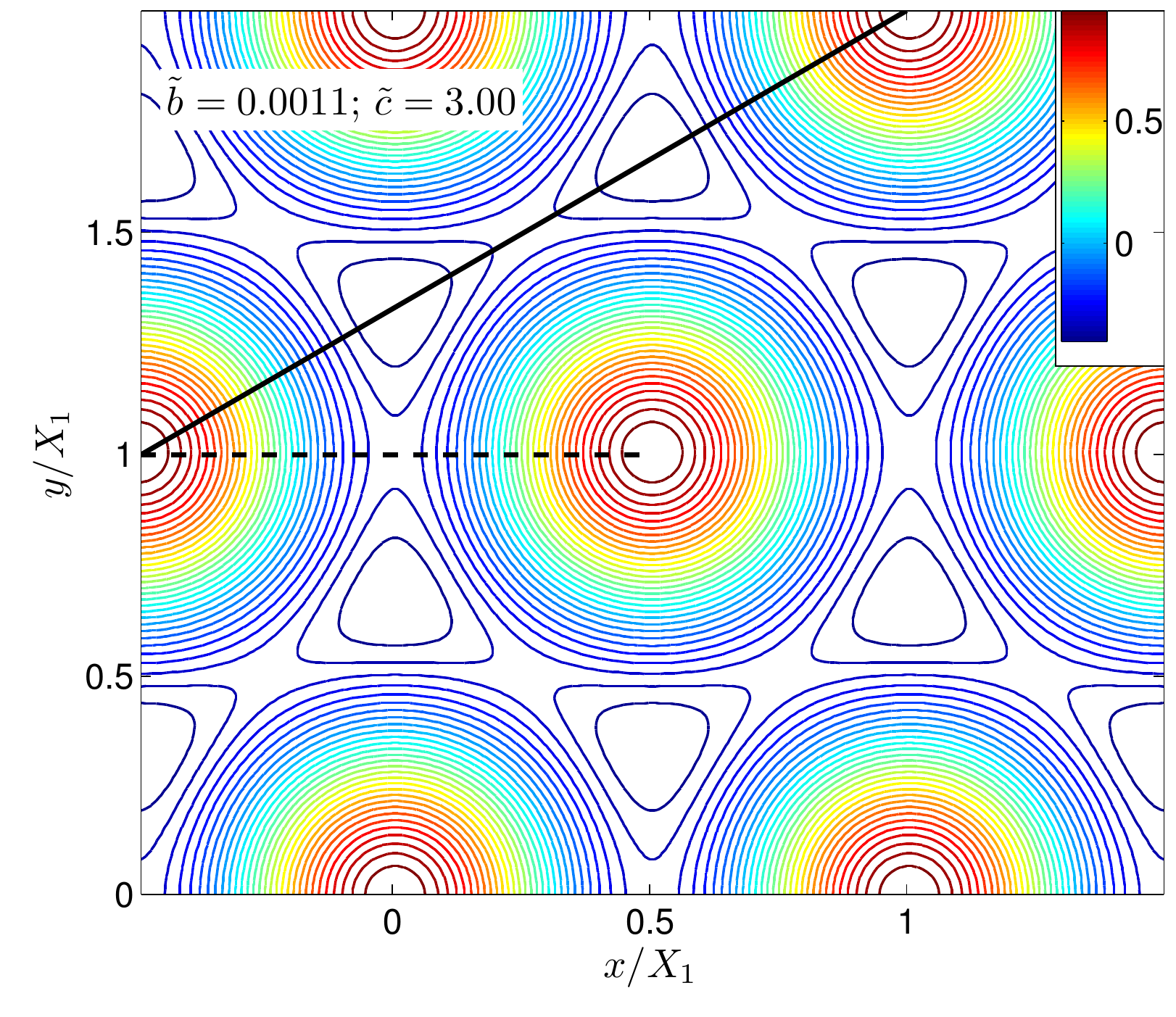}
\includegraphics[width=0.3\linewidth]{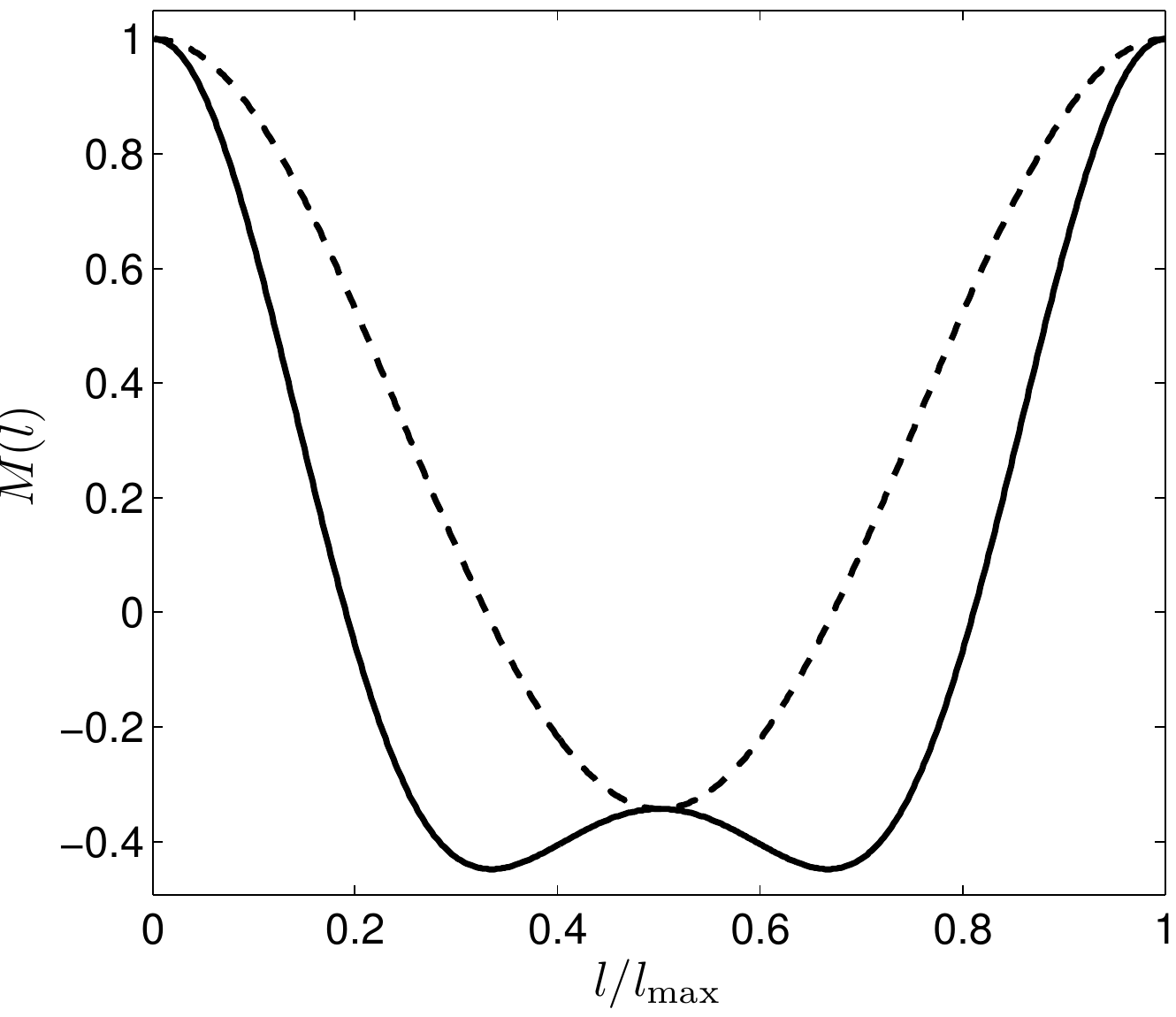}
\includegraphics[width=0.3\linewidth]{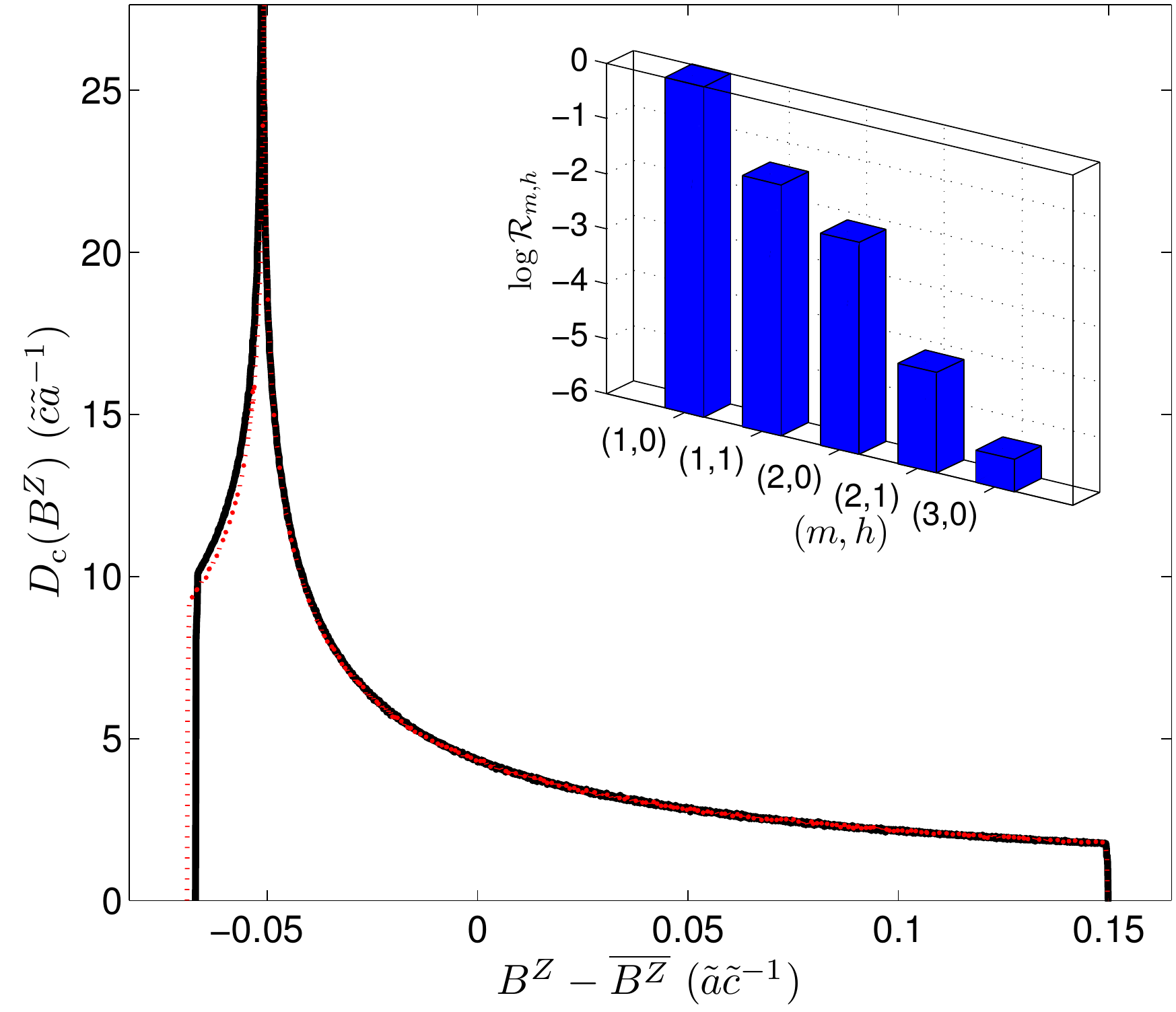}\\
\includegraphics[width=0.3\linewidth]{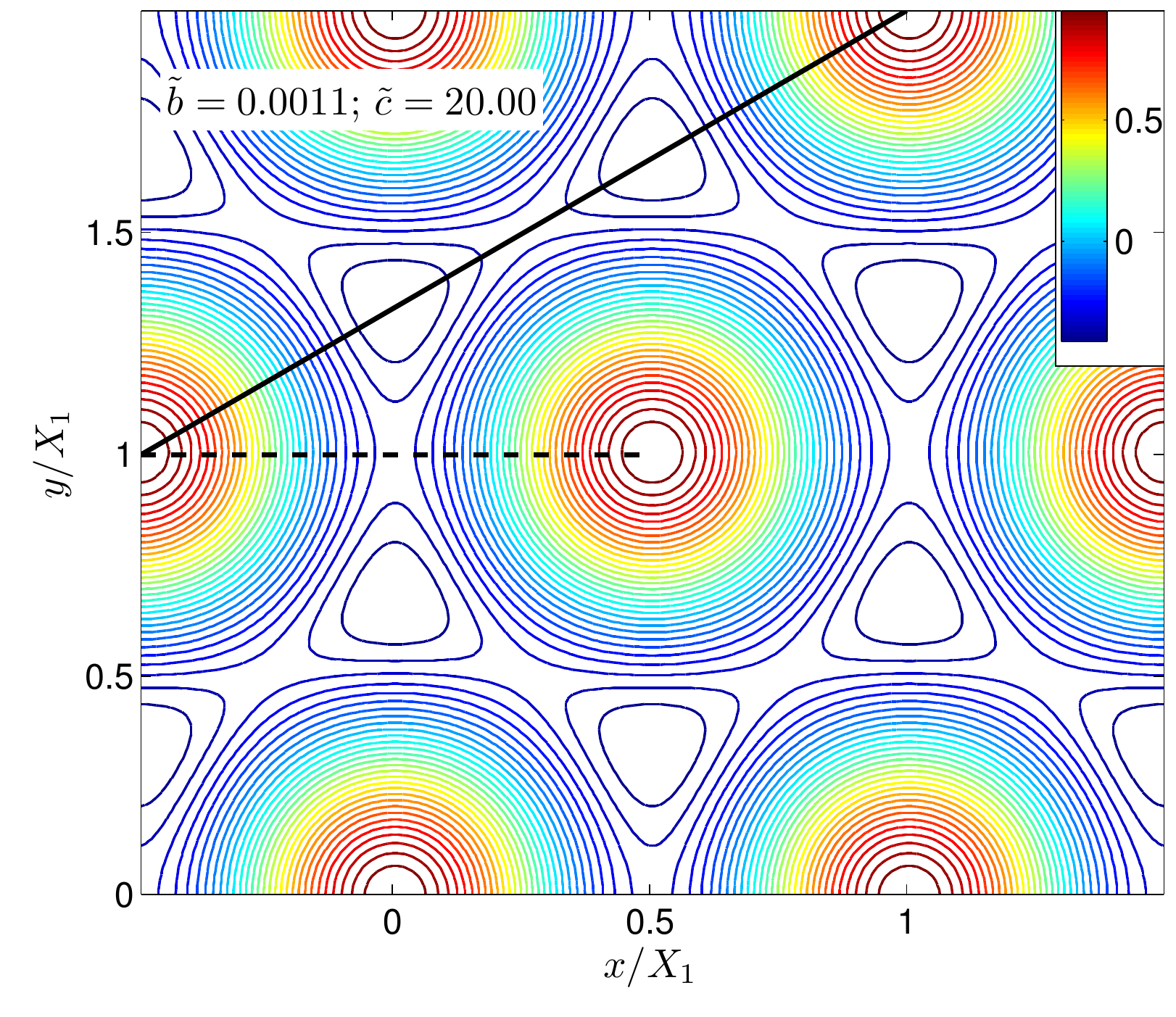}
\includegraphics[width=0.3\linewidth]{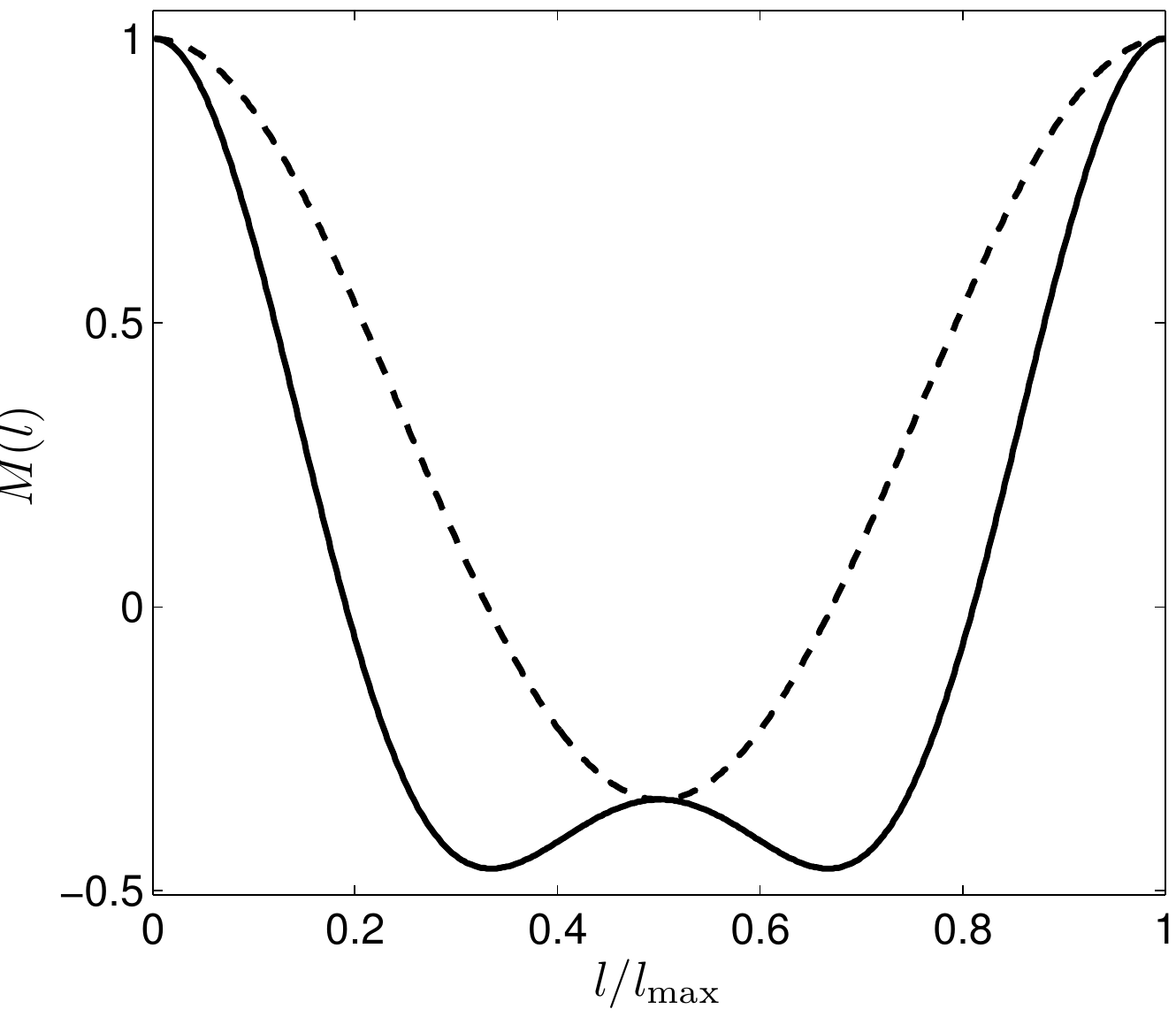}
\includegraphics[width=0.3\linewidth]{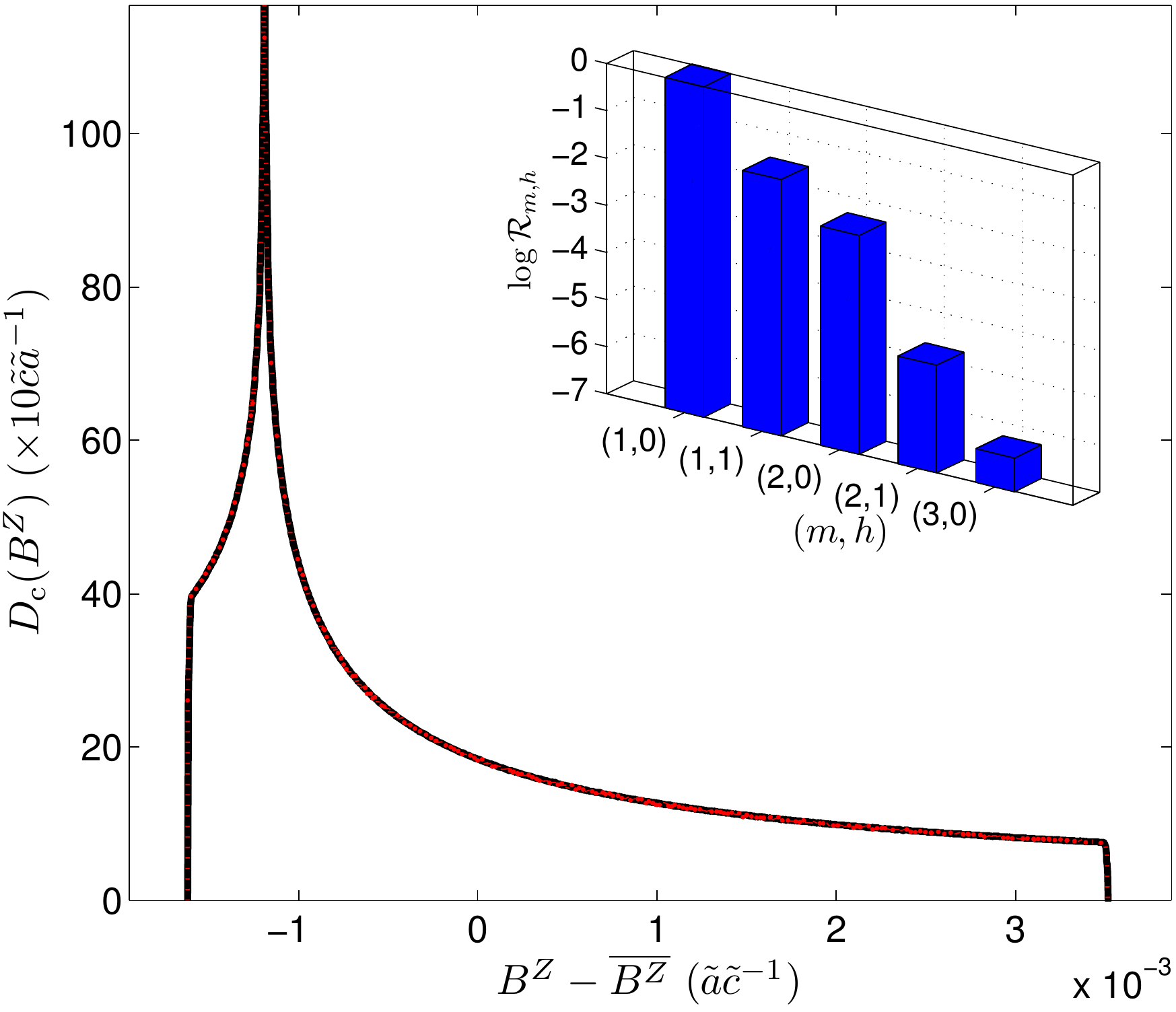}\\
\caption{ (Color online) The same as in Fig.~\ref{fig1}, but for
$\tilde{c}=0.90$, $1.30$, $3.00$, and  $20.00$, from top to bottom, respectively.
}
\label{fig2}
\end{figure*}

The analysis of $\tilde{b}$ and $\tilde{c}$ suggests to start our discussion of the
VL properties focusing on its $\tilde c$ dependence
near the $\tilde{b}\rightarrow 0$ limit. In Figs.~\ref{fig1} and \ref{fig2}
we show the contour plots of the spatial field distribution for
$\tilde{b}=0.0011$ ({\it i.e.} $b=0.99$),
and $\tilde{c} = 0.01$, 0.10, 0.30, 0.70, 0.90, 1.30, 3.00,  and 20.00.
Comparing the data at the top of Fig.~\ref{fig1} and bottom of Fig.~\ref{fig2}, the field map, profile,
field distribution, and form factor intensity are drastically different at low temperature
and near $T_{\rm c0}$. At $\tilde{c} = \tilde{c}_{\rm cross} \simeq 0.9$ a crossover occurs.
This value of $\tilde{c}_{\rm cross}$ corresponds to $T\simeq 0.6 T_{\rm c0}$ in agreement with
a report of Brandt (this value depends on $v_{\rm F}$).\cite{Brandt95}

For $\tilde{c}> \tilde {c}_{\rm cross}$ we do observe behaviours
expected in the GL regime.\cite{Yaouanc11,Annett04}
A minimum for the spatial field distribution is found in between three neighboring vortices,
while in  between two neighboring vortices, i.e. at midpoint on
the line connecting two neighboring vortices, a field saddle point is located.
The signature of the minimum of the field is obvious in $D_{\rm c}(B^Z)$.
The saddle point corresponds to the maximum in
$D_{\rm c}(B^Z)$.
As shown in the inserts of the field distributions, the intensity of the
form factor is expected to be reduced for large indices.

For $\tilde{c}<\tilde{c}_{\rm cross}$ the positions of the minimum and saddle point are
reversed. The shape of the profiles are substantially different.
In the limit $\tilde{c}\rightarrow 0$ the conical shape of the vortex profiles near the positions of the
minimum and saddle point are particularly pronounced. This property is even more clearly seen
at the minimum-field point as shown by the dashed line in the field profile.
This feature reflects directly the weak power-law decay of the form factor which is
a consequence of the Cooper's pair diffraction by the vortex cores. This can be qualitatively
understood if we recall that the weight of the $k$ harmonic for the infinite
Fourier series of a triangle wave -- a wave with a profile similar to the dashed line --
 is proportional to $(-1)^k/(2 k +1)^2$. Note
the alternating sign as in the $B^Z_{{\bf K}_{m,h}}$ expression in the $T=0$ limit,
as well as the power law decay. As reflected by the different exponents of the
power-law decays, this comparison is only qualitative. This is not surprising given the two
dimensional nature of the field map. The triangle-wave model has only one
dimension. For $\tilde{c}\geq 0.70$ the BCS solution
and the GL limit,  {\it i.e.}
$T \rightarrow T_{\rm c0}$, predict similar high-field $D_{\rm c}(B^Z)$.
But the low field features are still different.
As noticed from Fig.~\ref{fig1},
it is not required to go to extremely low temperature to observe Bragg's spots of
large indices when $B_{\rm ext} \rightarrow B_{\rm c2}$.

A linear high-field tail is predicted in Fig. \ref{fig1} for $D_{\rm c}(B^Z)$.
However, it is only expected
at really low temperature, i.e. for ${ \tilde c} \leq 0.10$, in contrast to observation.\cite{Herlach90}
The $D_{\rm c} (B^Z)$ shape near the minimum field is really different near
$T=0$ and $T_{\rm c0}$. This difference remains to be seen experimentally.
A signature of the crossover for $D_{\rm c}(B^Z)$ is in its sharp rise at low
field without any shoulder. Although never reported, it could be observed experimentally.

Up to now we have focused our attention on the physics very near the $B_{\rm c2}$ phase boundary
when the parameter $\tilde{b}$ is negligible.
Now we study it as we go out of the boundary.

\begin{figure*}
\includegraphics[width=0.3\linewidth]{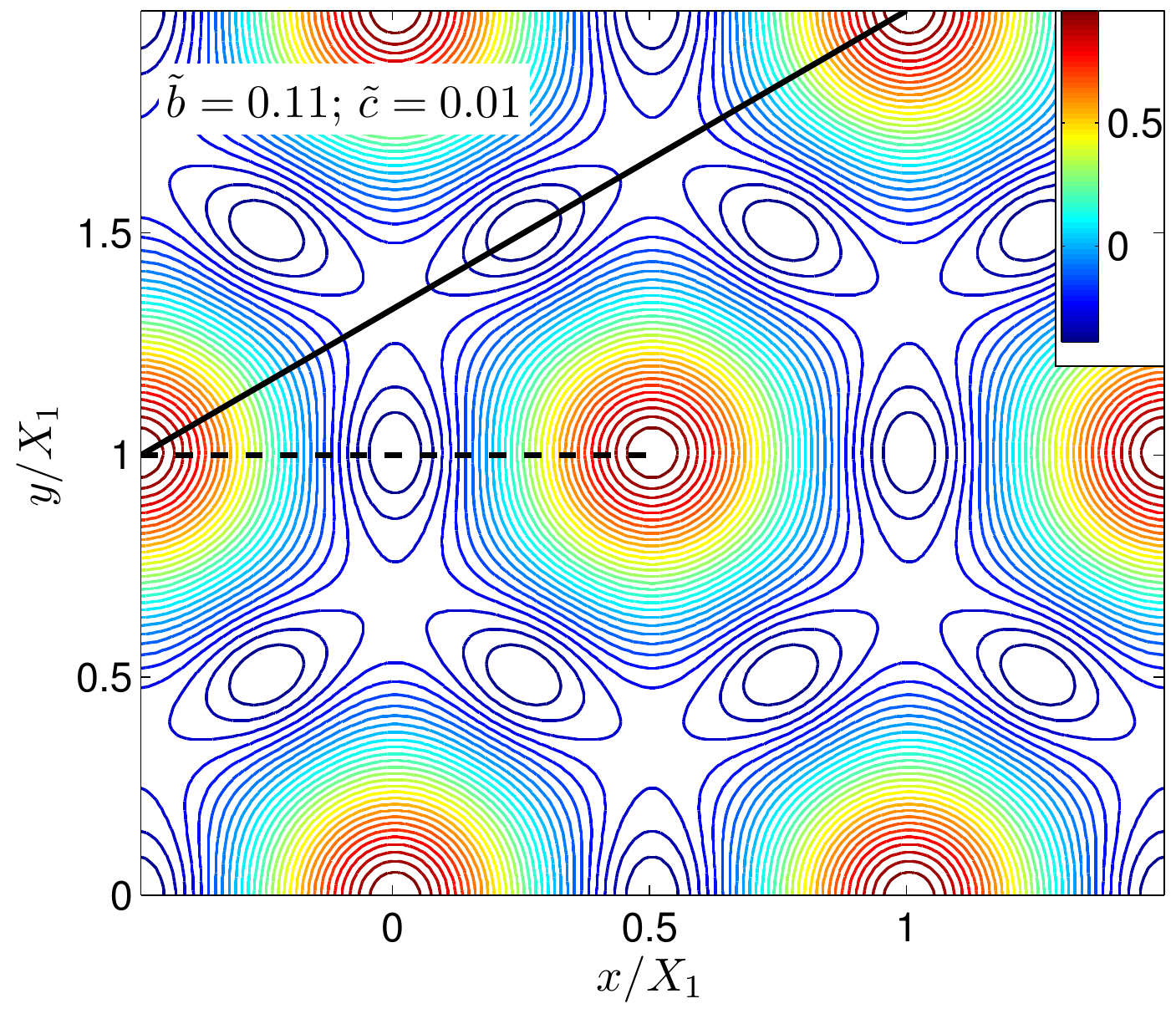}
\includegraphics[width=0.3\linewidth]{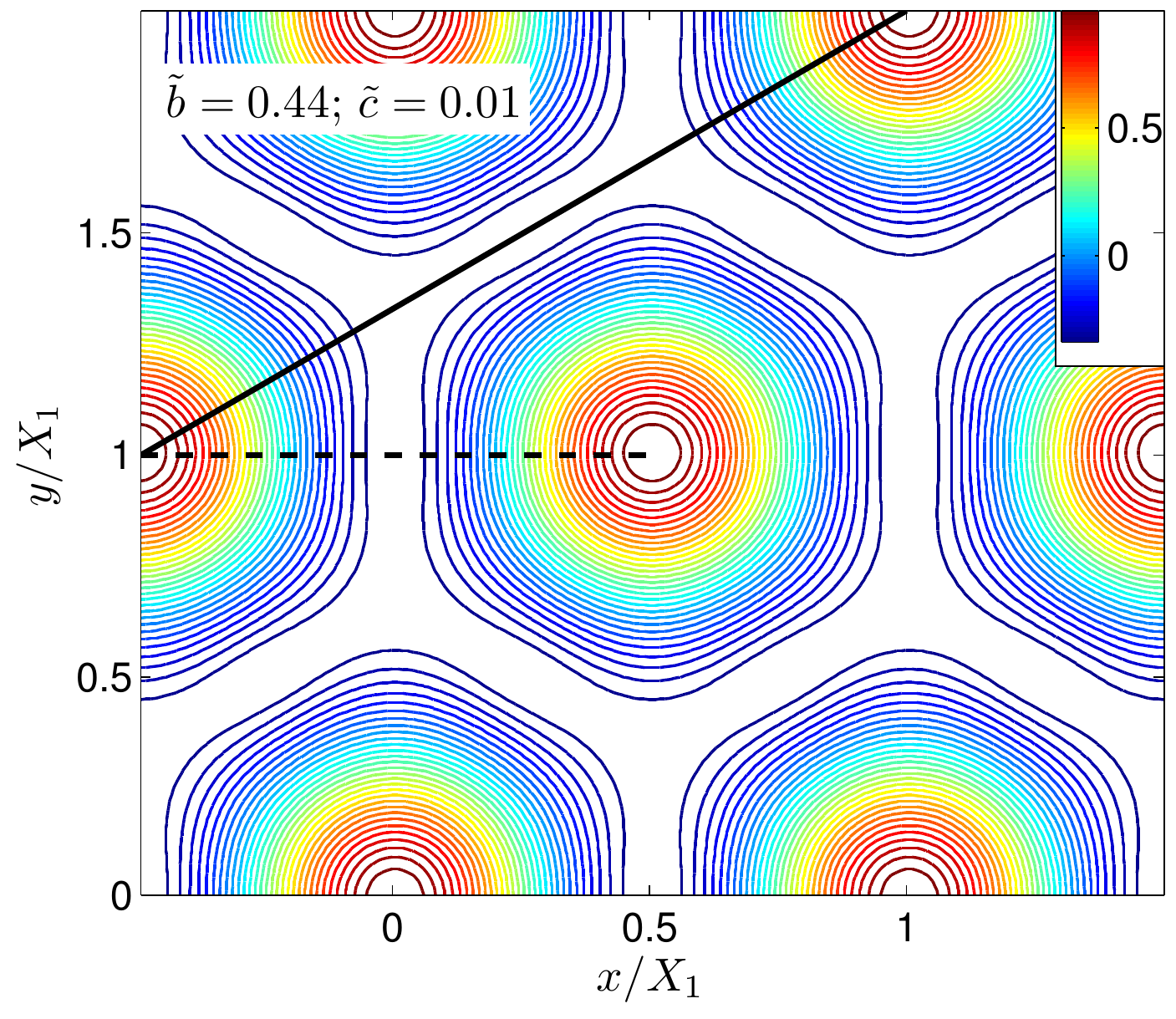}
\includegraphics[width=0.3\linewidth]{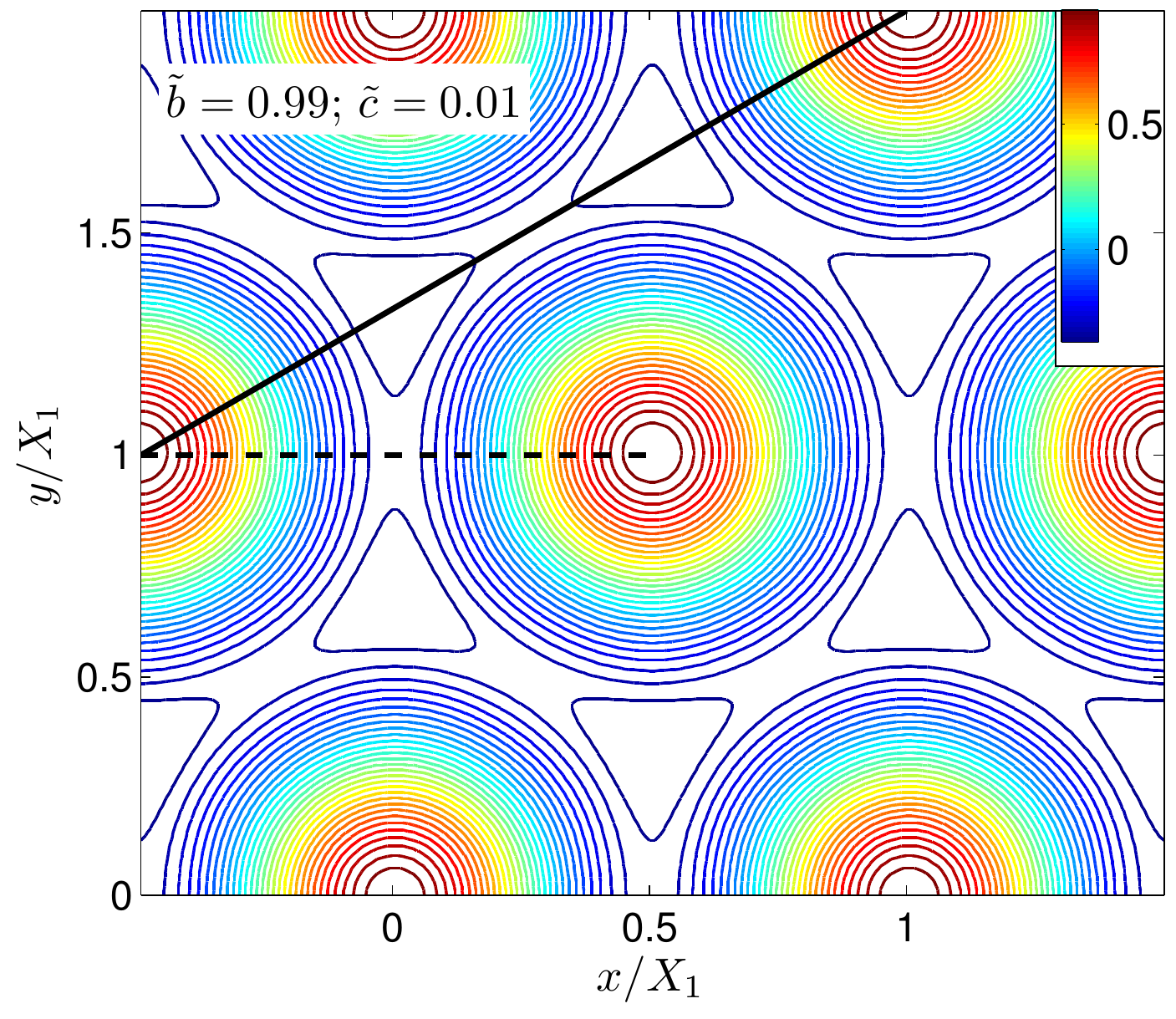}\\
\includegraphics[width=0.3\linewidth]{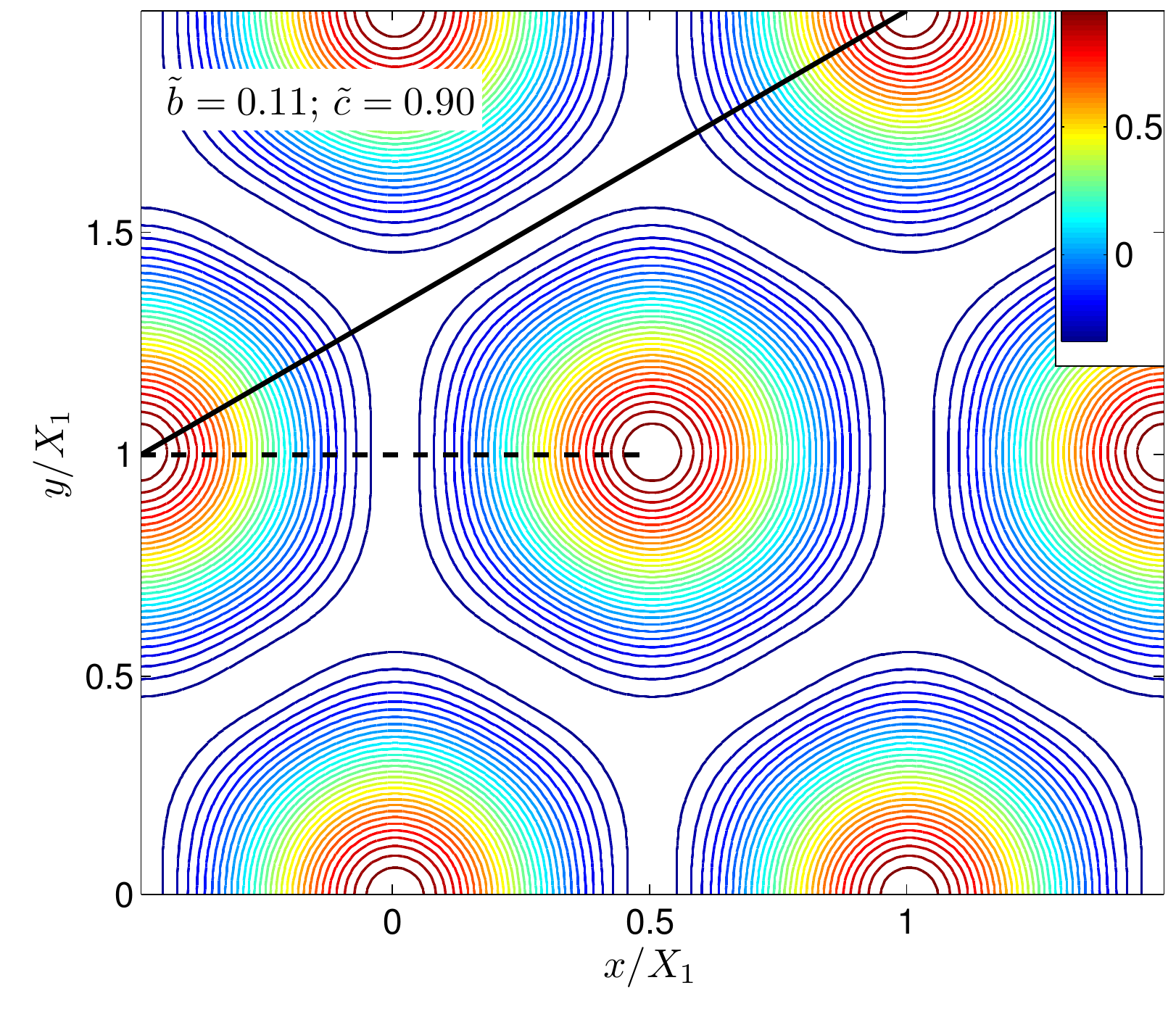}
\includegraphics[width=0.3\linewidth]{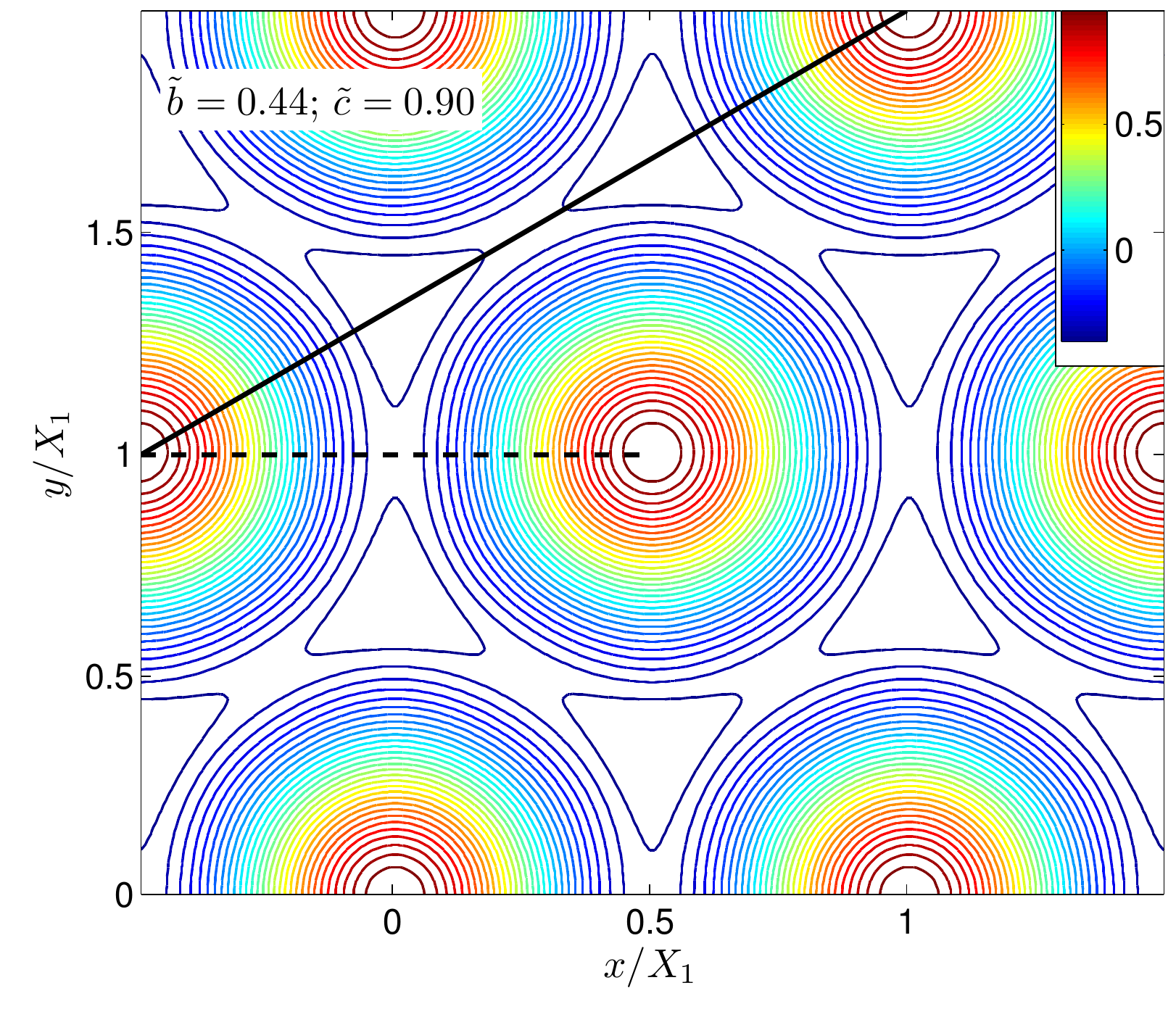}
\includegraphics[width=0.3\linewidth]{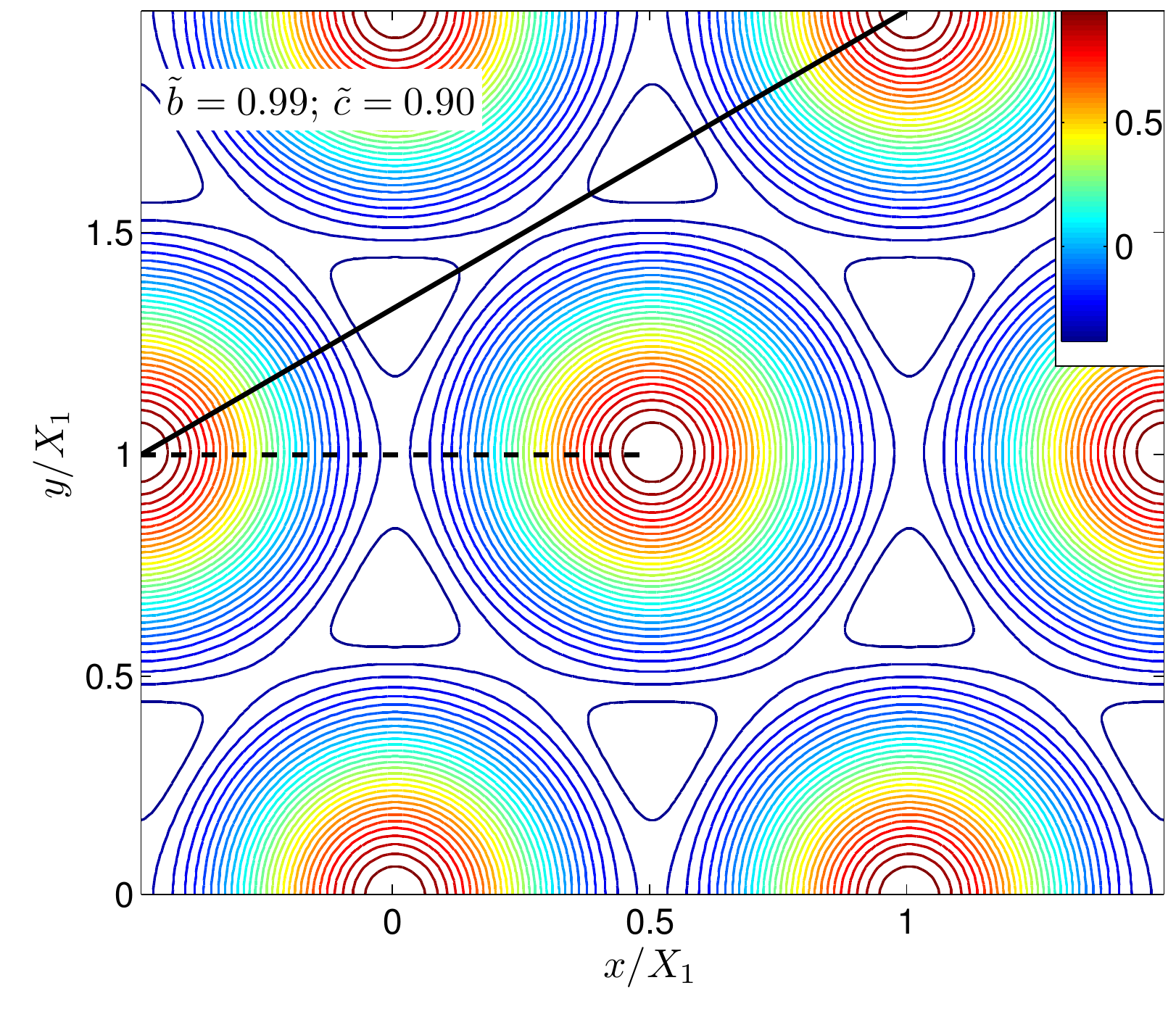}\\
\includegraphics[width=0.3\linewidth]{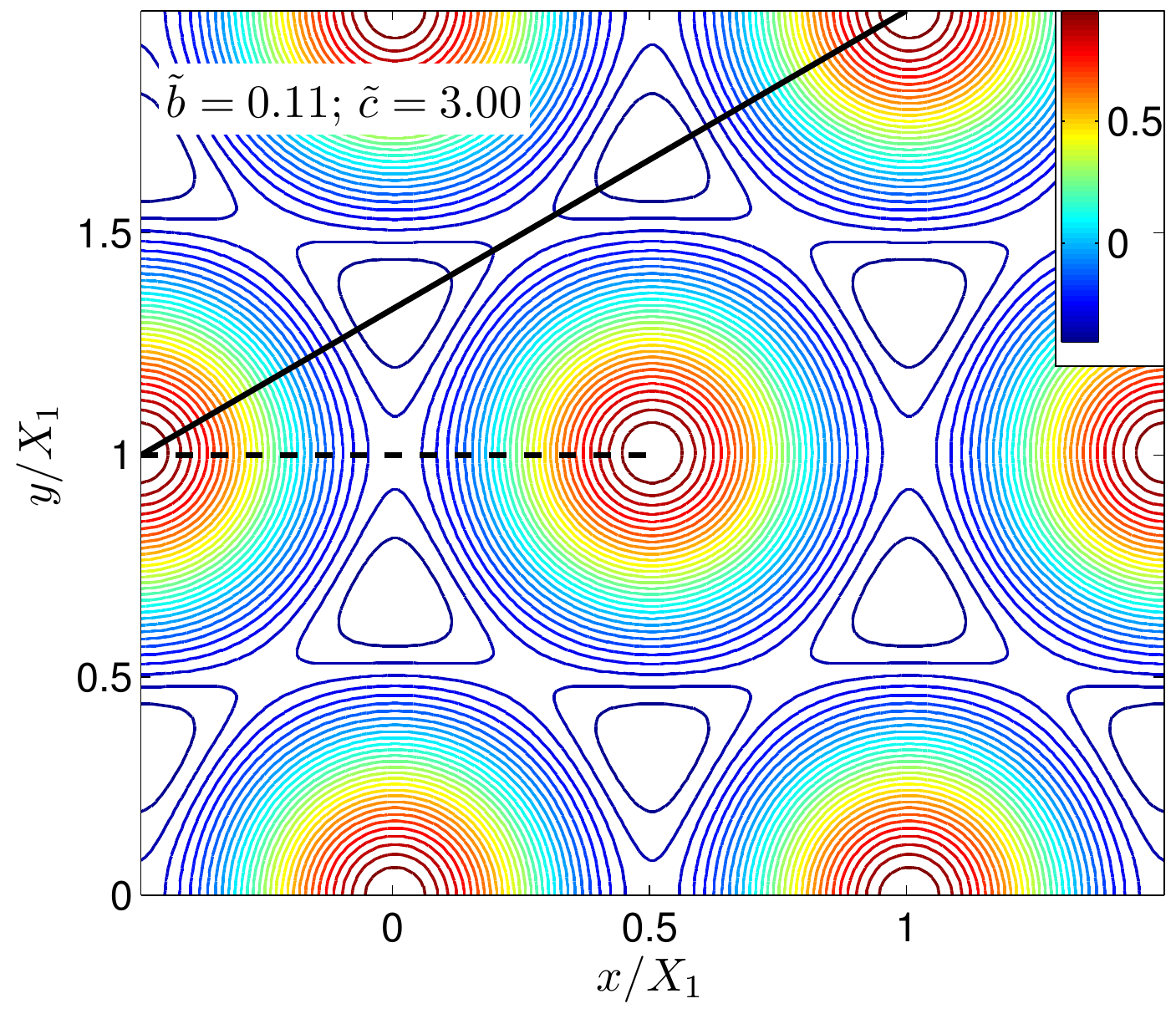}
\includegraphics[width=0.3\linewidth]{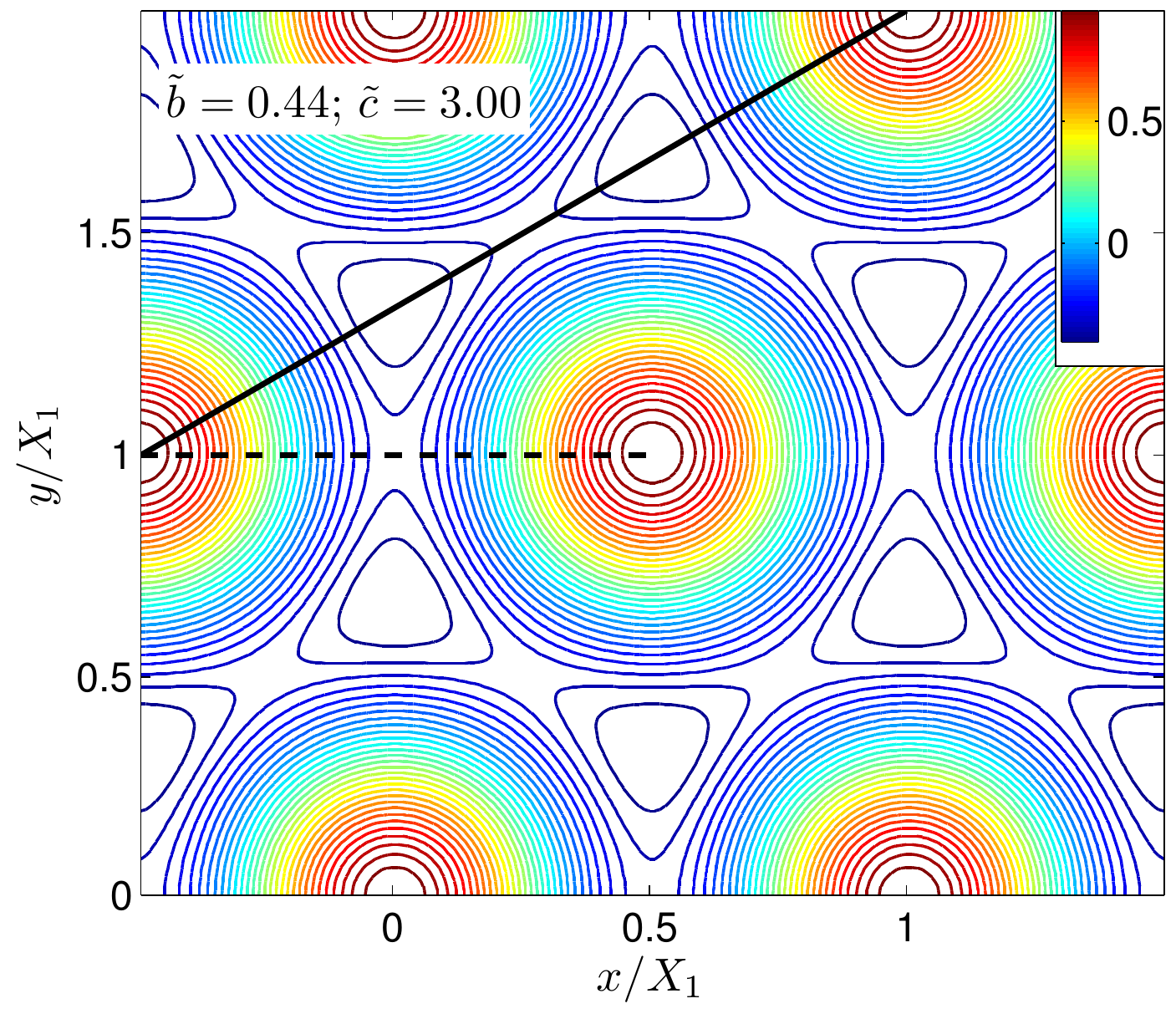}
\includegraphics[width=0.3\linewidth]{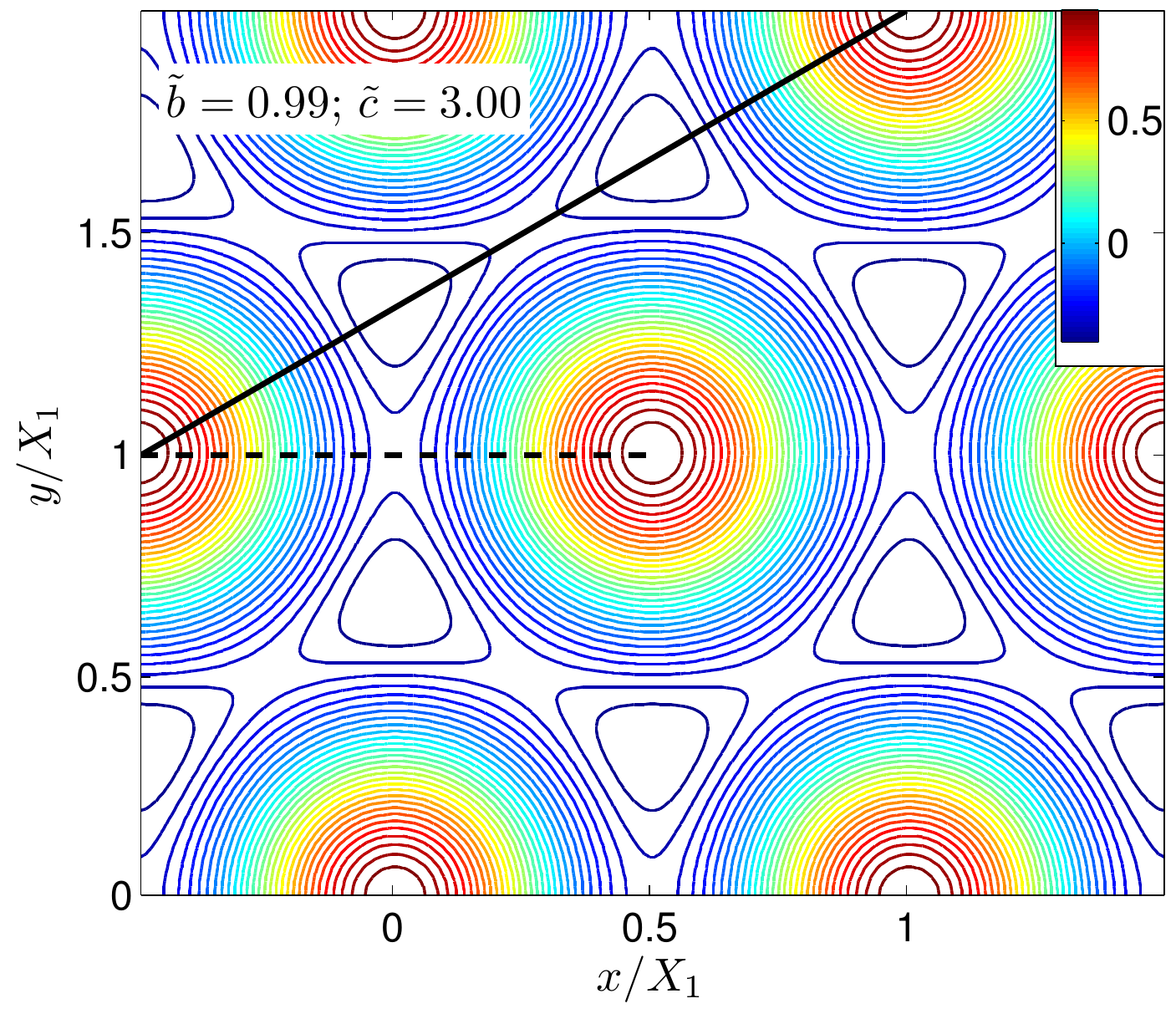}\\
\caption{ (Color online) Contour plots of the field distribution for
three values of $\tilde c$ ($\tilde{c} = 0.01$, 0.90, and 3.00), and for $\tilde{b}$
significantly larger than zero: $\tilde{b} = 0.11$, 0.44, and 0.99
(corresponding to reduced fields $b = 0.5$, 0.2, and 0.1, respectively).
}
\label{fig3}
\end{figure*}

In Fig.~\ref{fig3}
we present the maps near $T=0$ ($\tilde{c}=0.01$), for $\tilde{c} = 0.90$, i.e. at the
crossover temperature, as well as for $\tilde{c} = 3.00$.
They are computed for
$\tilde{b} = 0.11$, 0.44, and 0.99, corresponding to the reduced fields 0.5, 0.2, and 0.1,
respectively. Focusing first on the maps at the top of the figure, i.e. for $\tilde{c}=0.01$,
we note that the BCS regime, {\it i.e.} when the Cooper's pair diffraction matters,
disappears when leaving the $B_{\rm c2}$ phase boundary.
This is clearly seen as the saddle point moves in between two vortex cores, as expected
in the GL regime. Physically the distance between the cores becomes so large that
the Cooper's pair diffraction is no more operative. The recovery of the GL features of the VL
appears at a lower field if the temperature is increased, as seen from the maps at the
crossover temperature. As noted from the maps at the bottom, i.e. at high temperature,
their properties are field independent.

\begin{figure*}
\includegraphics[width=0.3\linewidth]{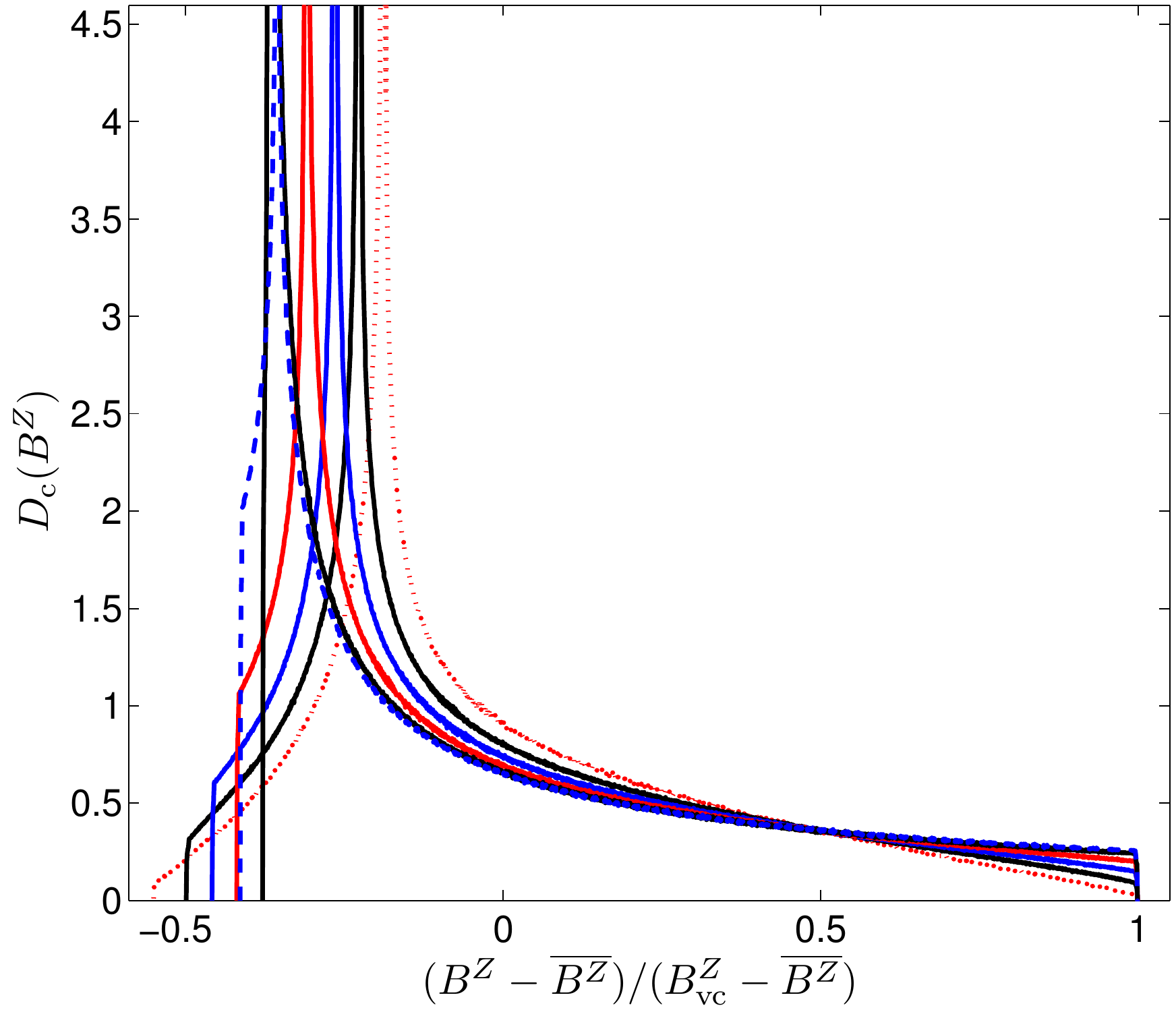}
\includegraphics[width=0.3\linewidth]{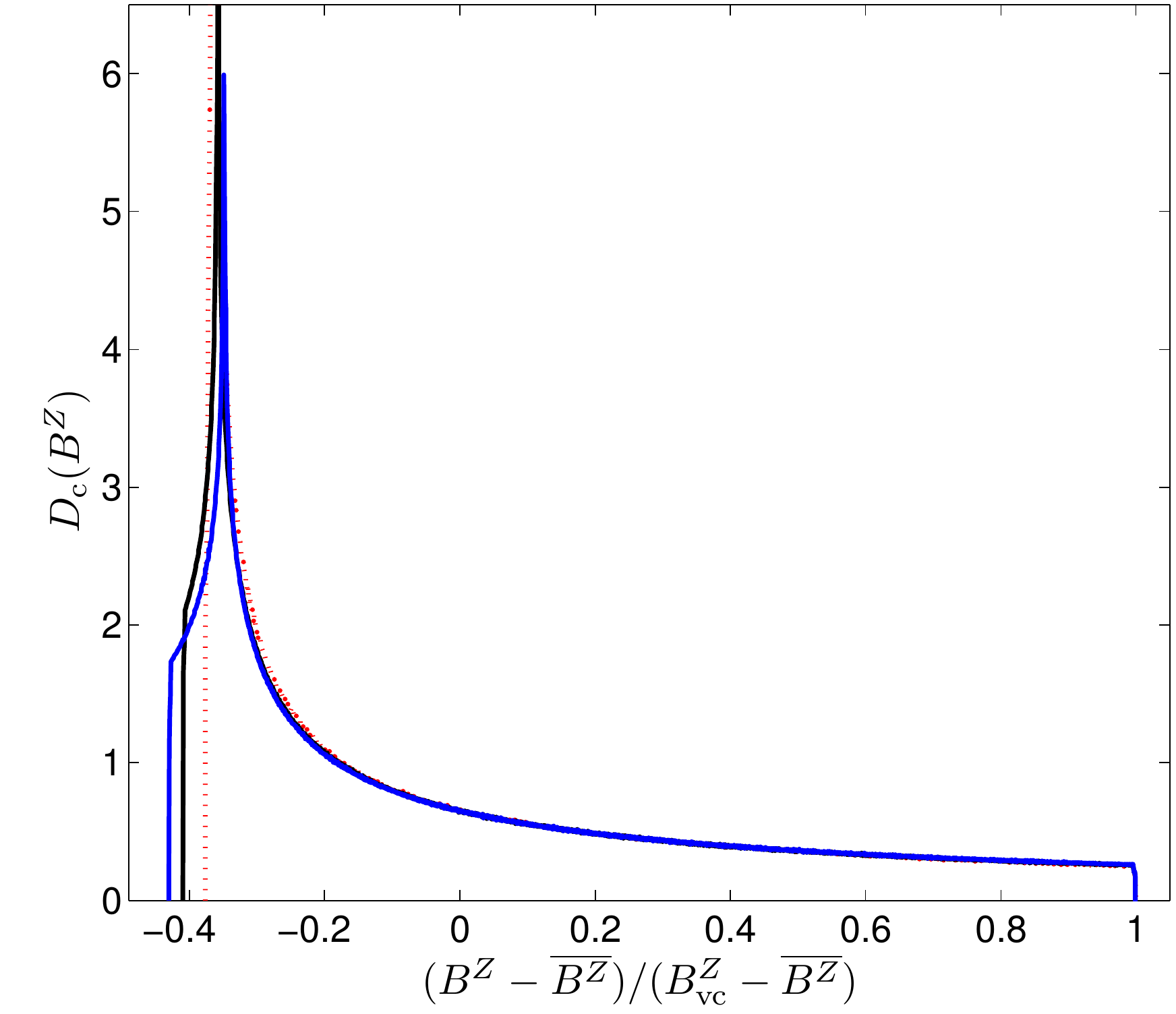}
\includegraphics[width=0.3\linewidth]{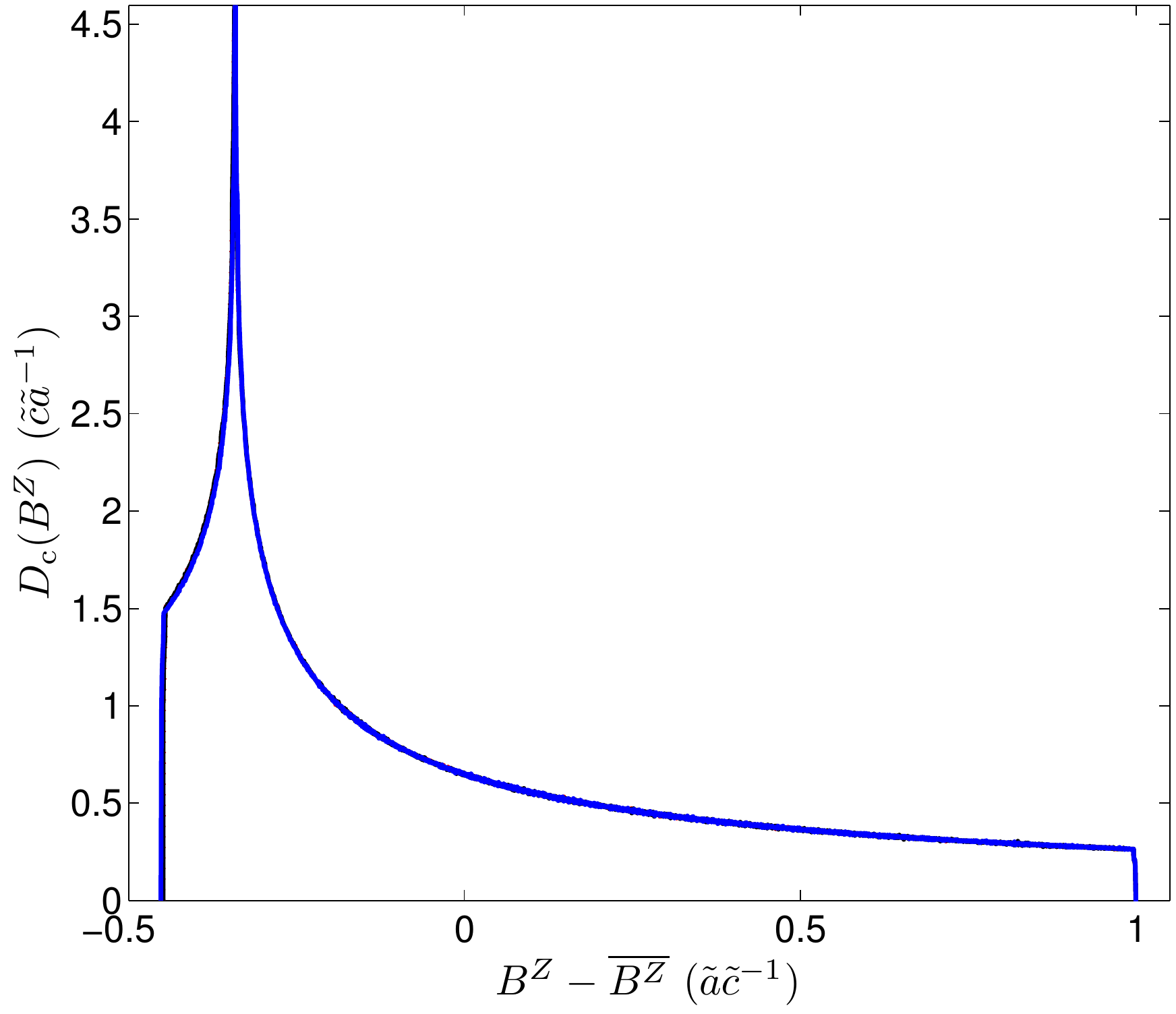}\\
\includegraphics[width=0.3\linewidth]{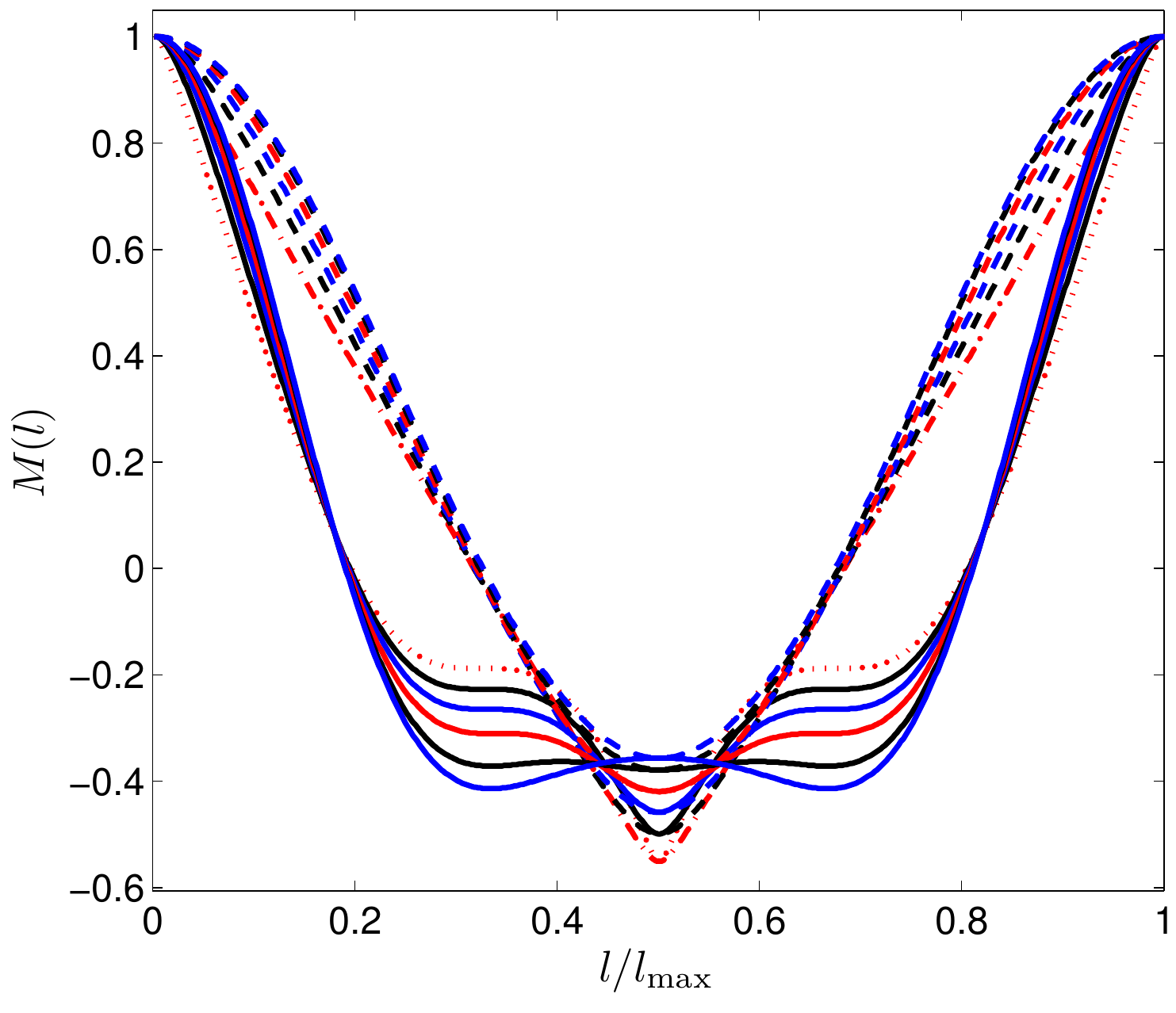}
\includegraphics[width=0.3\linewidth]{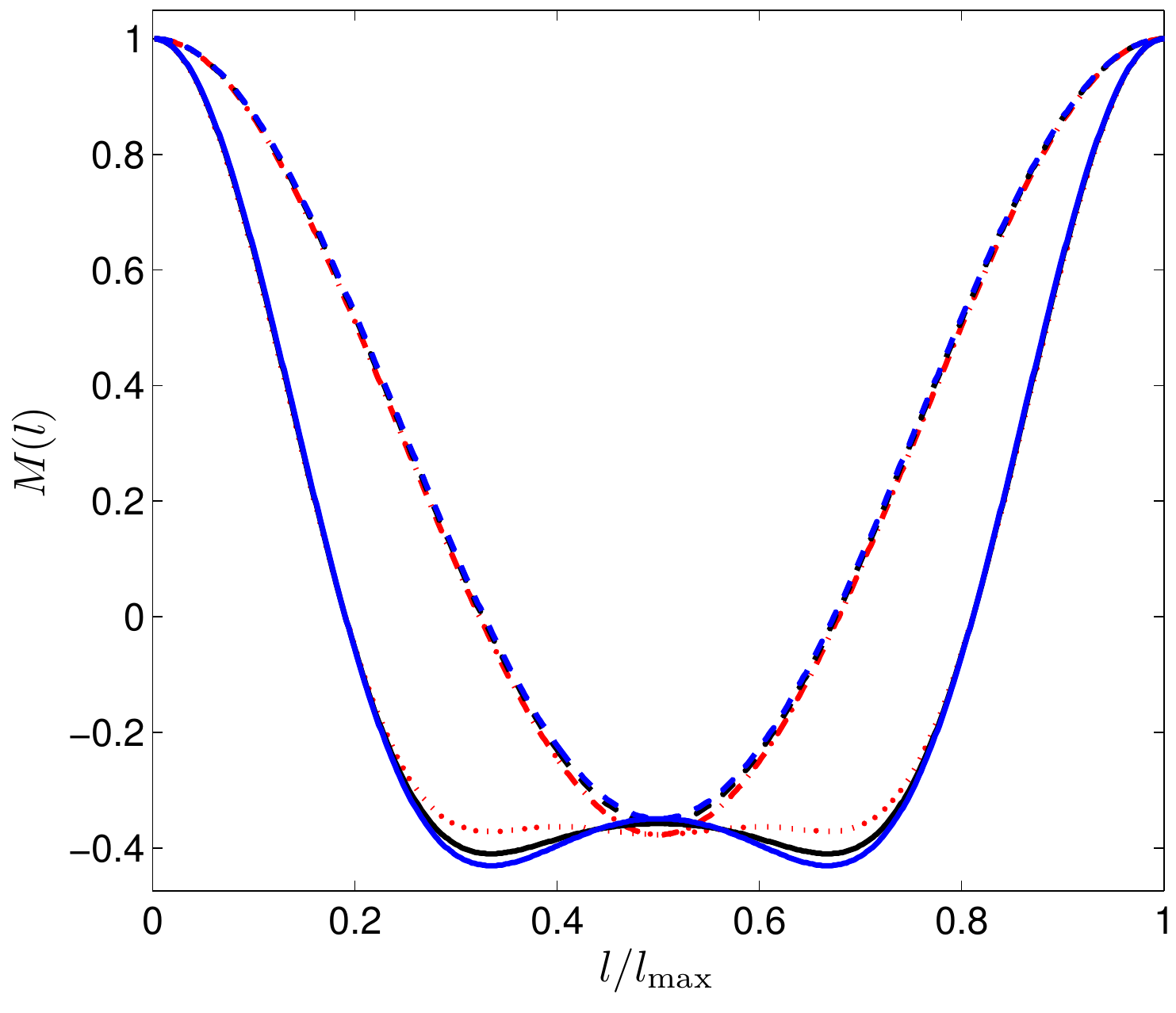}
\includegraphics[width=0.3\linewidth]{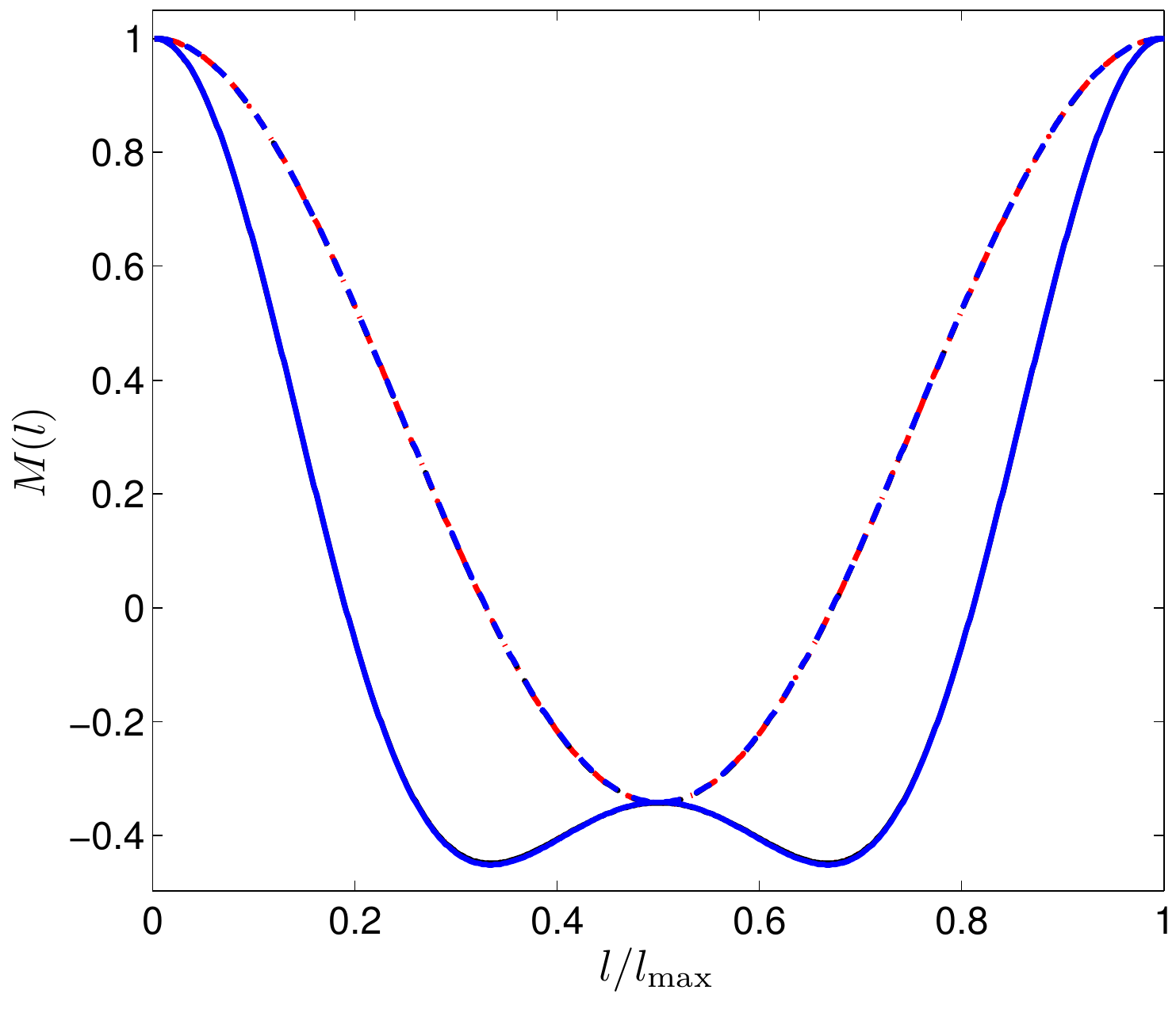}\\
\includegraphics[width=0.3\linewidth]{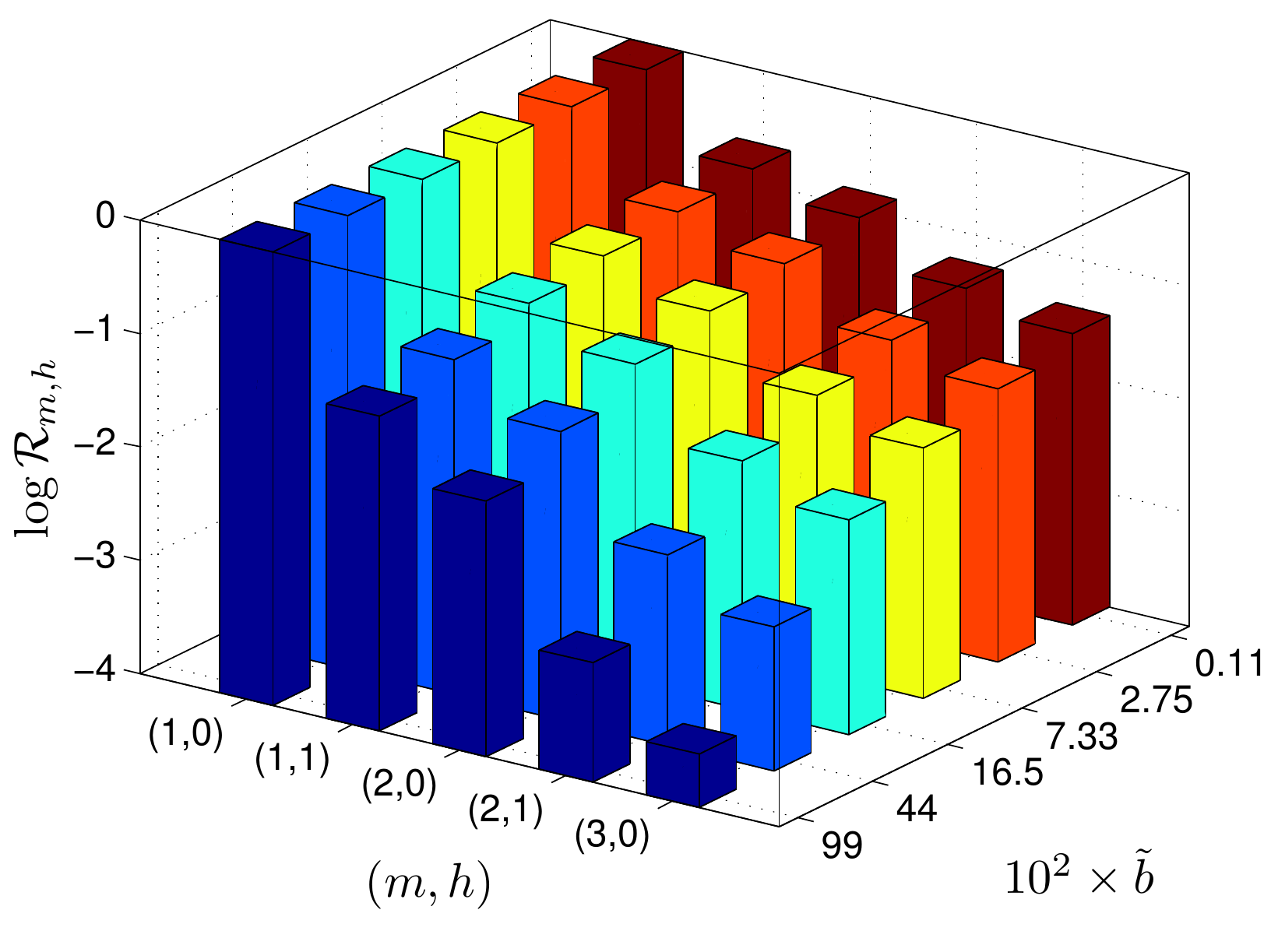}
\includegraphics[width=0.3\linewidth]{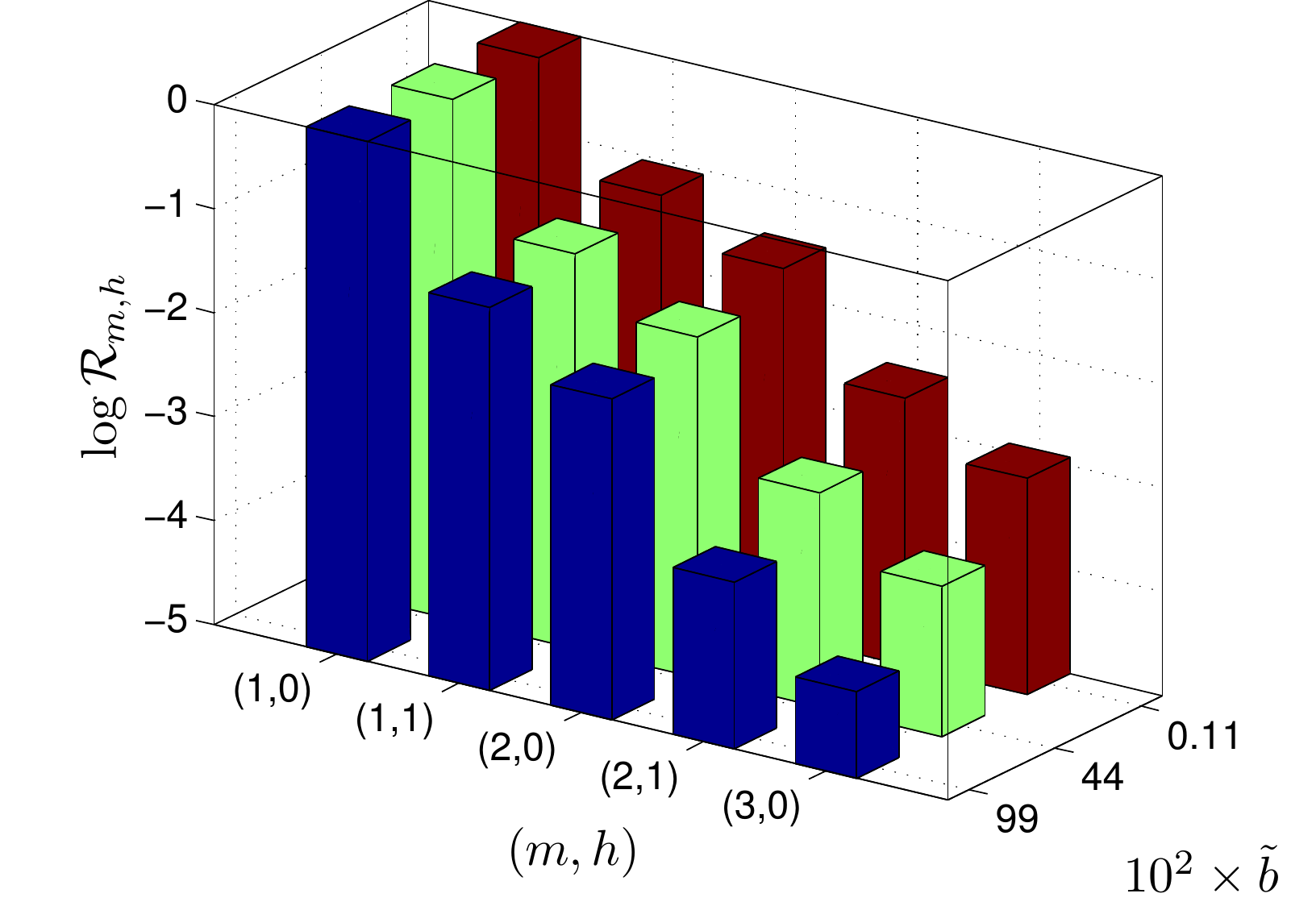}
\includegraphics[width=0.3\linewidth]{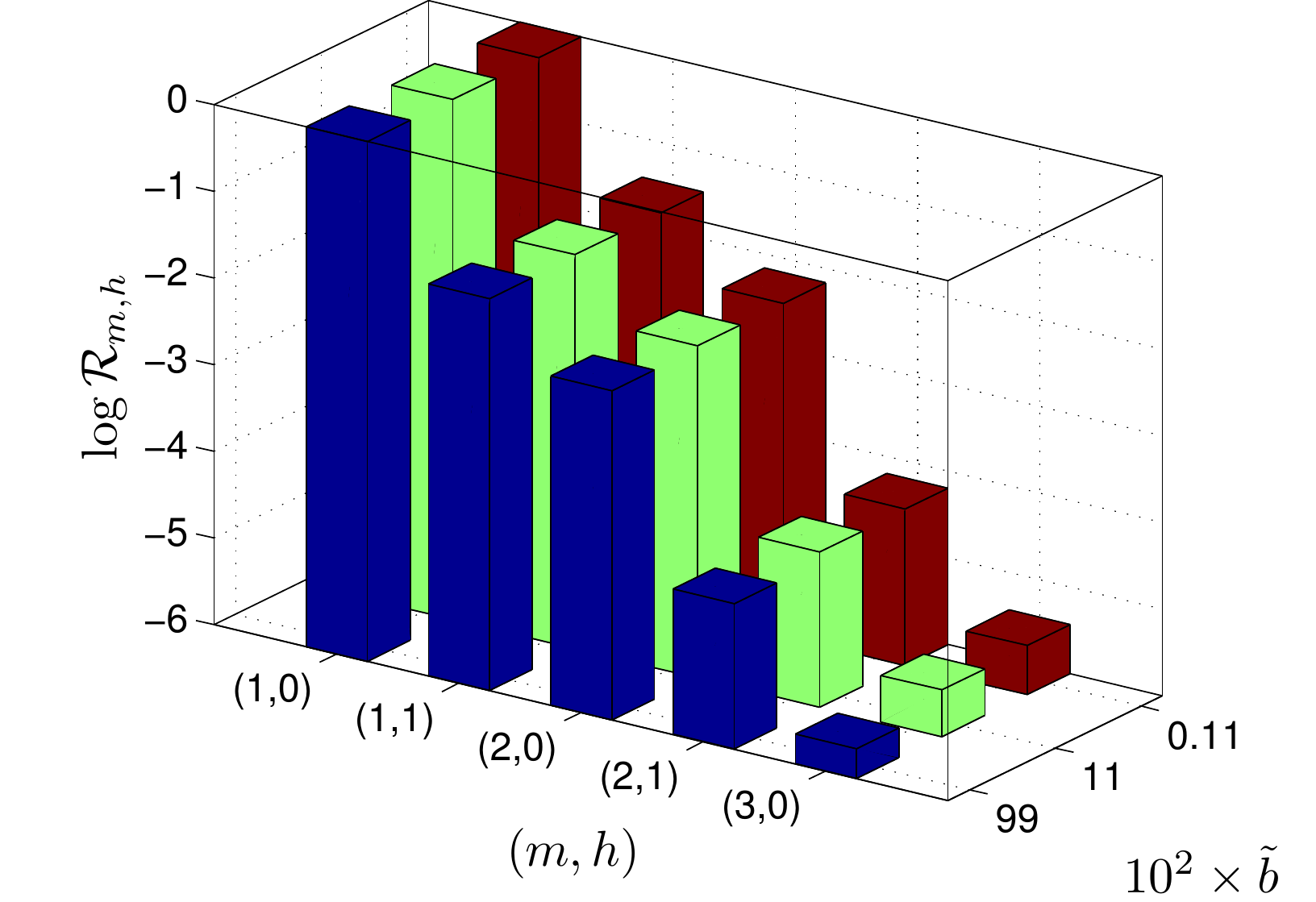}\\
\caption{ (Color online)
Field dependence of the component field distributions, field profiles,
and form factors at low
temperature --
$\tilde{c}=0.01$ and $\tilde{b} = 0.0011$, 0.0275, 0.0733, 0.165, 0.44, and 0.99
(corresponding to $b=0.99$, 0.8, 0.6, 0.4, 0.2, and 0.1, respectively) --
for the left column, at the crossover temperature
-- $\tilde{c}=0.90$ and  $\tilde{b}=0.0011$, 0.44, and 0.99
($b=0.99$, 0.2, and 0.1, respectively) -- in the middle column, and at high
temperature -- $\tilde{c}=3.00$ and  $\tilde{b}=0.0011$, 0.11, and 0.99
($b=0.99$, 0.5, and 0.1, respectively) -- in the right column.
}
\label{fig4}
\end{figure*}

In Fig. \ref{fig4} we show  $D_c(B^Z)$, the field profiles along the dashed and solid lines
in Fig. \ref{fig3}, and the form factors. While at ${\tilde c} = {\tilde c}_{\rm cross}$
they are still weakly $\tilde b$ dependent,
this is no more the case when  $\tilde c  = 3.00$. This is obviously consistent with
the field behaviour of the maps, as seen in Fig.~\ref{fig3}.
At low temperature, i.e. at $\tilde c  = 0.01$,
the shape of $D_c(B^Z)$, the field profiles, as well as the amplitude of the
renormalized form factors, are strongly field dependent,
confirming the results shown in Fig. \ref{fig3}.
Only at high field is the behaviour in the BCS regime observed.

\begin{figure*}
\includegraphics[width=0.3\linewidth]{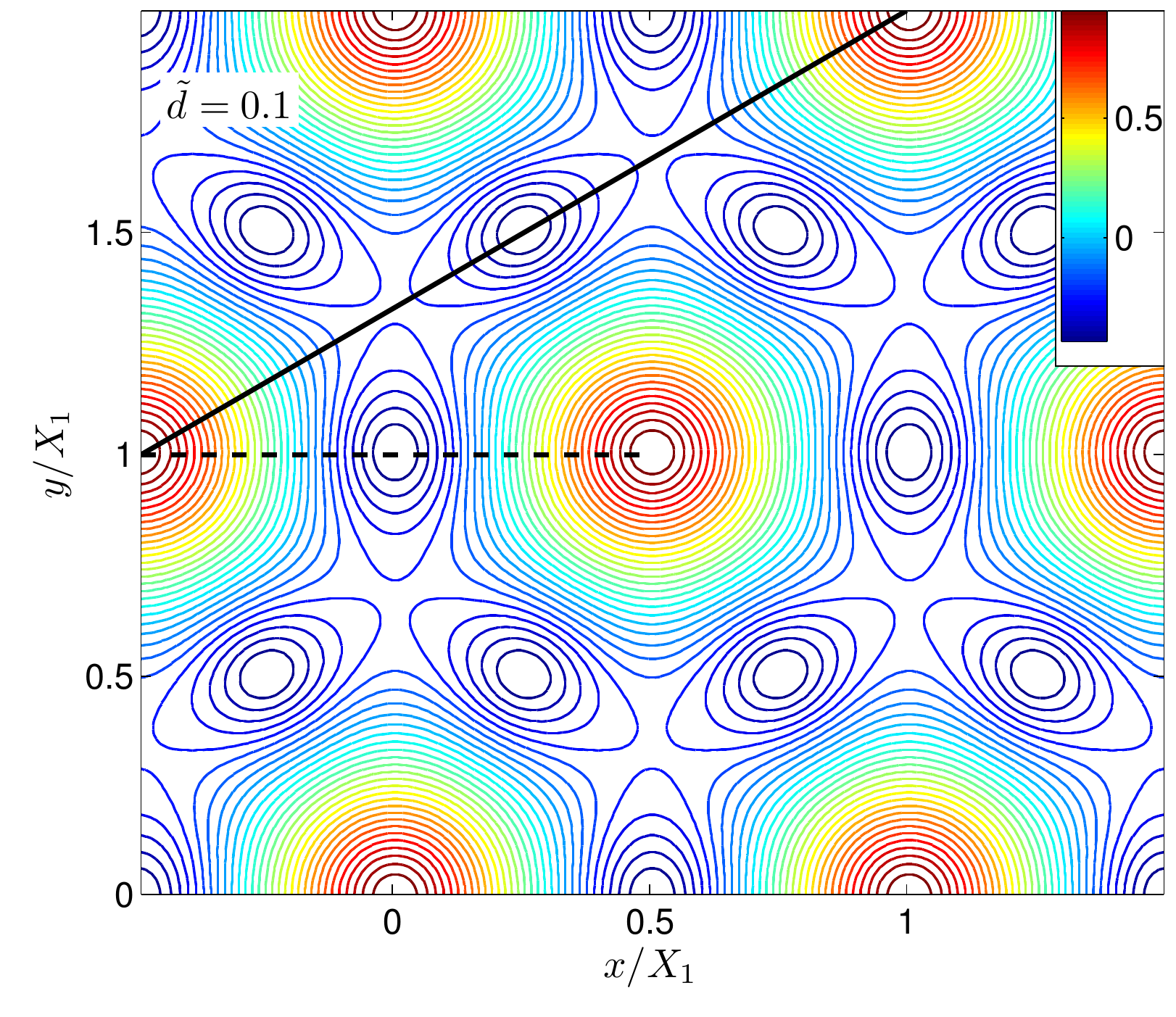}
\includegraphics[width=0.3\linewidth]{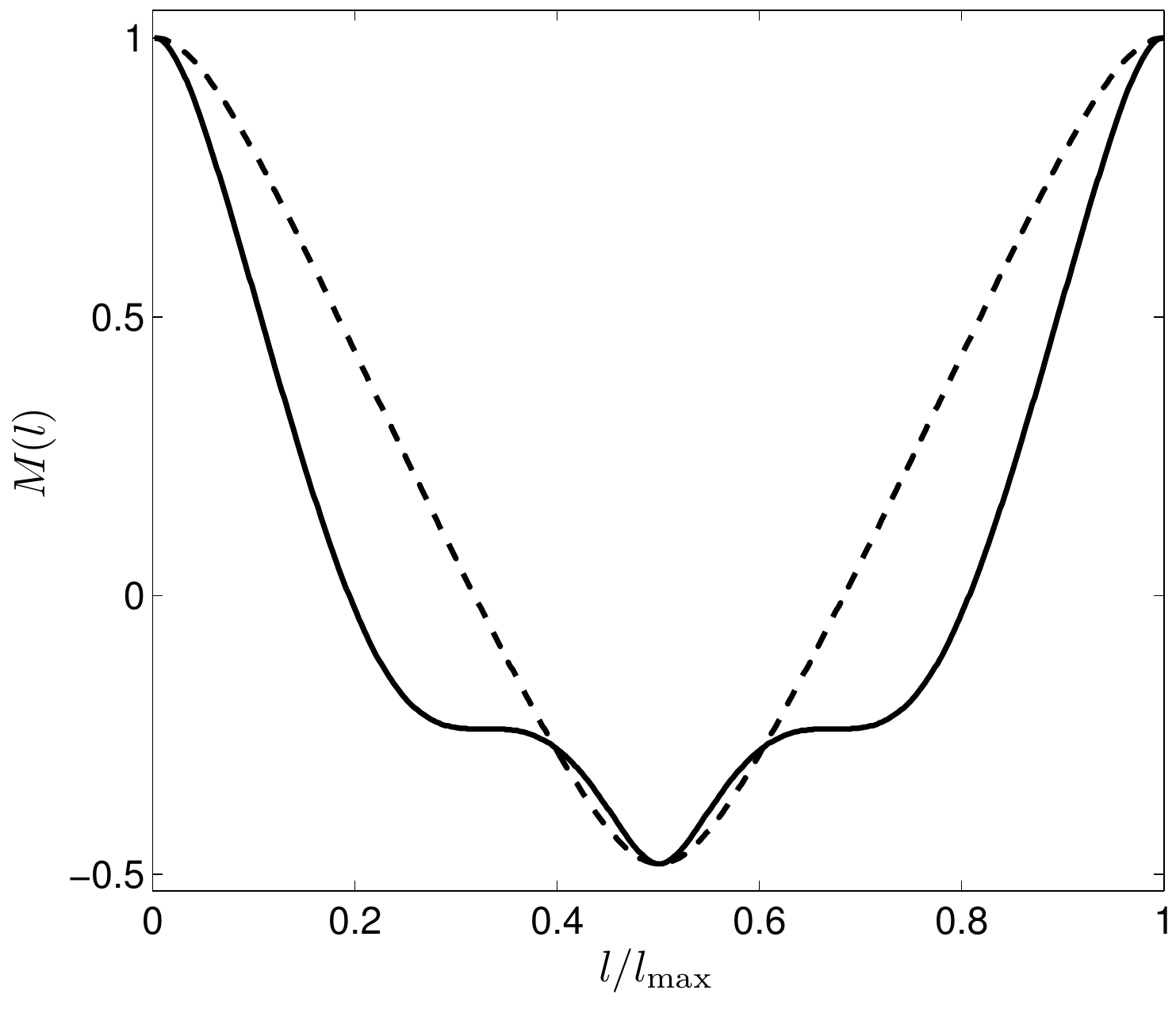}
\includegraphics[width=0.3\linewidth]{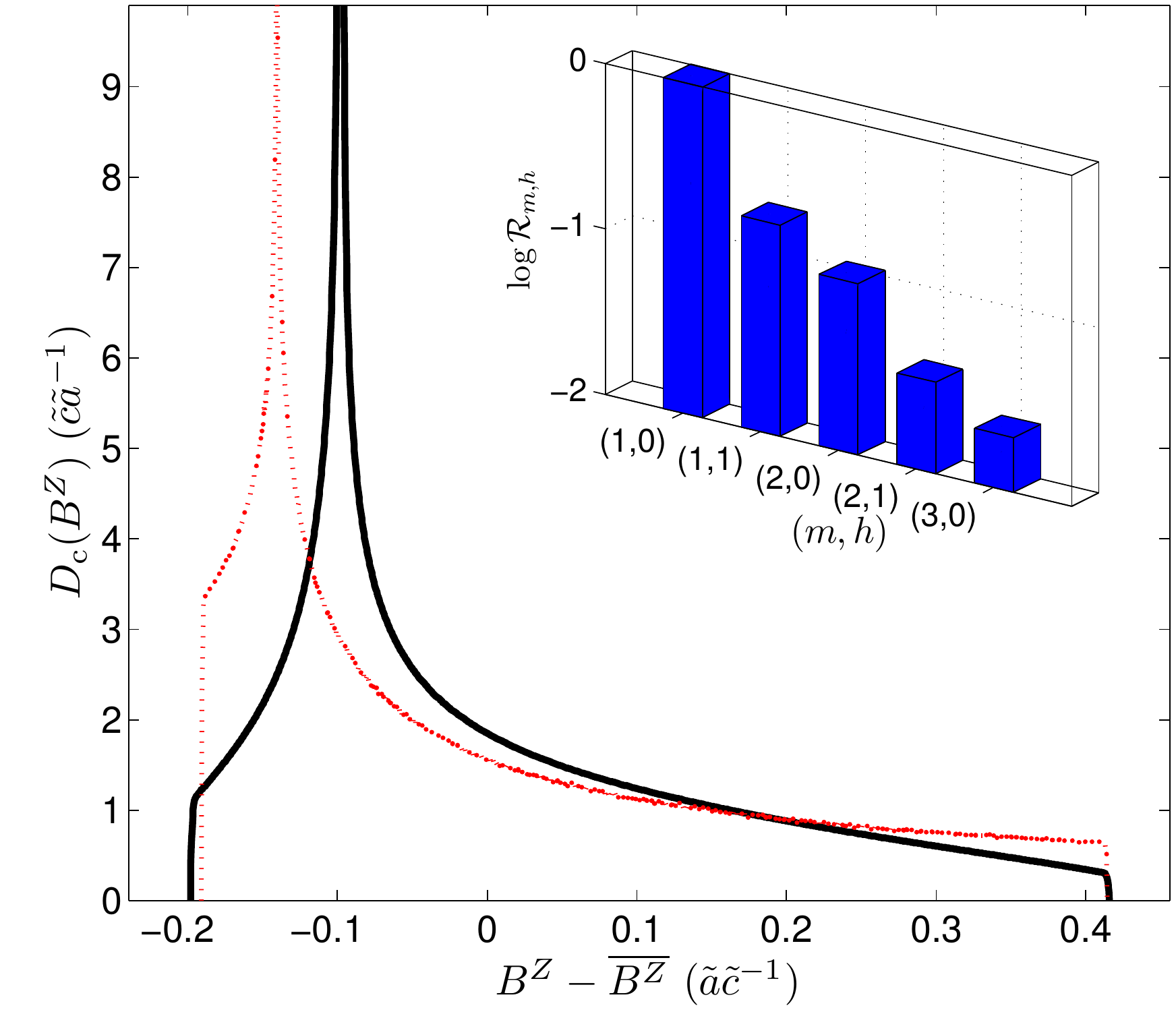}\\
\includegraphics[width=0.3\linewidth]{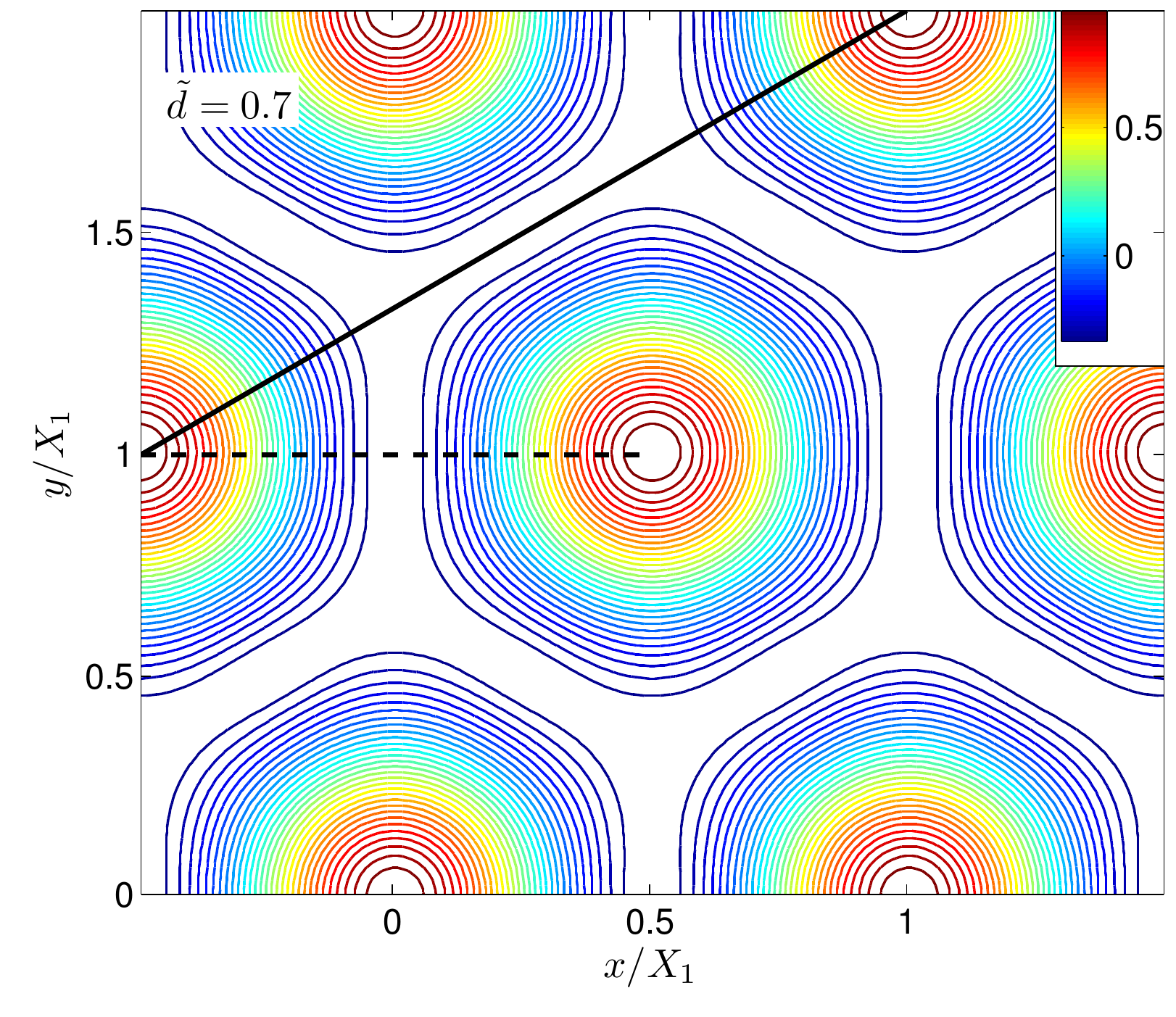}
\includegraphics[width=0.3\linewidth]{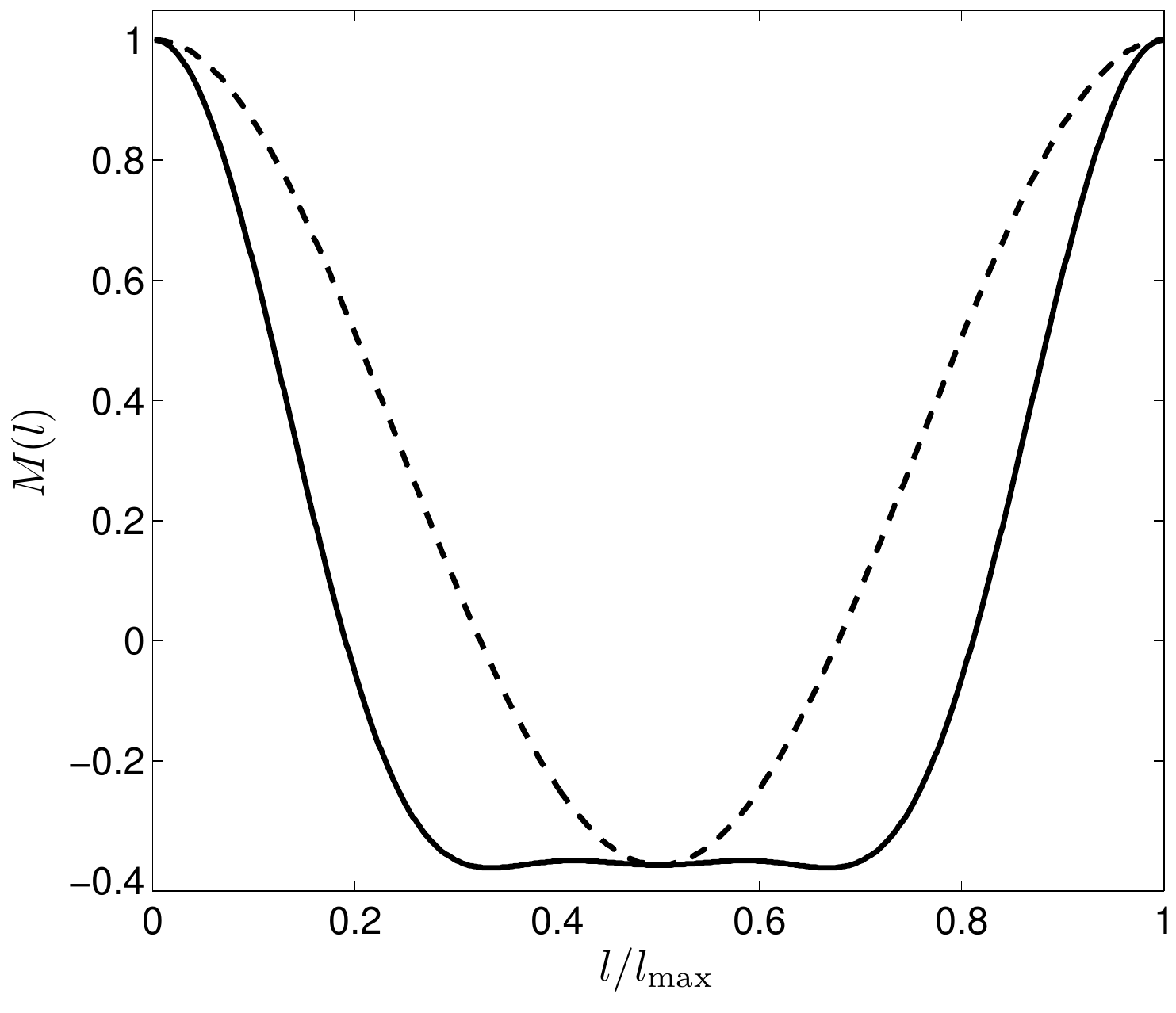}
\includegraphics[width=0.3\linewidth]{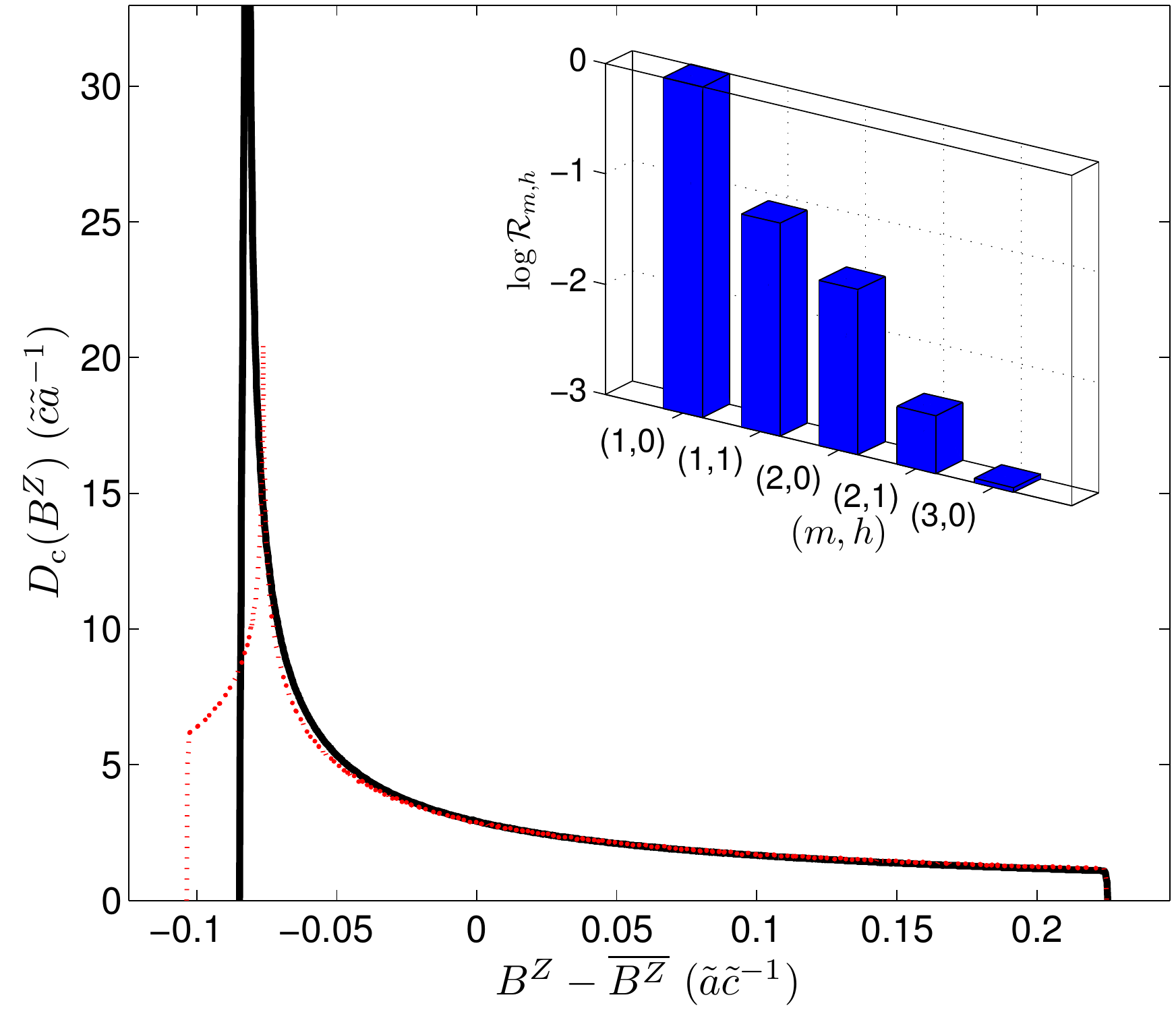}
\caption{ (Color online) The same as in Fig.~\ref{fig1}, but for
$\tilde{d}=0.1$ and 0.7 and fixed $\tilde{b}=0.0011$ and $\tilde{c}=0.001$.
The different plots show clearly  when the conditions for observation of a BCS-type of
VL are achieved ({\it i.e.} for $\tilde{b}\simeq 0$
and $\tilde{c}\simeq 0$) the transition from the BCS to the GL vortex lattice occurs
at $\tilde{d} \simeq 0.7$. For $\tilde{d}\rightarrow 0$
the VL is in clean limit (see also Fig. \ref{fig1}) while for $\tilde{d}\rightarrow \infty$
it is of the GL type as in the high temperature range, {\it i.e.} $\tilde{c}\rightarrow \infty$
(see Fig. \ref{fig2} and Appendix \ref{Reduced_dirty}).
}
\label{fig5}
\end{figure*}
\begin{figure}
\includegraphics[width=0.9\linewidth]{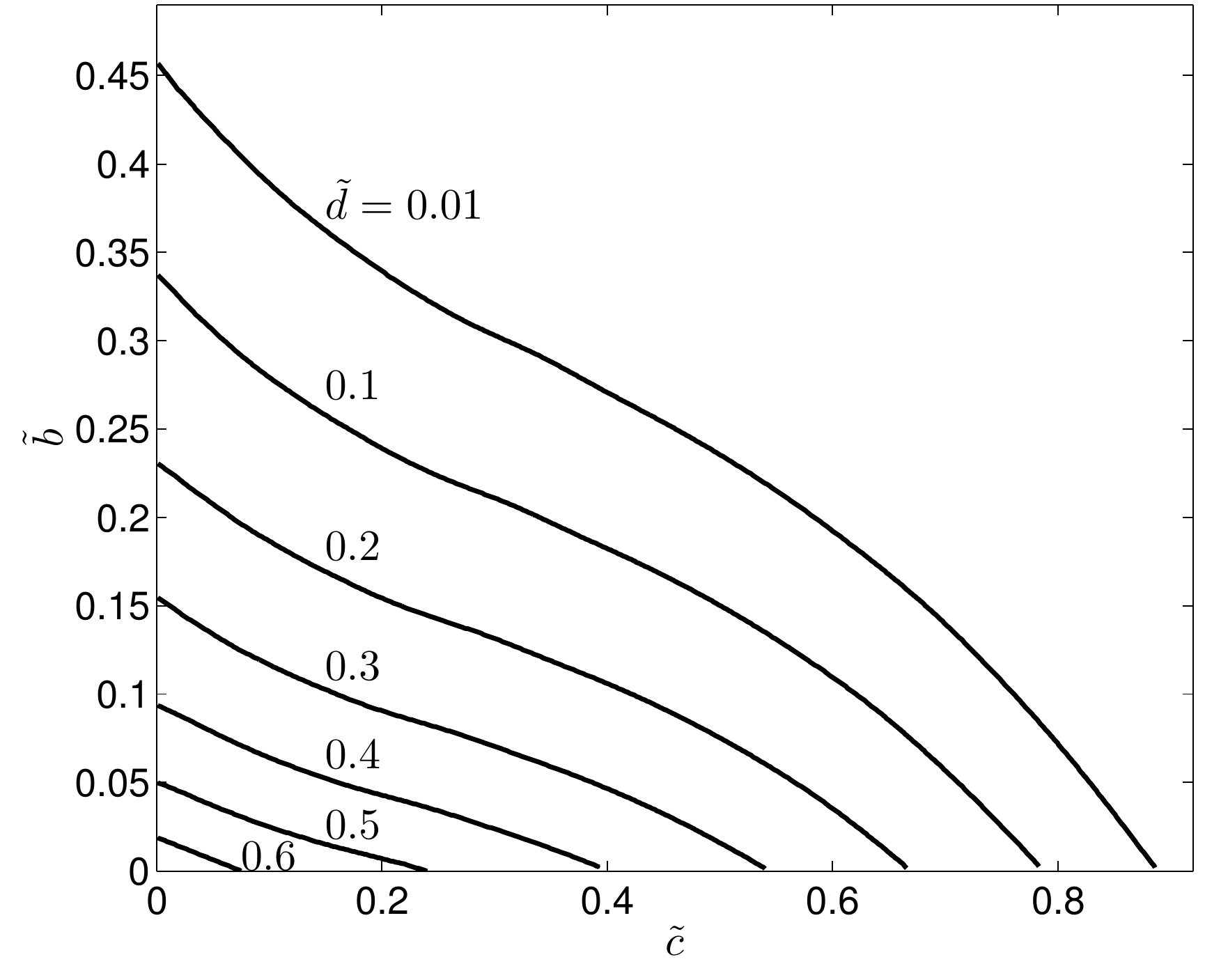}
\caption{ (Color online) Plots of $\tilde{b}$ vs $\tilde{c}$ for $\tilde{d}=0.01$,
0.1, 0.2, 0.3, 0.4, 0.5, and  0.6 where the condition of  equal fields at saddle and minimal points
is satisfied (the crossover condition; see {\it e.g.} top panel of Fig. \ref{fig2}). }
\label{fig6}
\end{figure}

\section{Discussion and conclusions}
\label{Discussion_conclusions}

As pointed out by U.~Brandt {\it et al.} the approximate Gorkov's equation used above is valid
in the region of fields near $B_{\rm c2}(T)$ where the magnetization vs field curve does not deviate appreciably
from the straight line.\cite{Brandt67} For superconductors with a GL parameter
$\kappa\sim 1$ this corresponds to a rather limited
field range, however for $\kappa \gg 1$ it covers a substantial part of the VL phase.
The validity of the model is related to two types of approximations which have been done for the
derivations of the magnetization and form factor.
We shall focus our discussion on the latter quantity.
The first type is inherent to the method and the second can be taken out if necessary.

We first recall the two approximations of the first type. The form factor is computed with an approximate
Green's function. First, the effect of the field
on the function is only described with a phase integral. This is the widely used semiclassical approximation.
The ratio defined at Eq.~\ref{eq:Correlation:3} has to be smaller than 1 for this approximation to be valid.
Second, the spatial variation of the induction is neglected.
Therefore this cannot be valid if $B_{\rm ext}$ is too close to $B_{\rm c1}$.

We now discuss the second type of approximation. First, a
spherical Fermi surface  has been chosen. It should not be a problem to describe
a superconductor with an anisotropic Fermi surface. However, it is
probably possible only numerically. Second, up to now the conduction electron mean-free-path
$\ell_{\rm mfp}$ has been  assumed to be infinite. Here we describe a method to account for the finite
$\ell_{\rm mfp}$ value.

In the case of an isotropic impurity diffusion, the effect of these impurities
can approximately be taken into account in the following way. First the
Matsubara angular precession frequency
$\omega_\ell$ in Eq.~\ref{eq:Correlation:2:1} is substituted by $\omega_\ell + 1/(2 \tau_{\rm imp})$ where
$\tau_{\rm imp} = \ell_{\rm mfp}/v_{\rm F}$. We identify $\tau_{\rm imp}$ introduced here with  $\tau_{\rm life}$
used in Sec.~\ref{Correlation}.
Hence in the formula for the form factor $\omega_\ell$ has the meaning\cite{Eilenberger67}
\begin{eqnarray}
\omega_\ell = (2 \ell +1 ) \pi k_{\rm B} T/\hbar + 1/(2 \tau_{\rm imp}).
\label{eq:Discus:concl:1}
\end{eqnarray}
Secondly $\Delta_0$ has to be renormalised.\cite{Delrieu74} It is substituted by
\begin{eqnarray}
{\Delta_0 \over 1 - \varepsilon (\omega_\ell)}
\label{eq:Discussion:concl:2a}
\end{eqnarray}
with
\begin{eqnarray}
\varepsilon (\omega_\ell) = \frac{1}{2}{\Lambda \over  \tau_{\rm imp} v_{\rm F}}
\int_0^{\pi/2} i v \left ( {i\omega_\ell \Lambda \over v_{\rm F} \sin \theta} \right ) {\rm d} \theta.
\label{eq:Discussion:concl:2b}
\end{eqnarray}
Note, $\lim_{\tau_{\rm imp}\rightarrow 0}\varepsilon(\omega_\ell)=1$.
Since $\varepsilon (\omega_\ell) > 0$, when the scattering is not too strong, i.e. when
$\tau_{\rm imp}$ is sufficiently long, the renomalisation
increases $\Delta_0$. This means that the Pippard-BCS coherence length decreases, as expected. Consistent with the
region of validity of the form factor expression given at Eqs.~\ref{eq:Dlr:FF:Expr1} and
\ref{eq:Dlr:FF:Expr2}, the proposed renormalisation is only valid if $B_{\rm ext}$
is not too far from $B_{\rm c2}$. In the $B^Z_{{\bf K}_{m,h}}$ expression
the renormalisation occurs three times: twice explicitly and once through the variable $u_\ell$.
As it has been done in the clean limit, it is possible to write the $B^Z_{{\bf K}_{m,h}}$ expression
in terms of a reduced number of parameters; see Appendix.~\ref{Reduced_dirty}.

Eilenberger has derived  approximate equations for the Green's functions for
which a numerical method has been developed to solve
them; see Refs.~[\onlinecite{Klein87,Miranovic04}]  and references therein.
With this formalism the form factor can be computed with the effect of a finite
$\ell_{\rm mfp}$ accounted for.\cite{Belova11}
It would be worthwhile to reproduce the analytical results described here with the numerical method
and extend this study outside the region of validity of the analytical solution, i.e. far below
$B_{\rm c2}(T)$.

The linear $D_{\rm c}(B^Z)$ tail at low and high field and the relative large amplitude ratio
${\mathcal R}_{m,h} = \left| B^Z_{{\bf K}_{m,h}}/B^Z_{{\bf K}_{1,0}}\right|$ for Bragg's spots
far from the center of the reciprocal space are consequences of the diffraction
of the Cooper's pairs on the vortex cores. Hence, {\it a priori } they should also be observed even for
non BCS s-wave superconductors. For these exotic properties to be found, measurements
have to be performed on very clean single crystal superconductors (which is usually the case
for high temperature superconductors) at low temperature
and for $B_{\rm ext}$ sufficiently close to $B_{\rm c2}$.

For the diffraction of carriers on a periodic VL to matter at low temperature
 $\ell_{\rm mfp}$ should be substantially larger than the intervortex distance,
{\it i.e.} $\tilde{d}=\Lambda/\ell_{\rm mfp}\ll 1$.
For $\tilde{d}\gg 1$ no diffraction takes place and the VL has the GL profile
similar to that shown in Fig. \ref{fig2}
(see Appendix \ref{Reduced_dirty}) while for $\tilde{d}=0$ the superconductor is in clean limit
which was discussed above. In Fig. \ref{fig5} we show contour plots of the spatial field
distributions, field profiles, component field distributions, and form factors for intermediate
values of $\tilde{d} = 0.1$ and 0.7 with fixed $\tilde{c}=0.001$ and $\tilde{b}=0.0011$
(since $\tilde{b}\simeq 0$ and $\tilde{c}\simeq 0$ are optimal for the observation of the diffraction effects).
For $\tilde{d}<0.1$ the characteristics of the VL are similar to that of clean VL. With increasing
$\tilde{d}$ the exotic behaviour of VL gradually vanishes and at $\tilde{d}\simeq0.7$
the crossover takes place. This result differs only slightly from that obtained by
E. H. Brandt using a nonlocal theory.\cite{Brandt74}
This crossover depends on the combination of $\tilde{b}$, $\tilde{c}$, and
$\tilde{d}$. In Fig. \ref{fig6} plots for the crossover condition are given
as a function of $\tilde{b}$, $\tilde{c}$ and $\tilde{d}$   ($\tilde{b}$ vs $\tilde{c}$ for different
$\tilde{d}$ values).
This conclusive figure illustrates the following natural condition for the observation of the
exotic VL behaviour due to diffraction:
the three length scales $\xi^T$, $\xi^B$, and $\ell_{\rm mfp}$ should be significantly larger
than the intervortex distance (here, $\xi^T =\hbar v_{\rm F}/ (2 \pi k_{\rm B}T)$,
$\xi^B={\hbar v_{\rm F} / \pi \Delta_0}$, see Appendix \ref{Reduced_clean}).

In conclusion, we have reviewed the previous works of Delrieu on the exotic behaviour of
the vortex lattice (VL) at high field and low temperature. It is the consequence of the Cooper's pair
diffraction on the periodic VL potential.
Analytical and numerical results for the magnetization and form factors are derived using the Green's
function formalism. In agreement with previous works of E. H. Brandt where a nonlocal
theory of superconductivity was utilized (see {\it e.g.} Refs.~[\onlinecite{Brandt74,Brandt95}] and
references therein), we find a set of conditions
for the observation of this exotic behaviour of the VL.  Namely, the intervortex distance
should be significantly smaller than each three length parameters:
$\xi^T$, $\xi^B$, and $\ell_{\rm mfp}$ (see the text).
An expression for the standard deviation of the component
field distribution has been derived. The results of a numerical study of the
form factors ($ B^Z_{{\bf K}_{m,h}}$), field map and field distribution [$D_{\rm c} (B^Z)$]
have been presented for a broad range of applied field $B_{\rm ext}$
and covering the whole range of temperatures from $T=0$ up to $T_{\rm c0}$.
In addition, effect of impurities was studied.
This has enabled us to determine features which distinguish GL from low-temperature BCS vortex lattices.
The behaviours of the experimentally accessible $D_{\rm c} (B^Z)$ and
$\left| B^Z_{{\bf K}_{m,h}}/B^Z_{{\bf K}_{1,0}}\right|$ quantities
versus the normalized temperature and external
field  have been exposed.
These results should at least apply to niobium for
${\bf B}_{\rm ext} \parallel [111]$, and maybe other classical BCS superconductors
such as vanadium.
This analysis will help in searching the exotic VL behaviour using the
SANS, $\mu$SR, and NMR techniques, since the Cooper's pair diffraction is
not restricted to the BCS theory and the conditions of the diffraction presented
above can well be satisfied by most of the high temperature superconductors.

\section*{Acknowledgments}
We are grateful to P. Dalmas de R\'eotier for helpful discussions and a careful
reviewing of this manuscript and M. Houzet for a useful discussion on the validity of the GL model.

\appendix

\section{The $v(z)$ function and the related Dawson's integral}
\label{Math}

The magnetization and the form factor for the field near $B_{\rm c2}$ are expressed in terms of the function $v(z)$ which is related to the
so-called complementary error function ${\rm erfc}(z)$:\cite{Abramowitz70}
\begin{equation}
v(z) = {1 \over \sqrt{\pi}} \int_{- \infty}^\infty {\exp(-t^2) \over z - t} {\rm d}t
= {\sqrt{\pi} \over i} \exp(-z^2) {\rm erfc}(-i z).
\label{eq:Math:1}
\end{equation}
We have the relation
\begin{eqnarray}
v(-z) = -2 \sqrt{\pi} i \exp(-z^2) - v(z).
\label{eq:Math:5}
\end{eqnarray}
Here $z$ and $t$ are complex and real variables, respectively. In our case $z = i x$, where $x$ is real, and in the asymptotic large $x$ limit
\begin{eqnarray}
iv(ix) = {1 \over x} \left [1 - {1 \over 2 x^2} + \mathcal {O}\left ({1 \over x^4} \right ) \right].
\label{eq:Math:2}
\end{eqnarray}
In general, we have the relation
\begin{eqnarray}
v^\prime(z) = -2 z v(z) + 2,
\label{eq:Math:3}
\end{eqnarray}
where $v^\prime(z) = {\rm d} v(z)/ {\rm d} z $. Combining the two previous equations, we derive
\begin{eqnarray}
v^\prime(ix) = 1/x^2, \, \, x \longrightarrow  \infty,
\label{eq:Math:4}
\end{eqnarray}
and
\begin{eqnarray}
i v^{\prime \prime} (ix) = -2/x^3, \, \, x \longrightarrow  \infty.
\label{eq:Math:4:1}
\end{eqnarray}

The  $iv (ix)$ function is bounded as follows:
\begin{equation}
- \left (x - \sqrt{x^2 + 2}  \right ) < iv (ix) < -{\pi \over 2 } \left (x - \sqrt{x^2 + {4 / \pi}} \right ).
\label{eq:Math:6}
\end{equation}

We note the asymptotic limit of the Dawson's integral:
\begin{eqnarray}
\exp(-x^2) \int_0^x \exp(t^2) {\rm d} t = {1 \over 2x},  \, \, x \longrightarrow  \infty.
\label{eq:Math:7}
\end{eqnarray}

\section{Evaluation of $a(\partial/\partial a) (u_\ell/a)$}
\label{Cal}

Here we evaluate
\begin{eqnarray}
A = a{ \partial \over \partial a} \left ( { u_\ell \over a} \right ),
\label{eq:Cal:1}
\end{eqnarray}
which is required  to derive the magnetization from the free energy.
It is easily found that
\begin{eqnarray}
A = { \partial u_\ell \over \partial a}  - { u_\ell \over a}.
\label{eq:Cal:2}
\end{eqnarray}
To compute the $\partial u_\ell / \partial a$, we first note that
according to Eq.~\ref{Delrieu_m_expression_3} we can write
\begin{eqnarray}
&- u_\ell  + 2 \hbar \omega_\ell a + \Delta^2_0 a^2 i v(i u_\ell)= f(u_\ell, a) = 0.
\label{eq:Cal:3}
\end{eqnarray}
This implies that
\begin{eqnarray}
{ \partial u_\ell \over \partial a} = - {\partial f/ \partial a \over \partial f / \partial u_\ell} =
2{\hbar \omega_\ell + \Delta^2_0 a i v(i u_\ell) \over 1 + \Delta_0^2 a^2 v^\prime(i u_\ell)}
\label{eq:Cal:4}
\end{eqnarray}
where $v^\prime(z) = {\rm d} v(z)/ {\rm d} z $. This means that
\begin{eqnarray}
A = \Delta^2_0 a {- u_\ell    v^\prime(i u_\ell)  +    i  v(i u_\ell) \over 1 + \Delta_0^2 a^2 v^\prime(i u_\ell)}.
\label{eq:Cal:5}
\end{eqnarray}
Now we note the relation
\begin{eqnarray}
{\partial \left [ u_\ell i v(iu_\ell) \right ] \over \partial (\hbar \omega_\ell)} =
{\partial u_\ell \over \partial (\hbar \omega_\ell)} \left [i v(i u_\ell) - u_\ell v^\prime (i u_\ell)  \right ].
\label{eq:Cal:6}
\end{eqnarray}
Since
\begin{eqnarray}
{\partial u_\ell \over \partial (\hbar \omega_\ell)} = -{\partial f/\partial (\hbar \omega_\ell) \over  \partial f /\partial u_\ell} =
{2 a \over 1 + \Delta_0^2 a^2 v^\prime (i u_\ell)},
\label{eq:Cal:7}
\end{eqnarray}
we derive
\begin{eqnarray}
{\partial \left [ u_\ell i v(iu_\ell) \right ] \over \partial (\hbar \omega_\ell)} =
{ 2 a \left [ i v(i u_\ell) - u_\ell v^\prime(i u_\ell) \right ] \over  1 + \Delta_0^2 a^2 v^\prime (i u_\ell)}.
\label{eq:Cal:8}
\end{eqnarray}
%
Combining the previous equation with Eq.~\ref{eq:Cal:5}, we obtain
\begin{eqnarray}
A = {\Delta_0^2 \over 2} {\partial \left [ u_\ell i v(iu_\ell) \right ] \over \partial (\hbar \omega_\ell)}.
\label{eq:Cal:9}
\end{eqnarray}
Recalling the relation given at Eq.~\ref{eq:Math:3}, we finally derive
\begin{eqnarray}
A = - {\Delta_0^2 \over 4} i v^{\prime \prime}(iu_\ell) {\partial u_\ell \over \partial (\hbar \omega_\ell) }.
\label{eq:Cal:10}
\end{eqnarray}
\section{Low temperature asymtotic limit of the magnetization}
\label{Delrieu_m_asym}

We start from Eq.~\ref{Delrieu_m_4}. When the temperature is very small, according to Eq.~\ref{eq:Correlation:2:1}
it is justified to replace the sum
$2 \pi k_{\rm B} T \sum_\ell$ by the integral  $\int_{u_0}^\infty {\rm d} (\hbar \omega)$. Since, according to  Eq.~\ref{eq:Math:4},
$v^\prime (i u_\ell)$ vanishes when $\ell \longrightarrow \infty$,
\begin{eqnarray}
M = -{ N_0 \Delta^2_0 \over 4 {\overline {B^Z}}} \int_0^{\pi/2} \sin (\theta)  v^{\prime}(iu_0)  {\rm d}\theta .
\label{eq:Delr:m:asym:1}
\end{eqnarray}
As a  Matsubara frequency vanishes with the temperature,
\begin{eqnarray}
u_0 =
{\Delta^2_0 \Lambda^2 \over \hbar^2 v^2_{\rm F} \sin^2 \theta} i v(i u_0).
\label{eq:Delr:m:asym:2}
\end{eqnarray}
Because we are focusing on the field region near $B_{\rm c2}$, except for a small domain for which $\theta$ can be small,
we can take $u_0 =0$. Using Eq.~\ref{eq:Math:3}, we then get $v^\prime (0) =2$, and finally derive the result written at
Eq.~\ref{Delrieu_m_6}.

\section{Asymptotic limits of the form factor}
\label{Delrieu_limits}

Here we determine analytically  two asymptotic limits of the form factor
$B^Z_{{\bf K}_{m,h}}$ starting from Eqs.~\ref{eq:Dlr:FF:Expr1}
and \ref{eq:Dlr:FF:Expr2}. We shall first study the high temperature limit.

\subsection{The behaviour near $T_{\rm c0}$}
\label{Delrieu_limits_Tc}

When approaching $T_{\rm c0}$ from below, $\overline {B^Z}$ vanishes as does $B_{\rm c2}$. Hence $\Lambda$ is getting
very large. Referring to Eq.~\ref{eq:Dlr:FF:Expr3}, this means that  $u_\ell$ is large. This has two consequences.

First, it is justified to consider the $u_\ell \gg n_{{\bf K}_{m,h}}$ limit for the numerator of the fraction in $g_\ell(\theta)$.
Recalling the Taylor expansion of a function,
\begin{align}
&iv\left (iu_\ell + i n_{{\bf K}_{m,h}} \right ) + iv\left (iu_\ell - i n_{{\bf K}_{m,h}} \right ) -2 iv\left (iu_\ell \right )\nonumber\\
&\approx - n^2_{{\bf K}_{m,h}} i v^{\prime \prime}(i u_\ell).
\label{eq:Delr:lim:Tc:1}
\end{align}
Secondly, let us now focus on the denominator, in particular on the second term. Because $u_\ell$ is large, for $iv(iu_\ell)$ we can
use the first term of its expansion given by Eq.~\ref{eq:Math:2}. From Eq.~\ref{eq:Dlr:FF:Expr3} we then get
\begin{eqnarray}
u_\ell = 2 \omega_\ell {\Lambda \over  v_{\rm F} \sin \theta} +
{\Delta^2_0 \Lambda^2 \over \hbar^2 v^2_{\rm F} \sin^2 \theta} { 1 \over u_\ell}.
\label{eq:Delr:lim:Tc:2}
\end{eqnarray}
Therefore, to a good approximation near $T_{\rm c0}$,
\begin{eqnarray}
u_\ell = 2 \omega_\ell {\Lambda \over v_{\rm F} \sin \theta}.
\label{eq:Delr:lim:Tc:3}
\end{eqnarray}
Using Eq.~\ref{eq:Math:4}, we deduce
\begin{equation}
{\Delta^2_0 \Lambda^2 \over \hbar^2 v^2_{\rm F}\sin^2 \theta} v^\prime \left (iu_\ell \right )=
\left ( { \Delta_0 \over 2 \hbar \omega_\ell } \right )^2 =
\left [ { \Delta_0 \over (2 \ell +1 ) 2 \pi k_{\rm B} T } \right ]^2.
\label{eq:Delr:lim:Tc:4}
\end{equation}
Hence, since $\Delta_0$ vanishes on approaching $T_{\rm c0}$, the second term of the denominator in Eq.~\ref{eq:Dlr:FF:Expr2}
becomes negligible relative to 1.

Using the two previous results, we derive
from Eqs.~\ref{eq:Dlr:FF:Expr1} and \ref{eq:Dlr:FF:Expr2} the asymptotic behaviour of $B^Z_{{\bf K}_{m,h}}$ near $T_{\rm c0}$:
%
%
\txtb{
\begin{align}
B^Z_{{\bf K}_{m,h}}  = &  {  \pi \mu_0 N_0  \Delta_0^2  k_{\rm B} T \over 2 \overline {B^Z}}  (-1)^{mh}
\exp \left ( - n^2_{{\bf K}_{m,h}}\right )\times  \nonumber \\
&\sum_{\ell =0}^\infty \int_0^{\pi/2}\sin (\theta) i v^{\prime \prime}(i u_\ell) {\partial u_\ell \over \partial (\hbar \omega_\ell)}{\rm d}\theta,
\label{eq:Delr:lim:Tc:5}
\end{align}
}
Recalling Eq.~\ref{Delrieu_m_4} we find, as expected, that $B^Z_{{\bf K}_{m,h}}$ is proportional to the magnetization as written explictly at
Eq.~\ref{eq:Dlr:FF:Expr4}.

\subsection{Low temperature limit}
\label{Delrieu_limits_zero}

Since we are interested in this work by the limit for which the field is near $B_{\rm c2}$, $\Delta_0$ is small.
According to Appendix~\ref{Math},  $v^\prime (i u_\ell)$ is bounded. Therefore, except
for small $\theta$ values, we can neglect
the second term in the denominator of the $g_\ell(\theta)$ function relative to one. As done for the study of the low temperature
limit of the magnetization, we can substitute the sum $2 \pi k_{\rm B} T \sum_\ell$ by
the integral  $\int_{u_0}^\infty {\rm d} (\hbar \omega)$. This gives

%
%
\txtb{
\begin{equation}
B^Z_{{\bf K}_{m,h}}  =  -{  N_0 \mu_0 \Delta_0^2\over  4 \overline {B^Z}} { (-1)^{mh}
\over n^2_{{\bf K}_{m,h}}}
\int_0^{\pi/2}\sin (\theta ) h_\ell(\theta) {\rm d} \theta,
\label{eq:Delr:lim:zero:1}
\end{equation} }
with
\begin{multline}
h_\ell(\theta)  =  \exp \left ( - n^2_{{\bf K}_{m,h}}\right )
   \left[\int_{u_0}^{\infty} i v(ix + in_{{\bf K}_{m,h}})  {\rm d} x +\right.\\
\left.\int_{u_0}^{\infty} i v(ix - i n_{{\bf K}_{m,h}})   {\rm d} x  - 2 \int_{u_0}^{\infty} i v(ix) {\rm d} x  \right]
\label{eq:Delr:lim:zero:2}
\end{multline}
In the first and second terms we use the new variables
$t=x + n_{{\bf K}_{m,h}}$ and $t=x - n_{{\bf K}_{m,h}}$, respectively, and split the integration
$\int_{u_{a}}^{\infty}i v(it){\rm d}t$ into $\int_{u_a}^{u_0}i v(it){\rm d}t + \int_{u_0}^{\infty}i v(it){\rm d}t$,
where $u_a = u_0+n_{{\bf K}_{m,h}}$ and $u_a = u_0-n_{{\bf K}_{m,h}}$ for the first and the second
terms, respectively. The two integrals $\int_{u_0}^{\infty}i v(it) {\rm d} t$ cancel the third
term. As a result we obtain the following relation:
\begin{multline}
h_\ell(\theta)  =  \exp \left ( - n^2_{{\bf K}_{m,h}}\right ) \times \\\left [\int_{u_0 +  n_{{\bf K}_{m,h}}}^{u_0} i v(it)  {\rm d} t +
\int_{u_0 -  n_{{\bf K}_{m,h}}}^{u_0} i v(it)   {\rm d} t \right ].
\label{eq:Delr:lim:zero:2:1}
\end{multline}

Let us study the function $h_\ell(\theta)$. Since the field is near $B_{\rm c2}$ and we are at low temperature, $u_0 \simeq 0$.
Then setting $u_0 = 0$ and using Eq.~\ref{eq:Math:5}, we get
\begin{multline}
h_\ell(\theta) = \exp \left ( - n^2_{{\bf K}_{m,h}}\right )\times \\
\left [2 \sqrt{\pi} \int_0^{n_{{\bf K}_{m,h}}} \exp(t^2) {\rm d} t  - 2\int_0^{n_{{\bf K}_{m,h}}} i v(it)   {\rm d} t \right ].
\label{eq:Delr:lim:zero:3}
\end{multline}
From Eq.~\ref{eq:Math:6},
\begin{equation}
\int_0^{n_{{\bf K}_{m,h}}} i v(it)   {\rm d} t < -{\pi \over 2 } \int_0^{n^2_{{\bf K}_{m,h}}}
\left (t - \sqrt{t^2 + {4 / \pi}} \right ) {\rm d} t.
\label{eq:Delr:lim:zero:4}
\end{equation}
Hence the second term of $h_\ell(\theta )$, i.e. $2\exp \left ( - n^2_{{\bf K}_{m,h}}\right )\int_0^{n_{{\bf K}_{m,h}}} i v(it)   {\rm d} t $,
is negligible if $n^2_{{\bf K}_{m,h}}$ is sufficiently large. Since the first term is proportional to the Dawson's integral, using
Eq.~\ref{eq:Math:7} we finally derive the $h_\ell(\theta )$ asymptotic limit:
\begin{eqnarray}
h_\ell(\theta ) = {\sqrt{ \pi} /  n_{{\bf K}_{m,h}}}  \, \, {\rm when} \, \,  n_{{\bf K}_{m,h}} \longrightarrow  \infty.
\label{eq:Delr:lim:zero:5}
\end{eqnarray}

Combining this result with Eq.~\ref{eq:Delr:lim:zero:1}, we derive the asymptotic limit written at Eq.~\ref{eq:Dlr:FF:Expr5}.

\section{The form factor in terms of a reduced number of parameters}
\label{Reduced}

The original $B^Z_{{\bf K}_{m,h}}$ expression  given by
Eqs. \ref{eq:Dlr:FF:Expr1}-\ref{eq:Dlr:FF:Expr3}
depends on three  material parameters $\Delta_0$, $N_0$, $v_F$ and two experimental
parameters $T$ and $B_{\rm ext}$ since $\overline{B^Z} \simeq B_{\rm ext}$.
In the next subsection we show that in fact it is a function of
only three independent parameters. Even more interesting, only one of these parameters
has to be varied to study the region close to the $B_{\rm c2}(T)$ boundary.
The second subsection gives a formula for $B^Z_{{\bf K}_{m,h}}$
when the electonic mean-free-path is a finite. The dirty limit is studied.

\subsection{The form factor in the clean limit}
\label{Reduced_clean}

It is easily shown that the formula for  $B^Z_{{\bf K}_{m,h}}$ can be written as follows:
\begin{equation}
B^Z_{{\bf K}_{m,h}} = \tilde{a}{ (-1)^{mh} \exp \left ( - n^2_{{\bf K}_{m,h}}\right )
\over  n^2_{{\bf K}_{m,h}}      }
\sum_{\ell =0}^\infty \int_0^{\pi/2}  f_\ell(\theta) {\rm d} \theta,
\label{eq:DelrBk}
\end{equation}
where
\begin{equation}
f_\ell(\theta) =
{iv\left (iu_\ell + i n_{{\bf K}_{m,h}} \right ) + iv\left (iu_\ell - i n_{{\bf K}_{m,h}} \right ) -2 iv\left (iu_\ell \right )
\over  \left [1 + \tilde{b}{ v^\prime\left (iu_\ell \right ) \over \sin^2 \theta} \right ]^3}.
\label{eq:DelrFell}
\end{equation}
We have used the analytical expression of
${\partial u_\ell / \partial (\hbar \omega_\ell)}$ written at Eq. \ref{eq:Cal:7}.
The $u_{\ell}$'s are found to be the solution of the equation
\begin{eqnarray}
u_\ell = \tilde{c}{(2 \ell+1)\over \sin\theta}  +
\tilde{b}{i v(i u_\ell) \over \sin^2 \theta} .
\label{eq:DelrUell}
\end{eqnarray}
The proportionality coefficient $\tilde{a}$ is in magnetic induction units. It is
written as follows:
%
\txtb{
\begin{equation}
\tilde{a}=-{ \mu_0 \pi N_0  \Delta_0^2} k_{\rm B} T \frac{\Lambda}{\overline{B^Z} \hbar v_{\rm F}} =
-\mu_0  N_0  \Delta_0^2  \frac{\tilde{c}}{2 \overline{B^Z} }. 
\end{equation}
}
We have also introduced the dimensionless parameters: 
\begin{equation}
\tilde{b} =  (  \Lambda / \pi\xi^B )^2,
\end{equation}
and
\begin{equation}
\tilde{c} =  \Lambda / \xi^T ,
\end{equation}
where we have defined the temperature and field dependent length scale parameters
$\xi^T =\hbar v_{\rm F}/ (2 \pi k_{\rm B}T)$ and $\xi^B = \hbar v_{\rm F} / \pi \Delta_0$, respectively.
Therefore $B^Z_{{\bf K}_{m,h}}$ has been written in terms of three parameters: $\tilde a$,
$\tilde b$, and $\tilde c$.
Interestingly, ${\tilde b}$ is vanishingly small  when $\Delta_0 \rightarrow 0$. Then the second term
of the denominator of Eq.~\ref{eq:DelrFell} is negligible and ${\tilde a}$ becomes small.
As a consequence, as expected, the form factor is also small.

There is interest to study the temperature dependence of  $\tilde b$ and $\tilde c$ near
the phase boundary $B_{\rm c2}(T)$, in particular their asymptotes. We first note that
$\Delta_0\rightarrow 0$. In addition, by the definition of the second critical field given at
Eq.~\ref{Delrieu_m_expression_6}, $\Lambda \simeq \xi_{\rm GL}$ since
$\overline{B^Z}\simeq B_{\rm c2}$. Let us investigate the $T\rightarrow 0$ limit.
According to Eq.~\ref{Delrieu_m_expression_5_1} we derive $\Lambda \simeq \xi_0/0.96$.
Hence $\Lambda$ is finite, and therefore $\tilde{b}\rightarrow 0$. Since
$\xi^T$ diverges, we also derive  $\tilde{c}\rightarrow 0$. Concerning the
$T\rightarrow T_{\rm c0}$ limit, we note that $\Lambda$ diverges as does $\xi_{\rm GL}$.
This first means
that $\tilde{b}$ is the ratio of two large numbers. Numerically we find
$\tilde{b} \ll 1$. An example is given in Sec.~\ref{Numerics_th_clean}.  Secondly, as
$\xi^T$ is finite, $\tilde{c} \rightarrow \infty$.

According to this discussion,  in the limit $T\rightarrow T_{\rm c0}$ and near $B_{\rm c2}$,
we can set $\tilde{c}\rightarrow \infty$ and $\tilde{b}\rightarrow 0$. This means that
\begin{eqnarray}
u_\ell = \tilde{c}{(2 \ell+1)\over \sin\theta},
\label{eq:DelrUell:GL}
\end{eqnarray}
and using Eqs. \ref{eq:Delr:lim:Tc:1} and \ref{eq:Math:4:1},
\begin{eqnarray}
f_\ell(\theta) \simeq - n_{{\bf K}_{m,h}}^2 iv''(iu_\ell) \simeq 2n_{{\bf K}_{m,h}}^2
\frac{\sin^3\theta}{\tilde{c}^3(2\ell+1)^3}.
\end{eqnarray}
Taking these results into account, we derive
\begin{multline}
B^Z_{{\bf K}_{m,h}} =  2{\tilde{a}\over \tilde{c}^3} (-1)^{mh} \exp \left ( - n^2_{{\bf K}_{m,h}}\right )\times\\
\left[ \sum_{\ell =0}^\infty (2\ell +1)^{-3} \right] \times \left[ \int_0^{\pi/2} \sin^3\theta {\rm d} \theta\right] \\
=  1.4024 {\tilde{a}\over \tilde{c}^3} (-1)^{mh} \exp \left ( - n^2_{{\bf K}_{m,h}}\right ).
\label{eq:DelrBk:GL}
\end{multline}
Here, we have used the results $\int_0^{\pi/2}\sin^3\theta {\rm d}\theta = 2/3$ and
$\sum_0^{\infty}(2\ell+1)^{-3}=7\zeta(3)/8=1.0518$, with $\zeta(s)$ denoting the Riemann Zeta function.
As can be seen from the last equation, for $T\rightarrow T_{\rm c0}$ the form factor converges to
the GL solution. It is  proportional to the factor $\tilde{a}/\tilde{c}^3$.
On the other hand, as it can be seen from Eq. \ref{eq:Dlr:FF:Expr5},
in the low temperature limit near $B_{\rm ext}\simeq B_{\rm c2}$
({\it i.e.} $\tilde{c}\rightarrow 0$ and $\tilde{b}\rightarrow 0$)
$B^Z_{{\bf K}_{m,h}}$ is proportionnal to $\tilde{a}/\tilde{c}$:
\begin{eqnarray}
B^Z_{{\bf K}_{m,h}} =
{\sqrt {\pi} \over 2} {{\tilde a} \over {\tilde c}}
{(-1)^{mh} \over n^3_{{\bf K}_{m,h}}}.
\label{eq:DelrBk:GLextra}
\end{eqnarray}
Thus, the two limits being proportionnal to  $\tilde{a}/\tilde{c}$, it is convenient to
use this field scale as units of field.

In conclusion, considering the region very close to $B_{\rm c2}(T)$ and the limits
near $T_{\rm c0}$ and $T = 0$, we find that $\tilde b$ is  very small.
On the other hand, $\tilde{c}$ is large in the first limit and negligible in the second.
Consequently $B^Z_{{\bf K}_{m,h}}$, and therefore the field map and distribution,
is expected to strongly depend on $\tilde{c}$. This fact is used
in Sec.~\ref{Numerics_th_clean} for the study of the crossover from GL to BCS vortex structures.

\subsection{The form factor for a finite mean-free-path}
\label{Reduced_dirty}

In the case of a finite electronic mean-free-path, i.e. of a finite scattering rate $1/\tau_{\rm imp}$,
the  form factor
depend on the four parameters $\tilde{a}$, $\tilde{b}$, $\tilde{c}$, and $\tilde{d}$ as follows:
\begin{equation}
B^Z_{{\bf K}_{m,h}} =  \tilde{a}{ (-1)^{mh} \exp \left ( - n^2_{{\bf K}_{m,h}}\right )
\over  n^2_{{\bf K}_{m,h}}      }
\sum_{\ell =0}^\infty \int_0^{\pi/2}  {f_\ell(\theta) \over (1-{\varepsilon}_{\ell})^2} {\rm d} \theta,
\label{eq:DelrBk:dirty}
\end{equation}
where
\begin{equation}
f_\ell(\theta) =
{iv\left (iu_\ell + i n_{{\bf K}_{m,h}} \right ) + iv\left (iu_\ell - i n_{{\bf K}_{m,h}} \right ) -2 iv\left (iu_\ell \right )
\over  \left [1 + \tilde{b}{ v^\prime\left (iu_\ell \right ) \over (1-\varepsilon_{\ell})^2\sin^2 \theta} \right ]^3}.
\label{eq:DelrFell:dirty}
\end{equation}
The $u_{\ell}$'s are found to be the solution of the equation:
\begin{equation}
u_\ell = {\tilde{c}(2 \ell+1) + \tilde{d} \over \sin\theta}  +
{ \tilde{b} \over (1-{\varepsilon}_{\ell})^2 \sin^2 \theta} i v(i u_\ell),
\label{eq:DelrUell:dirty}
\end{equation}
with
\begin{eqnarray}
{\varepsilon}_{\ell} =\frac{1}{2}
\tilde{d}\int_0^{\pi /2} iv\left(i{\tilde{c}(2 \ell+1) + \tilde{d} \over 2\sin\theta} \right)d\theta.
\label{eq:DelrEpsEll:dirty}
\end{eqnarray}
Note, $\lim_{\tilde{d}\rightarrow\infty} {\varepsilon}_{\ell} = 1$.
The parameters $\tilde{a}$, $\tilde{b}$ and $\tilde{c}$ are the same as in the clean limit case.
We have defined the dimensionless scattering parameter,
\begin{equation}
\tilde{d} = {\Lambda \over v_{\rm F} \tau_{\rm imp} } \simeq {1 \over b^{1/2}}
{\xi_{\rm GL} \over v_{\rm F} \tau_{\rm imp}},
\end{equation}
which is approximately the ratio of intervortex distance to the
electronic mean free path.
Hence, relative to the clean limit case, the effect of impurities and defects
is taken into account with only a single new parameter, i.e.~$\tilde{d}$.

In the dirty limit we have $\tilde{d}\gg 1$. This implies
\begin{eqnarray}
u_\ell \simeq {\tilde{c}(2 \ell+1) + \tilde{d} \over \sin\theta }.
\label{eq:DelrUell:dirty:GL}
\end{eqnarray}
Eq. \ref{eq:DelrEpsEll:dirty} converges to:
\begin{eqnarray}
{\varepsilon}_{\ell} =\frac{1}{2}
\tilde{d}\int_0^{\pi /2} {2\sin\theta \over \tilde{c}(2 \ell+1) + \tilde{d}} d\theta = {\tilde{d} \over \tilde{c}(2 \ell+1) + \tilde{d}}.
\end{eqnarray}
Here we used Eq. \ref{eq:Math:2} and $\int_0^{\pi/2}\sin x {\rm d}x=1$.
Therefore,
\begin{eqnarray}
(1-{\varepsilon}_{\ell})^{-2} =\left( {2 \ell+1 + \tilde{d}/\tilde{c} \over 2\ell +1  }\right)^2.
\end{eqnarray}
The denominator of Eq. \ref{eq:DelrFell:dirty} converges to:
\begin{equation}
 \left [1 + \tilde{b}{ v^\prime\left (iu_\ell \right ) \over (1-\varepsilon_{\ell})^2\sin^2 \theta} \right ]^3\simeq \left[1+
 \frac{\tilde{b}}{\tilde{c}^2}{1\over(2\ell+1)^2}\right]^3,
\end{equation}
and using Eqs. \ref{eq:Delr:lim:Tc:1} and \ref{eq:Math:4:1},
\begin{equation}
f_\ell(\theta) \simeq 2 n_{{\bf K}_{m,h}}^2 {\sin^3\theta \over
\left[\left( \tilde{c} (2\ell +1) +\tilde{d}  \right) \left(1+  \frac{\tilde{b}}{\tilde{c}^2}{1\over(2\ell+1)^2}\right)\right]^{3}}.
\label{eq:DelrFell:dirty:GL}
\end{equation}
This means for the form factor:
\begin{equation}
B^Z_{{\bf K}_{m,h}} \simeq  \tilde{A} (-1)^{mh} \exp \left ( - n^2_{{\bf K}_{m,h}}\right ),
\label{eq:DelrBk:dirty:GL}
\end{equation}
with
\begin{align}
\tilde{A}={4\over 3}\tilde{a}\tilde{c}^3 \times
&\sum_{\ell =0}^\infty\left[ (2\ell +1+\tilde{d}/\tilde{c})^{-1}(2\ell+1)^{-2}\times \right. \nonumber \\
& \left. (\tilde{c}^2+  \tilde{b}{/(2\ell+1)^2})^{-3} \right] .
\end{align}

Thus, as expected, no matter the temperature, if the scattering parameter $\tilde{d}$ is large the
VL properties are described by the GL model.


\end{document}